\documentclass[a4,11pt]{article}

\usepackage{soul}

\usepackage{amsmath}  
\usepackage{cancel}
\usepackage{hyperref}
\usepackage{url}

\usepackage{amssymb}  
\usepackage{float} 
\usepackage{xcolor} 
\usepackage{graphicx} 
\usepackage{cite} 
\usepackage[normalem]{ulem}
\usepackage{bbm}
\usepackage{dsfont}
\usepackage[mathscr]{euscript}
\usepackage{eufrak}
\usepackage{ntheorem}
\usepackage{wrapfig}
\usepackage{array}
\usepackage{hhline}
\usepackage{tikz}
\usetikzlibrary{shapes,snakes,decorations.pathreplacing,calligraphy}
\newcommand{\widl}{0.3}

 \newcommand{\Yvec}{{\delta Y}}
 \newcommand{\rot}{{\operatorname{rot}}}

 \newcommand{\newH}{{\bf H}}
 
 \newcommand{\Hvol}{{\frak h}}
 \newcommand{\Hbd}{{ \hat{\Hvol}}}
 
 \newcommand{\nomu}{{\alpha}}
 \newcommand{\nonealpha}{{\beta}}
 
 \newcommand{\omcL}{{\,\,\overline{\!\!{\mcL}}}}

 \newcommand{\paeqref}[1]{{\eqref{#1}, p.~\pageref{#1}}}
 
 \newcommand{\prefc}[1]{{\ref{#1}, p.~\pageref{#1},}}

 \newcommand{\threeg}{{\gamma}}

 \DeclareMathOperator{\sech}{sech}
 \newcommand{\spherem}{{\check \gamma}}
 
 \newcommand{\cphi}{{\chi}}
 
 \newcommand{\sD}{{\check{D}}}
 
 \newcommand{\xdS}{{x}}

 \newcommand{\mcF}{{\mcH^0}}
 
 \newcommand{\wtx}{{n}}
 \newcommand{\myLie}{{\mathbf{L}}}

 \newcommand{\pmass}{{+m^2}}

 \newcommand{\omtwo}[1]{\overset{\mbox{\tiny (2)}}{#1}}
 \newcommand{\omone}[1]{\overset{\mbox{\tiny (1)}}{#1}}
 \newcommand{\ozero}[1]{\overset{\mbox{\tiny (0)}}{#1}}

 \newcommand{\otwo}[1]{\overset{\mbox{\tiny (-2)}}{#1}}
 \newcommand{\othree}[1]{\overset{\mbox{\tiny (-3)}}{#1}}
 
 \newcommand{\oone}[1]{\overset{\mbox{\tiny (-1)}}{#1}}

 \newcommand{\mypi}{{\pi}}
 \newcommand{\tpi}{{\tilde\mypi}}

 \newcommand{\R}{\mathbb{R}}

 \newcommand{\lapseTB}{{N}}

 \newcommand{\zhTBW}{{\gamma}}
 \newcommand{\zzhTBW}{{\mathring{\zhTBW}}}

 \newcommand{\zh}{{\zzhTBW }}

 \newcommand{\mcM}{{\mycal M}}
 \newcommand{\mcN}{{\mycal N}}

 \newcommand{\sectionofScri}%
 {{ \,\,\,\,\mathring{\!\!\!\!\mcN}}}
 
 \newcommand{\hyp}{{\mycal S}}
 \newcommand{\mcL}{{\mycal L}}

 \newcommand{\scri}{{\mycal I}}%
 \newcommand{\scrip}{\scri^{+}}%
 \newcommand{\Scri}{\scri}

 \newcommand{\eq}[1]{(\ref{#1})}

 \newcommand{\eeal}[1]{\label{#1}\end{eqnarray}}

\newtheorem{theorem}{\sc  Theorem\rm}[section]

\newtheorem{proposition}[theorem]{\sc Proposition\rm}

\DeclareFontFamily{OT1}{rsfs}{}
\DeclareFontShape{OT1}{rsfs}{m}{n}{ <-7> rsfs5 <7-10> rsfs7 <10-> rsfs10}{}
\DeclareMathAlphabet{\mycal}{OT1}{rsfs}{m}{n}

\definecolor{applegreen}{rgb}{0.55, 0.71, 0.0}
\definecolor{armygreen}{rgb}{0.29, 0.33, 0.13}

\definecolor{caribbeangreen}{rgb}{0.0, 0.8, 0.6}

\newcounter{mnotecount}[section]

\renewcommand{\themnotecount}{\thesection.\arabic{mnotecount}}

\newcommand{\mnotex}[1]
{\protect{\stepcounter{mnotecount}}$^{\mbox{\footnotesize
	$
	\bullet$\themnotecount}}$ \marginpar{
\raggedright\tiny\em
$\!\!\!\!\!\!\,\bullet$\themnotecount: #1} }

\newcommand{\ptc}[1]{\mnotex{{\bf ptc:} {  #1}}}

\newcommand{\bel}[1]{\begin{equation}\label{#1}}
\newcommand{\bea}{\begin{eqnarray}}
\newcommand{\bean}{\begin{eqnarray}\nonumber}
	\newcommand{\beal}[1]{\begin{eqnarray}\label{#1}}
		\newcommand{\eea}{\end{eqnarray}}

	\newcommand{\qed}{\hfill $\Box$}
	\newcommand{\qedskip}{\hfill $\Box$\medskip}
	\newcommand{\proof}{\noindent {\sc Proof:\ }}
	\newcommand{\be}{\begin{equation}}
		\newcommand{\eeq}{\end{equation}}
	\newcommand{\ee}{\end{equation}}
\newcommand{\beqa}{\begin{eqnarray}}
	\newcommand{\eeqa}{\end{eqnarray}}
\newcommand{\beqan}{\begin{eqnarray*}}
	\newcommand{\eeqan}{\end{eqnarray*}}
\newcommand{\ba}{\begin{array}}
	\newcommand{\ea}{\end{array}}

\newcommand{\const}{\mathrm{const}} 

\newcommand{\ptcheck}[1]{\ptc{checked on #1}}

\def\beq{\begin{eqnarray}}
	\def\eeq{\end{eqnarray}}

\def \C{{\mathbb{C}}}

\def\a{\alpha}
\def\b{\beta}

\def\be{\begin{equation}}
	\def\ee{\end{equation}}
\def\bea{\begin{eqnarray}}
	\def\eea{\end{eqnarray}}

\newcommand{\zspaceD}{ {\mathring D}}

\newcommand{\mcC}{\mycal C}
\newcommand{\mcE}{\mycal E}
\newcommand{\mcH}{\mycal H}

\newcommand{\nobarg}{{g}}

\newcommand{\Lie}{{\mathcal{L}}}

\newcommand{\difH}{\Delta \mcH}

\newcommand{\TSF}{{\mathcal{F}}}

\newcommand{\cbk}{\color{black}}
\newcommand{\TSred}{\color{black}}

\newcommand{\Keig}{\lambda}
\newcommand{\Hfuna}{\mathtt{h}}
\newcommand{\Hfunb}{\mathtt{F}^2}
\newcommand{\HfunaI}{\mathtt{h}_{I}}
\newcommand{\HfunaII}{\mathtt{h}_{II}}
\newcommand{\HfunA}{\mathtt{h}_{A}}
\newcommand{\Hscaa}{\mathtt{h}}
\newcommand{\Hscab}{\mathtt{h}_{L}}
\newcommand{\HscaaI}{\mathtt{h}_{I}}
\newcommand{\HscaaII}{\mathtt{h}_{II}}
\newcommand{\HscaA}{\mathtt{h}_{A}}
\newcommand{\SField}{{\phi}}
\newcommand{\LKmom}{\mathcal{P}_{d S}}
\newcommand{\LKbst}{\mathcal{L}_{d S}}
\newcommand{\LKmcon}{p}
\newcommand{\LKbcon}{l}
\newcommand{\PadS}{\mathcal{P}_{a d S}}
\newcommand{\LadS}{\mathcal{L}_{a d S}}
\newcommand{\PadSc}{\widetilde{p}}
\newcommand{\LadSc}{\widetilde{l}}
\newcommand{\Sinf}{S_{\infty}}
\newcommand{\aldS}{\widetilde{\alpha}}

\newcommand{\amcL}{\widetilde{\mcL}}
\newcommand{\aSField}{\widetilde{\SField}}
\newcommand{\api}{\widetilde{\pi}}
\newcommand{\amcH}{\widetilde{\mcH}}
\newcommand{\amcE}{\widetilde{E}}
\newcommand{\amcJ}{\widetilde{J}}
\newcommand{\amcP}{\widetilde{P}}
\newcommand{\amcC}{\widetilde{C}}
\newcommand{\amcT}{\widetilde{T}}

\newcommand{\atanh}{\mathop {\rm areatanh}\nolimits }

    \usepackage{tikz}
    \usetikzlibrary{shapes,snakes}

{\catcode `\@=11 \global\let\AddToReset=\@addtoreset}
\AddToReset{equation}{section}

{\catcode `\@=11 \global\let\AddToReset=\@addtoreset}
\AddToReset{figure}{section}

{\catcode `\@=11 \global\let\AddToReset=\@addtoreset}
\AddToReset{table}{section}

\renewcommand{\ptcheck}[1]{}

\begin{document}

\title{{ Hamiltonian charges on light cones for linear field theories on (A)dS backgrounds\thanks{Preprint UWThPh-2023-7}
}
}

\author{Piotr T. Chru\'sciel\\
Gravitational Physics\\
University of Vienna
\\
\\
Tomasz Smo\l ka
\\
Max Planck Institute for Gravitational Physics
}

\maketitle

\begin{abstract}
We analyse the Noether charges for scalar and Maxwell fields on light cones on a de Sitter, Minkowski, and anti-de Sitter backgrounds. Somewhat surprisingly, under natural asymptotic conditions all charges for the Maxwell fields on both the de Sitter and anti-de Sitter backgrounds are finite. On the other hand, one needs to renormalise the charges for the conformally-covariant scalar field when the cosmological constant does not vanish. In both cases well-defined  renormalised charges, with well-defined fluxes, are obtained. Again surprisingly,   a Hamiltonian analysis of a suitably rescaled scalar field leads to finite charges, without the need to renormalise.  Last but not least, we indicate natural phase spaces where the Poisson algebra  of charges is well defined.
\end{abstract}

\tableofcontents

\section{Introduction and summary}
 \label{s30VII21.1}

In field theory it is commonplace to identify the total energy of a field configuration with the Hamiltonian charge, also known as the Noether charge, associated with time translations. Consider, then, a field theory on a Minkowski, de Sitter, or anti-de Sitter background. When the cosmological constant $\Lambda$ is negative the notion of time translation is somewhat muddled by the fact that there are no globally timelike Killing vector fields. However, in all the above spacetimes, given a light-cone, there   exists a family of  Killing vectors which are timelike at its tip,  generating flows which move  isometrically the whole light-cone to its future. The associated Hamiltonian provides  a good candidate for the definition of total energy contained in the light-cone; the resulting formula coincides with the usual definition of energy when the cosmological constant vanishes.

It should be kept in mind that the problem of real interest  is the   full nonlinear theory, including the gravitational field, in the presence of a cosmological constant. While some progress towards the understanding of that problem has been done~\cite{Compere,ChIfsits,Freidel:2021dxw,PooleSkenderisTaylor},
 there remain ambiguities which are far from understood. Therefore a systematic analysis of the simpler problem, of linear fields on a fixed background, appears in order.

In recent work~\cite{ChHMS} we analysed the Hamiltonian charge associated with a timelike translation of tips of light cones for the scalar field and the linearised gravitational field on the  backgrounds just listed. Much to our surprise, we found that the charge integrals diverge when the cosmological constant does not vanish. We proposed a renormalisation procedure that led to finite  charges, with well defined flux integrals. The aim of this work is to analyse similar charges associated with the flow of the remaining Killing vector fields for the conformally-covariant scalar field  and for the Maxwell field on these backgrounds.

Somewhat surprisingly, we find that all  resulting charges for the Maxwell field are finite. On the other hand, when $\Lambda\ne 0$  the charges for the scalar field need to be renormalised again, except for angular momentum where the divergent terms in the integrand integrate-out to zero on spheres. After renormalisation one obtains a well defined set of charges, with well defined flux formulae.

As a byproduct, we find an alternative Lagrangian for the scalar field which leads to finite charges, without need for renormalisation. The alternative Lagrangian depends explicitly upon the coordinates, and leads to different global charges. This raises the question of physical significance and relevance of the resulting expressions, and we do not have an answer for this.
 The point of view advocated by Kijowski \cite{KijowskiGRG,JKW},
  that different energy expressions correspond to different sets of boundary conditions, does not seem to be helpful for radiating systems.

Given a full set of charges of the scalar field and the Maxwell field, it is tempting to enquire about their algebra. One is then faced with the problem of boundary terms in the variational formulae, which appear to obstruct a meaningful definition of a Poisson bracket. One way out is to work in phase-space sectors where the boundary terms vanish by choice of boundary conditions. But then the charges are defined only up to a functional which depends upon the boundary data, and there does not exist a clear principle to single-out a preferred one.
Here we propose a simple solution, to extend the phase space to include the boundary degrees of freedom.
For yet another proposal, see~\cite{SolovyevI}.

 \bigskip

We now pass  to a more detailed summary of our results.


\subsection{Scalar fields}

On Minkowski, de Sitter and Anti-de Sitter spacetime  we consider a scalar field with Lagrangian
\begin{equation}\label{S9VII21.1}
	\mcL =- \frac 12  \sqrt{|\det g|}
	\big(
	g^{\mu\nu}\partial_\mu \SField \, \partial_\nu \SField
	+ \frac{2 \Lambda}3 \SField^2
	\big)
	\,,
\end{equation}
where $\Lambda$ is the cosmological constant. The mass term is chosen so that the resulting field equation is conformally covariant. We consider fields with the following asymptotic behaviour, for large $r$,
\begin{equation}\label{S13VIII21.t1}
	\SField(u,r, x^A) = \frac{\oone \SField(u,x^A)}{r}
	+ \frac{\otwo \SField(u,x^A)}{r^2}
	+ \frac{\othree \SField(u,x^A)}{r^3} + ...
	\, ,
\end{equation}
which can be justified by an analysis of the Cauchy problem for the field equations.
Indeed, if the asymptotic expansion \eqref{S13VIII21.t1} is imposed on an initial light cone, it is preserved by evolution when $\Lambda \ge 0$. This is also the case when $\Lambda < 0$ after requiring that the associated solutions vanish at the conformal boundary at infinity.

Consider the Noether charge associated with a vector field $X$ and a hypersurface $\hyp$ (we follow the formalism of \cite{KijowskiTulczyjew}):
\begin{eqnarray}
	\mcH [X, \hyp ]
	& := &
	\int_\hyp
	\big(
	\underbrace{
		{\pi}{}^\mu \Lie_X {\phi} - X^\mu \mcL}
	_{=:\mcH^\mu[X]}
	\big)
	dS_\mu
	\label{7X22.1h}
	\\
	&
	= &
	\frac 12
	\left( \int_\hyp
	\omega^\mu({\phi}, \Lie_X {\phi})
	\, dS_\mu
	{-}
	\int_{\partial \hyp }
	X^{[{{\sigma}}}
	{\pi}{}^{\mu]}  {\phi}
	dS_{{{\sigma}}\mu}
	\right)
	\,,
	\phantom{xxxxxxx}
	\label{S6IX18.5+}
\end{eqnarray}
with
\begin{eqnarray}
	\pi^\mu = \frac{\partial \mcL}{\partial \phi_{,\mu}}
	\,,
	\qquad
	\omega^\mu({\phi}, \Lie_X {\phi})=
	{\pi} {}^\mu
	\Lie_X {\phi}
	-
	{\phi}
	\,
	\Lie_X {\pi} {}^\mu
	\,.
	\nonumber
\end{eqnarray}	
Here \eqref{7X22.1h} is a definition while \eqref{S6IX18.5+} is an identity for linear field theories, compare~\cite[Proposition~1]{ChHMS}; the reader is referred to the text below for notation that has not been defined so far.

\subsubsection{The energy}

Let us denote by $\mcC_{u}$ the light cone of constant retarded time $u$. We use the symbol $\mcC_{u,R} $ to denote the truncation of $\mcC_u$ to $r\le R$:
$$\mcC_{u,R}= \mcC_{u}\cap \{r\le R\}
\,.
$$

It turns out that the Noether charge on $\mcC_{u,R}$, associated with translations in $u$, diverges as $R$ tends to infinity. A direct analysis of the integrand gives  (cf.~\eqref{24V21.t1as} below)
\begin{eqnarray}
	\nonumber
	\lefteqn{
		{E _\mcH }[  \mcC_{u ,R}]  :=
		\int_{\mcC_{u,R }}  \mcH^\mu [\partial_u] dS_\mu
	}
	&&
	\\
	&=&
	\frac{1}{2}
	\int_{\mcC_{u,R }}
	\Big(
	\zh^{A B}\zspaceD_{A} \SField \zspaceD_{B} \SField
	+
	m^{2} r^{2} \SField^{2}
	\nonumber
	\\
	& &
	+
	\partial_{r}
	\Big[
	\big(r^{2} - \alpha^{2} r^{4}\big) \SField \big(\partial_{r} \SField\big)
	\Big]
	-
	\SField
	\partial_{r}
	\big[
	\big(r^{2} - \alpha^{2} r^{4}\big) \big(\partial_{r} \SField\big)
	\big]
	\Big)
	\,dr \, d\mu_{\zzhTBW}
	\nonumber
	\\
	&=&
	\frac{\alpha^{2} R }{2}  \int_{S_R}
	(\oone \SField)^{2}  \, d\mu_{\zzhTBW}
	+ O(1)
	\,,
	\phantom{xxxxx}
	\label{S24V21.t1as}
\end{eqnarray}
where $S_R$ denotes a sphere  $r=R$ within $\mcC_u$,
with
$$
\alpha^2 = \frac{\Lambda}{3}
\,,
$$
and
where $O(1)$ here denotes a volume integral which has a finite limit as $R\to\infty$. A finite renormalised Noether charge can then be obtained by discarding the divergent boundary integral:
\begin{equation}\label{8X22.3}
	\hat
	E _\mcH  [  \mcC_{u }]  :=
	\lim_{R\to\infty}
	\left\{
	\int_{\mcC_{u,R }}  \mcH^\mu [\partial_u] dS_\mu -
	\frac{\alpha^{2} R }{2}  \int_{S_R}
	(\oone \SField)^{2}  \, d\mu_{\zzhTBW}
	\right\}
	\,.
\end{equation}
The divergent term  evolves on its own, so that the renormalised Noether charge  follows a well defined evolution law, as derived below in \eqref{8X22.1}:
\begin{eqnarray}
	\frac{d \hat E_\mcH  [ \mcC_{u  }]}{d u }
	& = &
	\int_{S_{R}}
	\Big[ \alpha^{2} {\oone \SField}\partial_{u} {\otwo \SField}+\left(2 \alpha^{2} {\otwo \SField}-\partial_{u} {\oone \SField}\right)\partial_{u} {\oone \SField}
	\Big]
	d\mu_{\zzhTBW}
	\, .
	\phantom{xxxxx}
	\label{S8X22.1}
\end{eqnarray}

Now, some authors discard the divergence term in \eqref{S6IX18.5+} and use directly
\begin{eqnarray}
	E_\omega[ \hyp]
	:= \frac 12
	\int_\hyp
	\omega^\mu({\phi}, \Lie_X {\phi})
	\, dS_\mu
	\label{S6IX18.5+a}
\end{eqnarray}
as a definition of Noether charge; this gives of course the same total energy as the original formula \eqref{7X22.1h} for field configurations for which the boundary term vanishes.
In  \cite{ChHMS} we observed that the integral \eqref{S6IX18.5+a} is finite, leading to a rewriting
(see Equation~(2.64) there):%
\footnote{References to numbering in \cite{ChHMS} are to the arXiv version.\label{F20II23}}
\begin{eqnarray}
	E_\mcH[ \mcC_{u,R}]
	& = &%
	E_\omega[\mcC_{u,R}]
	- \frac{1}{2}
	\int_{S^2}
	\oone \phi
	\big( \partial_u\!\oone \phi - \alpha^2 R  \oone \phi \big) \, d\mu_{\zzhTBW}
	+ o (1)
	\,.
	\phantom{xxxx}
	\label{S27XII19.519a}
\end{eqnarray}
So, the divergent part of \eqref{S24V21.t1as} appears directly in the boundary term in \eqref{S6IX18.5+}.
Equation~\eqref{S27XII19.519a} implies
\begin{eqnarray}
	\hat E_\mcH[ \mcC_{u }]
	& = &%
	E_\omega[\mcC_{u }]
	- \frac{1}{2}
	\int_{S^2}
	\oone \phi
	\partial_u\!\oone \phi  \, d\mu_{\zzhTBW}
	\,.
	\phantom{xxxx}
	\label{S27XII19.519b}
\end{eqnarray}
The energy $ E_\omega[\mcC_{u }]$ satisfies a flux formula (see~\paeqref{2XI22.t1} below)
\begin{equation}
	\frac{d  E_\omega[ \mcC_{u }]}{du}
	=
	\frac12\int_{S^2}
	\big(
	-\alpha^2 \oone \phi \partial_u \otwo \phi +\alpha^2 \otwo \phi \partial_{u} \oone \phi +\oone \phi \partial^{2}_{u}  \oone \phi-(\partial_u \oone \phi )^2
	\big)
	\, d\mu_{\zzhTBW}
	\,.
	\label{2XI22.t1S}
\end{equation}

The fact that $E_\omega[\mcC_{u,R}]$ has a finite limit as $R\to\infty$ suggests that the resulting Noether charge $E_\omega[\mcC_{u}]$ is more fundamental than $E_\mcH$. But one should keep in mind that the equality between \eqref{7X22.1h} and \eqref{S6IX18.5+} is only guaranteed for linear theories. In fact,  \eqref{7X22.1h} is defined for any theory, whether linear or not, while \eqref{S6IX18.5+} does not make sense for nonlinear theories, such as Yang-Mills or metric gravity.
Last but not least, $E_\omega[\mcC_{u}]$ is not monotonically decreasing in asymptotically Minkowskian spacetimes, as it should; see \paeqref{2XI22.t1} below.
Therefore we view \eqref{7X22.1h} as a more  fundamental equation.

To make things even more confusing,
it turns out that the field equations for the field
$$
\aSField=r \SField
\,,
$$
where $r$ is an affine parameter along the generators of the light cone, can be derived from the Lagrangian%
\footnote{Note that the singularity at $r=0$ in \eqref{S28I2022.t3} is integrable for fields $\phi$ which are smooth at the origin. While the presence of an unbounded integrand might be aesthetically unpleasant, it does not present difficulties as far as calculus of variations is concerned.} 
\begin{equation}
	\label{S28I2022.t3}
	\mcL =- \frac{1}{2 r^2} \sqrt{|\det g|}
	\,
	g^{\mu\nu}\nabla_\mu   \aSField   \, \nabla_\nu   \aSField
	\, ,
\end{equation}
which differs from \eqref{S9VII21.1} by a boundary term, compare \eqref{28I2022.t2} below.
Somewhat surprisingly, again under the asymptotic conditions \eqref{S13VIII21.t1}. the Noether charge associated with translations in $u$ turns out to be finite (cf. \eqref{31I22.t1} below):
\begin{equation}
	\tilde E_\mcH  [  \mcC_{u }]:= {\tilde E }[\partial_u, \mcC_{u  }]
	=
	\frac{1}{2}
	\int_{\mcC_{u  }}
	\underbrace{
		\Big( \frac{1}{r^2}
		\zh^{A B}\zspaceD_{A} \aSField \zspaceD_{B} \aSField
		+
		\big(1 - \alpha^{2} r^{2}\big)\big(\partial_{r} \aSField\big)^{2}
		\Big)
	}_ { O(r^{-2})}
	\,dr \, d\mu_{\zzhTBW}
	\,.
	\label{S31I22.t1}
\end{equation}
So we have a third candidate for the energy of the conformally-covariant scalar field, with flux (cf.\ \eqref{8X22.5} below)
\begin{eqnarray}
	\frac{d \tilde E_\mcH  [  \mcC_{u }]}{d u }
	& = &
	\int_{S}
	\big(\alpha^{2} {\oone \aSField}-\partial_{u} {\ozero \aSField}\big)\partial_{u} {\ozero \aSField}
	d\mu_{\zzhTBW}
	\, .
	\label{S8X22.5}
\end{eqnarray}
The fact that the numerical value of $\tilde E_\mcH $ differs from that of both $\hat E_\mcH$  and $E_\omega$
{
	when $\alpha\ne 0$
}
is made clear by comparing \eqref{S8X22.5} with \eqref{S8X22.1} and \eqref{2XI22.t1S}:  all three fluxes differ.

The question then arises, whether  the analysis of the Poisson algebra might give a hint, which of the energy-type expressions above have better properties.
This is addressed in Section \ref{s6II21.1}.
To answer this question one needs to have a well defined Poisson algebra, which seems to be a problem when ``charges are leaky'', i.e.\ when the variations of functionals lead to nonvanishing boundary integrals. We emphasise that in our setup such boundary terms are unavoidable, because the fields under consideration radiate along light cones.

Now, it was observed in~\cite{CJK} that the charge-leaking can be remedied, in the case of (fully nonlinear) gravitational fields with $\Lambda =0$, by extending the phase space of data on the light cone by adding suitable data on the portion of $\scrip$ to the past of the intersection of the light cone $\mcC_u$ with $\scrip$. In Section~\ref{s23IV22.1} we show how to generalise the procedure from~\cite{CJK} to the conformally-covariant scalar field with $\Lambda > 0$.
For this it is convenient to introduce a coordinate system  in which the de Sitter metric
takes the form
\begin{eqnarray}
	g
	&= &
	\underbrace{
		\cosh^{2}(\alpha \tau)
	}_{=:\xdS^{-2}}
	\big(
	-{\cosh^{-2}(\alpha \tau)} d \tau^{2}
	+  \alpha^{-2}
	\underbrace{
		\left(d \psi^{2}+\sin ^{2}\psi \left(d \theta^{2}+\sin ^{2}\theta d \varphi^{2}\right)\right)
	}_{=:\spherem}
	\big)
	\nonumber
	\\
	&= &
	(\alpha\xdS)^{-2}
	\big(
	-  \frac { d \xdS^{2}} {1-\xdS^2}
	+
	\spherem
	\big)
	\, ,
	\label{S30VII12.2}
\end{eqnarray}
with $x$ vanishes on $\scrip$. Under our asymptotic conditions above the field
\begin{equation}\label{1VIII22.1}
	\cphi:= \frac{\SField}{\xdS}
\end{equation}
extends smoothly to $\scrip$, and the expansion
\eqref{S13VIII21.t1}
translates to, for small $x$,
\begin{equation}\label{30VII22.9}
	\SField =   {\omone \SField} \xdS +  {\omtwo \SField} \xdS^2  + \ldots
	\qquad
	\Longleftrightarrow
	\qquad
	\cphi=   {\ozero \cphi}   +  {\omone \cphi} \xdS   + \ldots
	\,,
\end{equation}

Using the fact that $\mcH^\mu $
has vanishing divergence under the current conditions, together with \eqref{4XI22.3a} and \eqref{4XI22.5}, the following equivalent equations hold
\begin{eqnarray}
	\hat E_{\mcH}[\mcC_u]
	& = &
	\alpha^{-1}  \int_{\scrip\cap I^+(\mcC_u)}
	\omone \cphi  \partial_{\psi}
	(\sin \psi  \ozero \cphi)
	d\mu_{\spherem}
	\nonumber
	\\
	&&
	-\frac{1}{ \alpha \cosh^{3} (\alpha u ) }
	\int_{S_{0,u }}
	\ozero \cphi \Big(
	\ozero \cphi \sinh(\alpha u)
	-\omone \cphi
	\Big)
	d\mu_{\zzhTBW}
	\,.
\end{eqnarray}

In the phase space of Section~\ref{ss23II23.1} the dynamical system induced by translating in $u$  the tip of the light-cone is Hamiltonian, with Hamiltonian  equal to (see \eqref{20II23.11} with $\Hvol^x$   given   by \eqref{1VIII22.2}, ${\Hvol}^u $ by   \eqref{4XI22.6}
and $\Delta \mathcal{B}_{0,u }$  by \eqref{5III23.6}): 
\begin{eqnarray}
	\nonumber 
	\newH & =&
	\int_{\mcC_{u }} \frac{1}{2 r^2}
	\Big(
	\zh^{A B}\zspaceD_{A} \SField \zspaceD_{B} \SField
	+
	m^{2} r^{2} \SField^{2}
	-
	\SField
	\partial_{r}
	\big[
	\big(r^{2} - \alpha^{2} r^{4}\big) \big(\partial_{r} \SField\big)
	\big]
	\Big)
	r^2\,dr \, d\mu_{\zzhTBW}
	\nonumber
	\\
	&&
	+
	\frac{1}{2 \alpha \cosh^{3} (\alpha u ) }
	\int_{S_{0,u }}
	\ozero \cphi
	\Big(
	\omone \cphi
	-
	\partial_{\psi}\ozero \cphi
	-
	\ozero \cphi
	\sinh(\alpha u )
	\Big)
	\sin \theta
	\, d \theta \, d \varphi
	\nonumber
	\\
	&&
	+
	\alpha^{-1}
	\int_{\scrip_u}
	\Big\{    (1-\xdS^2)
	\sin \psi \partial_\xdS  \cphi  \partial_{\psi}  \cphi
	\nonumber
	\\
	&&
	+\frac    { 1}  2
	\cos \psi
	\Big(  (1-\xdS^2) (  2  \cphi  \partial_\xdS \cphi +   \xdS
	(\partial_\xdS \cphi)^2)
	+ 2 \xdS \cphi^2
	+ \xdS  |\sD    \cphi |^2_\spherem
	\Big)
	\Big\}
	d\mu_ \spherem
	\label{5III23.5a}
	\\
	& =: &
	\check E_\mcH[\mcC_u] + \check E_\mcH[\scrip_u]
	\, ,
	\label{5III23.5b}
\end{eqnarray}
where $\check E_\mcH[\scrip_u]$ is the volume integral over $\scrip_u\equiv \scrip\setminus I^+(\mcC_u)$ in \eqref{5III23.5a},   with all  integrals    finite.


\begin{table}[t]
	\centering
	\renewcommand{\arraystretch}{1.5}
	\begin{tabular}{|c|c|c|c|c|}
		\hhline{|=|=|=|=|=|}
		& \centering $\Lambda>0$  & $\Lambda=0$ & $\Lambda<0$&   \\
		\hhline{|=|=|=|=|=|}
		$E_{\mcH}$ & $\infty$  & $<\infty, \, \frac{dE_{\mcH}}{d u}\le0$  &$\infty$&    \\
		\hline
		$E_\omega$ & $<\infty$  & $<\infty,$ $\frac{d E_\omega}{d u}$ {\small can have any sign}  &$<\infty$ &
		\\
		\hline
		$\hat E_{\mcH}$
		& $<\infty$
		& $= E_\mcH$
		& $<\infty$
		& {\small corner term ad hoc }
		\\
		\hline
		$\check E_{\mcH}$
		&   $ <\infty$
		&  (already considered in \cite{CJK})
		&
		-
		& {\small corner term ad hoc }
		\\
		\hline
		$\tilde E $ & $<\infty$ & $
			{= E_\mcH} 
		$  &$<\infty$ & { \small Lagrangian explicitly}   \\
		&
		&
		&
		& { \small coordinate dependent } \\
		\hhline{|=|=|=|=|=|}
	\end{tabular}
	\caption{Various energies for the conformally-covariant scalar field. $E_{\mcH}$ is defined in \eqref{S24V21.t1as}; $E_\omega$ is defined in \eqref{S6IX18.5+a} and differs from \eqref{S24V21.t1as} by a total divergence; similarly for $\hat E_{\mcH}$, with yet another boundary term; $\check E_\mcH$ is defined in \eqref{5III23.5b}  based on phase-space considerations and differs again from $E_{\mcH}$ by a further boundary term;
		$\tilde E $ uses the canonical definition  as in  \eqref{7X22.1h}
		but with the alternative Lagrangian \eqref{S28I2022.t3}.
	}
	\label{T5III23.1}
\end{table}

Our analysis of the scalar field can be summarised as follows, compare
Table~\ref{T5III23.1}:

\begin{enumerate}
	\item The defining equation \eqref{S24V21.t1as} for $E_\mcH$ makes sense for any theory, including non-linear ones, has the right properties when $\Lambda=0$, but does not lead  to convergence integrals when $\Lambda \ne 0$. It needs to be ``renormalised'', with ambiguities concerning the finite part of the renormalising corrections.
	\item The ``energy'' $E_\omega$ defined in \eqref{S6IX18.5+a} leads directly to finite integrals for all $\Lambda$. However,  it does not  lead to a monotonously decreasing quantity when $\Lambda=0$. Moreover it does not have any obvious generalisation to nonlinear fields.
	\item  The ``energy'' $\hat E_{\mcH}[\mcC_u]$ of \eqref{8X22.3} has several desirable properties:
	\begin{enumerate}
		\item It is finite for all $\Lambda$.
		\item It is non-increasing when $\Lambda =0$
		(since it coincides
		with $E_\mcH$ then),
		and is conserved when $\Lambda< 0$ and the standard boundary condition $\oone \SField=0$ is imposed (cf.~\eqref{S8X22.1}).
		\item It has a reasonably natural derivation, namely one   removes a manifestly divergent term in  Bondi coordinates.
	\end{enumerate}
	However, the choice of Bondi coordinates is ad-hoc, and other similar prescriptions using different coordinate systems will lead to different expressions.
	
	\item The energy $\tilde E$ given by \eqref{S31I22.t1} has properties similar to $\hat E$ (cf.  \eqref{S8X22.5}) but arises from a coordinate-dependent Lagrangian, which does not have any obvious generalisations to non-conformally-covariant theories.
	\item
	The energy $\newH$ of \eqref{5III23.5a} appears naturally when extending the phase space to include the degrees of freedom at $\scrip$, its numerical value is the same for all cones $\mcC_u$, and thus carries only global information about the field. It splits  into a volume integral on a subset of $\scrip$ and a remainder which is determined by the fields on $\mcC_u$. However, the uniqueness of this splitting is not clear.
\end{enumerate}


\subsubsection{Further charges}
\label{ss8X22.1a}

The total angular-momentum is obtained from the following integral:
\begin{eqnarray}
	J[  \mcC_{u }]&:=&
	\int_{\mcC_{u  }}  \mcH^\mu [\mathcal{R}] dS_\mu
	\equiv  R_i J^i [\mcC_{u  }]
	\,,
	\label{S5VIII21.3GWs}
\end{eqnarray}
with (cf.\ \eqref{5VIII21.3GWbs} below)
\begin{eqnarray}
	J^i[\mcC_{u }]
	&  = &
	\int_{\mcC_{u  }} r  \varepsilon^{A B} \zspaceD_{B} x^i \zspaceD_{A} \SField
	\partial_{r} \SField
	\,dr\, d\mu_{\zzhTBW}
	\, ,
	\label{S5VIII21.3GWbs}
\end{eqnarray}
which converges because a potentially divergent terms in the asymptotics of the integrand integrates out to zero.

The alternative Lagrangian leads to the same integral, in a form which is manifestly convergent, as determined in \eqref{31I22.t3} below:
\begin{eqnarray}
	\tilde J^i[\mcC_{u }]
	&  = &
	\int_{\mcC_{u }}
	\underbrace{ r^{-1}
		\varepsilon^{A B} \zspaceD_{B} x^i \zspaceD_{A} \aSField
		\partial_{r} \aSField
	}_{
		O({r}^{-2}) }
	\,dr\, d\mu_{\zzhTBW}
	=
	J^i[\mcC_{u }]
	\, ,
	\label{S31I22.t3}
\end{eqnarray}

Explicit expressions for the remaining charges associated with Killing vector fields of the background, as well as their fluxes, can be found in Sections~\ref{s27II23.4} and \ref{s27II23.2}.

\subsection{Maxwell fields}

We consider Maxwell fields on  Minkowski, de Sitter and Anti-de Sitter spacetime. Each of these spacetimes has a conformal  boundary at infinity, and we consider  fields which smoothly extend through that boundary; a large class of such solutions of the sourceless Maxwell equations exists, which can be justified by an analysis of the Maxwell equations on the conformally rescaled manifolds. An elegant explicit family of such solutions is presented in Appendix~\ref{s17VI21}, essentially due to~\cite{BlanchetDamour}.

We use the field equations to derive  the  asymptotic behaviour of various components of the field along light cones in Section~\ref{Ms22XII19.1}; in Bondi coordinates (cf.~Equation~\paeqref{8VII20.11}):
\begin{eqnarray}
	F_{ur} &= & \frac{\otwo F_{ur}}{r^{2}}
	-\frac{\zspaceD^A \otwo F_{Ar}}{r^3}
	-\frac{\zspaceD^A \othree F_{Ar}}{2 r^4}
	+\ldots
	\,,
	\label{26VI20.t2MI}
	\\
	F_{AB}&=&
	\ozero F_{AB}
	+
	\frac{\partial_{A} \otwo F_{Br}-\partial_{B} \otwo F_{Ar}}{r}%
	+ \frac{\partial_{A} \othree F_{Br}
		-\partial_{B} \othree F_{Ar}}{2 r^2 }
	+ \ldots
	\, ,
	\phantom{xxx}
	\\
	F_{u A}
	&=&
	\ozero F_{u A}
	+
	\frac{
		\alpha^2 \othree F_{A r}
		-\zspaceD_{A} \otwo F_{u r}
		-\zspaceD^{B} \ozero F_{B A}
	}{2 r}
	+ \ldots \, ,
	\label{22V21.t2MI}
\end{eqnarray}
see \eqref{26VI20.2} and below for details. 

Section~\ref{s30VII21.2} starts with an analysis of Noether currents and their flux for Maxwell theory in a general background.
In order to obtain a gauge-independent Hamiltonian, following~\cite{KijowskiTulczyjew} we use a notion of Lie-derivatives of the Maxwell potential arising from the $U(1)$-principal-bundle formulation of the theory.  The results are applied   to the de Sitter background in Section~\ref{s6VIII21.1}.
Recalling that $\mcC_{u}$ denotes the light cone of constant $u$,  a calculation leads to the following formula for the Noether charge on light cones associated with $u$-translations of $\mcC_{u}$ (cf. \paeqref{24V21.t1a}):
\begin{eqnarray}
	\nonumber
	{E _\mcH [} \mcC_{u }]
	&  = &
	\int_{\mcC_{u }}  \mcH^\mu [\partial_u] dS_\mu
	\nonumber
	\\
	& = &
	\frac{1}{16 \pi}  \int_{\mcC_{u }}
	\Big(
	\frac{1}{r^2} \zh^{AC} \zh^{B D} F_{A B} F_{C D}
	+2 F_{u r}^2%
	-2 \epsilon N^2 \zh^{A B} F_{r A} F_{r B}
	\Big)
	\,dr\,  d\mu_{\zzhTBW}
	\,,
	\nonumber
	\\
	& &
	\label{24V21.t1aSumma}
\end{eqnarray}
where the convergence of the integral follows from \eqref{26VI20.t2MI}-\eqref{22V21.t2MI}.

Likewise the components of the total angular-momentum vector are given by
convergent integrals:
\begin{eqnarray}
	J[\mathcal{R}]&:=&
	\int_{\mcC_{u }}  \mcH^\mu [\mathcal{R}] dS_\mu
	\equiv  R_i J^i
	\,,
	\label{5VIII21.3GW}
\end{eqnarray}
with
\begin{eqnarray}
	J^i&=&
	\frac{1}{4 \pi}  \int_{\mcC_{u }}  \varepsilon^{A B} \zspaceD_{B} \wtx ^i \Big(
	r^2 F_{u r} F_{A r}
	+
	\zh^{BC} F_{B r}F_{C A}
	\Big)
	\,dr\,d\mu_{\zzhTBW}
	\,.
	\phantom{xxxxx}
	\label{5VIII21.3GWb}
\end{eqnarray}
Explicit formulae for the momentum and center of mass of the field can be found in \eqref{5VIII21.3GWa}-\eqref{5VIII21.3GWaWI}.

In Section~\ref{s6VIII21.2a} we apply the formalism to light-cones in Minkowski space-time, while Section~\ref{s6VIII21.2b} is concerned with anti-de Sitter spacetime.

In Section~\ref{ss10XII22.1} we consider the time-evolution of the charges, by which we mean the evolution of the charges when the tips of the light-cones are moved along the Killing vector $\partial_u$. In particular we find the following  formulae for the flux of the energy,
\begin{eqnarray}
	\frac{d  E _\mcH [\mcC_{u}]}{d u }
	&=&
	-\frac{1}{4 \pi}\int_{\Sinf}
	\Big[ \zh^{A B}\big(\alpha^2\otwo{F}_{A r} \ozero{F}_{B u}+ \ozero{F}_{A u} \ozero{F}_{B u}\big)
	\Big]
	d\mu_{\zzhTBW}
	\, ,
	\label{8III23.99}
\end{eqnarray}
and for that of angular-momentum:
\begin{eqnarray}
	\frac{d J^{i}}{d u }
	&=&
	-
	\frac{1}{4 \pi}\int_{\Sinf} \Big[
	{  \varepsilon^{A B} \zspaceD_{B}( \wtx ^i)} \Big(
	\zh^{BC} \big(\alpha^2 \otwo{F}_{B r}+\ozero{F}_{B u}
	\big) \ozero{F}_{C A}
	\nonumber
	\\
	& &
	\phantom{-
		\frac{1}{4 \pi}\int_{\Sinf} \Big[
	}
	-\otwo{F}_{u r} \ozero{F}_{A u}
	\Big)
	\Big]d\mu_{\zzhTBW}
	\, .
\end{eqnarray}

Similarly to the scalar field case,
one can  avoid phase-space leakage for the Maxwell field by considering jointly fields on $\mcC_u$ and $\scrip\setminus I^+(\mcC_u)$. This leads to a Hamiltonian dynamics, with $u$-independent Hamiltonian (cf.~\eqref{20II23.21cf}, with $\mcH^u{[\partial_u]}$ given by \eqref{24V21.t1a} and $\mcH^x{[\partial_u]}$ by \eqref{8III23.2}
\begin{eqnarray}
	\newH &=&
	\frac{1}{16 \pi}  \int_{\mcC_{u }}
	\Big(
	\frac{1}{r^2} \zh^{AC} \zh^{B D} F_{A B} F_{C D}
	+2 F_{u r}^2%
	-2 \epsilon N^2 \zh^{A B} F_{r A} F_{r B}
	\Big)
	\,dr\,  d\mu_{\zzhTBW}
	\nonumber
	\\
	&&
	-\frac{\alpha}{4 \pi}
	\int_{\scrip\setminus I^+(\mcC_u)}
	\Big\{
	\frac{1}{2} \xdS (1-\xdS^2) \cos \psi F_{x k} F_{x l} \spherem^{k l}
	\nonumber
	\\
	& &
	\phantom{ -\frac{\alpha}{4 \pi}
		\int_{\scrip_u} }
	+ (1-\xdS^2) \sin \psi F_{x k} F_{\psi l} \spherem^{k l}
	+\frac{1}{4} x \cos \psi F_{m k} F_{n l} \spherem^{m n} \spherem^{k l}
	\Big\}
	d\mu_{\spherem}
	\, ,
	\phantom{xxxxxx}
	\label{8III23.3}
\end{eqnarray}
where all the integrals are finite, without the need for any corrections. Since $\newH$ is $u$-independent,  formula \eqref{8III23.99} describes the flow  of energy between $\mcC_u$ and $\scrip\setminus I^+(\mcC_u)$.

In absence of a clear guiding principle for adding boundary terms to the Noether charges, we have not attempted to repeat the analysis of various alternative energies, as done for the scalar field, in the Maxwell case.

\subsection{Poisson brackets}
\label{ss8X22.1}

Section~\ref{s6II21.1} is devoted to an analysis of the Poisson brackets for unconstrained fields.
As already pointed-out, a direct calculation of Poisson brackets associated to initial data on characteristic surfaces is tricky. We circumvent this problem by using the fact that, for conserved quantities, the relevant Poisson brackets can be calculated by evolving the field to a spacelike hypersurface $\hyp$ and calculating the brackets there,  using the formula
advocated in \cite{BrownHenneaux}: for functionals of the form
\begin{equation}\label{4IV23.1-i}
	F = \int_\hyp f(\phi^A,\partial_i \phi^A, \pi^A) \, dS_0
	\,,
	\quad
	G = \int_\hyp g(\phi^A,\partial_i \phi^A, \pi^A) \, dS_0
	\,,
\end{equation}
one sets
\begin{equation}
	\{F, G \}_\hyp: =
	\int_{\hyp}\left(
	\frac{\delta f}{\delta {\SField^A}} \frac{\delta g}{\delta {\pi_A}}
	-\frac{\delta f }{\delta {\pi_A}} \frac{\delta g}{\delta {\SField^A}}
	\right) d S_{0}  \, .
	\label{16II22.2PB}
\end{equation}
In Proposition~\prefc{P10XII22.2} we list a series of conditions that guarantee the equality
\begin{eqnarray}
	\left\{H_{X}, H_{Y}\right\}_{\hyp}
	= H_{[X,Y]}
	\, .
	\label{8III23.1}
\end{eqnarray}
This leads to another problem, of boundary terms arising in variational identities, which might affect equations such as \eqref{8III23.1}, and leads us to propose alternative phase spaces for the problem at hand, already mentioned above.

We turn our attention to Poisson brackets for Maxwell field in Section~\ref{ss29III22.1}. The considerations of Section~\ref{ss24II23.1}
do not apply without further due because of gauge-invariance, and the resulting constraints. We start with an ab-initio analysis, on a general spacelike hypersurface in a general spacetime, using ADM notation: in adapted coordinates such that $\hyp=\{x^0=0\}$,
\begin{eqnarray}
	&
	\displaystyle
	\threeg_{ij}:=g_{ij}
	\,,
	\quad
	N :=\frac{1}{\sqrt{-g^{00}}} \, ,
	\quad
	N_{k}
	:=
	g_{0k}
	\, .
\end{eqnarray}
We define the electric field on $\hyp$ as
\begin{equation}
	E^k = F^{k\mu } T_\mu
	\,,
\end{equation}
	where $T^\mu$ is the field of unit normals to $\hyp$, with the orientation chosen so that
	\begin{equation}
		T_\mu dx^\mu = -N  dt
		\quad
		\Longleftrightarrow
		\quad
		T = N ^{-1} ( \partial_t  - N^k  \partial_k
		)
		\,.
	\end{equation}
	The
	canonical momentum is defined by the usual formula,
	\begin{equation}
		\label{31III22.t4aSPB}
		{\pi^{\mu}}:=
		{\pi}^{\mu 0}
		=
		\frac{\partial \mcL}{\partial\left(\partial_{0} A_{\mu}\right)}
		\, .
	\end{equation}
	When the Lagragian depends only upon $F_{\mu\nu}$ the zero-component of $\pi^\mu$ vanishes, so only its space-part $\pi^k$ remains of interest.
	In the standard Maxwell electrodynamics the field $\pi^k$ is the densitised equivalent of the electric field $E^k$:
	\begin{equation}\label{3XI22.1TB}
		\pi^k
		\equiv  - \frac{1}{4\pi}   \sqrt{\det  \threeg_{ij} }  N  F^{0k}
		=  - \frac{1}{4\pi}   \sqrt{\det  \threeg_{ij} } E^k
		\,,
	\end{equation}

	Now, functionals which depend only upon $F_{\mu\nu}$, such as the Noether current, can be expressed in terms of the space-part $A_i$ of the four-potential $A_\mu$ and of the electric field. 
	For instance, in the standard Maxwell electrodynamics we have, using the ADM notation for the metric (cf. \paeqref{20IX22.t11})
	\begin{eqnarray}
		\lefteqn{
			H[\hyp,X]
			=
			\int_{\hyp} \mcH^{0} d S_{0}
		}
		&&
		\nonumber
		\\
		&=&
		\frac{1}{8 \pi} \int_{\hyp}
		\Big[ N X^{0}
		\big(  E^{k} E^{l}\threeg_{lk}
		+
		\frac 12  \threeg ^{km}\threeg ^{ln} F_{mn} F_{k l}
		+  2  N^{-1}  E^{k}N^{l} F_{lk}
		\big)
		\nonumber
		\\
		& &
		+ 2 E^{k} X^{\l} F_{l k}
		\Big] \sqrt{\det \threeg}  \, d S_{0}
		\, .
		\label{10XII22.11SPB}
	\end{eqnarray}

	Since $\pi^0$ vanishes by antisymmetry of $F^{\mu\nu}$,
	we cannot define the Poisson bracket using  \eqref{16II22.2PB} with $(\phi^A)=(A_\mu)$. Instead we set
	\begin{equation}
		\{F, G \}_\hyp: =
		\int_{\hyp}\left(
		\frac{\delta f}{\delta A_{l}} \frac{\delta g}{\delta \pi^{l}}
		-\frac{\delta f}{\delta \pi^{l}} \frac{\delta g}{\delta A_{l}}
		\right) d S_{0}  \, .
		\label{10IV22.t1PB}
	\end{equation}
	In this formula $A_0$ has become irrelevant, though it has neither been gauged away nor discarded, being part of the $U(1)$-gauge potential $A_\mu dx^\mu$.
	
	When deriving the Hamilton equations for the Maxwell field, 
	or indeed when considering \eqref{10IV22.t1PB}, 
	there arises a difficulty related to the fact that the Maxwell momenta are not arbitrary, but satisfy the Gauss constraint equation $\partial_i \pi^i=0$.
	This is addressed in Section~\ref{ss29III22.1}, both in an approach where the Lie derivative of the Maxwell potential is that of a covector field on spacetime, 
	and where the Maxwell potential is treated as a connection form on a $U(1)$-bundle. One can implement the Gauss constraint by writing
	\begin{equation}\label{12IV23.2Intro}
		\delta \pi^k = \epsilon^{k\ell m }D_\ell \Yvec_m
		\,,
	\end{equation}
	where $\Yvec_m$ is an arbitrary covector density,
	leading to the following variational identity on the set of solutions of the field equations (cf.\ \eqref{14XII22.51comp}  and \eqref{12IV23.3} with $\mcE^\mu=0$)
	\begin{eqnarray}
		0 & = &
		\int_{\hyp} \Big[
		\epsilon^{k\ell m }D_\ell \Big(\frac{\delta \mcH^{0}}{\delta {\pi^{k}}} -\myLie_{X}A_k
		\Big) \Yvec_m
		+
		\Big(
		\frac{\delta \mcH^{0}}{\delta {A_{k}}}
		+
		\Lie_{X} {\pi^{k }}
		\Big)
		\delta A_{k}
		\Big] dS_0
		\nonumber
		\\
		&&
		+
		\int_{\partial\hyp}
		\Big[
		\Big(
		\frac{\partial \mcH^{0}}{\partial {A_{k,\ell}}}
		- (X^{\ell} {\pi^{k}}
		-X^{0} \pi^{k \ell}
		- X^{k} \pi^{\ell } )\Big)
		\delta   A_{k}
		\phantom{xxx}
		\nonumber
		\\
		&&
		\phantom{ \int_{\partial \hyp} \Big[}
		+
		\Big(\frac{\delta \mcH^{0}}{\delta {\pi^{k}}} -\myLie_{X}A_k
		\Big) \epsilon^{k\ell m }  \Yvec_m
		\Big]
		dS_{0\ell}
		\,,
		\label{12IV23.3Introu}
	\end{eqnarray}
	which can be seen to reproduce the standard form of Maxwell equations in Minkowski spacetime.%
	%
		
		Section~\ref{ss3V23.51} is devoted to the Poisson algebra of Hamiltonian charges. We prove the identity (cf.\ \paeqref{25VII22.t3})
		\begin{eqnarray}
			\left\{H_{X}, H_{Y}\right\}
			&=& H_{[X,Y]}
			\nonumber
			\\
			&& +
			\int_{\hyp} \Big \{
			Y^{\beta} \Big(
			\mcE^{\kappa}   \myLie_X A_{\kappa}
			- \pi^{\lambda \kappa} [\Lie_{X}, \nabla_{\kappa}] A_{\lambda}
			- \frac{\partial \mcL}{\partial g_{\kappa \lambda}} \Lie_{X} g_{\kappa \lambda}\Big)
			\nonumber
			\\
			& &
			-X^{\beta} \Big(
			\mcE^{\kappa}   \myLie_Y A_{\kappa}
			- \pi^{\lambda \kappa} [\Lie_{Y}, \nabla_{\kappa}] A_{\lambda}
			- \frac{\partial \mcL}{\partial g_{\kappa \lambda}} \Lie_{Y} g_{\kappa \lambda}\Big)
			\nonumber
			\\
			& &
			+\big(E_{X}^{k} Y^{\mu} -E_{Y}^{k} X^{\mu}  \big)F_{\mu k}
			\Big \}
			d S_{\beta}
			\nonumber
			\\
			&&
			+
			2 \int_{\partial \hyp}   \Big(X^{[\alpha} \mcH^{\beta]}_{Y}-Y^{[\alpha} \mcH^{\beta]}_{X}+X^{[ \alpha} Y^{\beta]} \mcL\Big) dS_{\alpha\beta}
			\, .
			\phantom{xxxx}
			\label{25VII22.t3Intro}
		\end{eqnarray}
		This makes clear what fields have to vanish to obtain a closed subalgebra.

\bigskip

We now pass to the details of the above.

\section{Asymptotics of Maxwell fields along light cones}
\label{Ms22XII19.1}

In the next  section we will   apply the formalism developed in~\cite{ChHMS} to Maxwell fields on Minkowski, de Sitter and anti-de Sitter spacetimes. For this it is first necessary to derive the asymptotic behaviour of the fields under natural conditions arising from conformal invariance of the equations.


We consider simultaneously the Minkowski space-time, the de Sitter and the anti-de Sitter space-times in Bondi coordinates. In these the metric takes the form
\be
{\nobarg}\equiv {\nobarg}_{\a \b} dx^\a dx^\b = \epsilon \lapseTB^2 du^2-2du \, dr
+ r^2
\underbrace{(d\theta^2+\sin^2 \theta \, {d\varphi^2})}_{=:\zzhTBW }
\,,
\label{8VII20.11}
\ee
where
$$
\lapseTB := \sqrt{|
	(1-\alpha^2 r^2)|}
\,,
\quad
\alpha \in \big\{0,\sqrt{\frac{\Lambda}{3}}
\big\} \subset \R\cup i \R
\,,
\quad
\epsilon \in \{\pm 1\}
\,,
$$
with  $\epsilon$ equal to one if $1-\alpha^2 r^2<0$, and minus one otherwise; note that  any  $\Lambda \in \R$, is allowed, and hence $\alpha\in \C$ but $\alpha^2\in \R$.

We have
$$
g^{\alpha\beta}\partial_\alpha\partial_\beta = -2 \partial_u\partial_r -  \epsilon \lapseTB^2 (\partial_r)^2
+ r^{-2}\zzhTBW^{AB}\partial_A\partial_B
\,.
$$

For $r\to \infty$ we replace the coordinate $r$ by a new coordinate
\begin{equation}\label{26VI20.1}
	x:= r^{-1}
	\,.
\end{equation}
In this coordinate system the de Sitter metric \eqref{8VII20.11} becomes
\begin{eqnarray}
	\nonumber
	g
	&= &
	-(1-\alpha^2 r^2)du^2-2du \, dr
	+ r^2
	\underbrace{(d\theta^2+\sin^2 \theta {d\varphi^2})}_{=:\zzhTBW }
	\,,
	\\
	&= & x^{-2}
	\big(
	-(x^2-\alpha^2 )du^2+2du \, dx
	+
	\zzhTBW
	\big)
	\,.
	\label{16I20.2a}
\end{eqnarray}
The volume element is equal to
\begin{equation}
	\sqrt{-g}= x^{-4}\sqrt{\det \zzhTBW}
	\,.
\end{equation}

Conformal invariance of the Maxwell equations shows that, for solutions that evolve out of smooth initial data on some spacelike Cauchy surface in de Sitter spacetime, the $(u,x,x^A)$-components of the Maxwell field are smooth functions of $(u,x,x^A)$:
\begin{eqnarray}
	F
	&= &
	F_{xu}dx\wedge du + F_{xA}dx\wedge dx^A + F_{uA} du\wedge dx^A + \frac 12 F_{AB} dx^A \wedge dx^B
	\nonumber
	\\
	&= &
	-r^{-2}( F_{xu}dr\wedge du + F_{xA}dr\wedge dx^A) + F_{uA} du\wedge dx^A
	\nonumber
	\\
	&&     + \frac 12 F_{AB} dx^A \wedge dx^B
	\,,
	\label{26VI20.2}
\end{eqnarray}
with $F_{xu}$, etc., having full Taylor expansions in $x\equiv 1/r$ around $x=0$.
In particular the  fields $F_{Ar}$ which are associated with a conformally smooth Maxwell field have  expansions of the form
\begin{eqnarray}
	\label{26VI20.3}%
	F_{Ar} & = & - \ozero F_{Ax} r^{-2} + \ldots= \otwo F_{Ar} r^{-2} + \ldots
	\,,
\end{eqnarray}
where the expansion coefficients are functions of $u$ and $x^A$.

Those sourceless Maxwell equations which involve $r$-derivatives read
\begin{eqnarray}
	\partial_r(r^2 {  \sqrt{\det \zzhTBW}}  F^{r\mu})
	& = &
	- r^2 {  \sqrt{\det \zzhTBW}} \partial_u F^{u\mu} - \partial_A(r^2 {  \sqrt{\det \zzhTBW}}  F^{A\mu} )
	\,, \phantom{xxxx}
	\label{26VI20.5}
	\\
	\partial_r F_{\mu\nu}
	&  = &
	-  \partial_\mu F_{\nu r}
	- \partial_\nu F_{r \mu}
	\,.
	\label{22V21.1}
\end{eqnarray}
Using
\begin{equation}\label{22V21.7}
	F^{ru}=  F_{ur}
	\,,
	\quad
	F^{rA} =
	- r^{-2} \zzhTBW^{AB}( F_{uB} + \epsilon N^2 F_{rB})
	\,,
	\quad
	F^{uA} =
	- r^{-2} \zzhTBW^{AB}  F_{rB}
	\,,
\end{equation}
we find
%
\begin{eqnarray}
	\partial_r(r^2 F_{ur}) &=&   \zspaceD^A  F_{Ar}
	\,,
	\label{26VI20.6}
	\\
	\partial_{r}\left( F_{u A} +\epsilon N^2 F_{rA}\right)
	&=&
	\partial_{u} F_{A r}+ r^{-2}\zspaceD^{B} F_{BA}
	\,,
	\label{26VI20.t1}
	\\
	\partial_r F_{AB} &=& -  \partial_A F_{Br} +  \partial_B F_{Ar}
	\,,
	\label{26VI20.8}
	\\
	\partial_u F_{Ar} &=& - \partial_r F_{uA} -  \partial_A F_{ru}
	\,,
	\label{26VI20.7}
\end{eqnarray}
where \eqref{26VI20.6} and \eqref{26VI20.t1} are special cases of \eqref{26VI20.5} with $\mu=u$ and $\mu=A$. Here, and elsewhere, $\zspaceD$ denotes the covariant derivative of the metric $\zzhTBW$.

Inserting \eqref{26VI20.7} in \eqref{26VI20.t1}
one obtains
%
\begin{equation}\label{22V21.5}
	\partial_{r}\left(   2 F_{uA}-  \epsilon N^2 F_{Ar}\right)
	=
	-  \partial_A F_{ru}
	+ r^{-2}\zspaceD^{B} F_{BA}
	\,,
\end{equation}
We conclude that prescribing $F_{Ar} dx^A$ on a cone $\{u=\const\}$ allows one to determine the remaining fields on this cone by successive integrations of \eqref{26VI20.6}, \eqref{26VI20.8} and \eqref{22V21.5}. We will refer to these equations as the \emph{characteristic constraint equations}.
One can then view \eqref{26VI20.7} as an equation which determines $F_{Ar}$ ``on the next cone".

The remaining Maxwell equations have an evolution character:
\begin{eqnarray} \partial_{u} F_{r u}&=&    r^{-2}\zspaceD^{A} \left( F_{A u} + \epsilon N^2 F_{A r} \right)
	\, ,
	\label{26VI20.t8}
	\\
	\partial_u F_{AB} &=& -  \partial_A F_{Bu} +  \partial_B F_{Au}
	\,.
	\label{26VI20.t9}
\end{eqnarray}
Another evolution  equation can be obtained by subtracting
\eqref{26VI20.t1} from \eqref{26VI20.7}:
%
\begin{equation}\label{22V21.6}
	2 \partial_u F_{Ar} =
	{\TSred-} \partial_{r}\left(  \epsilon N^2 F_{Ar}\right)
	-  \partial_A F_{ru}
	- r^{-2}\zspaceD^{B} F_{BA}
	\,.
\end{equation}

Integrating (\ref{26VI20.6}) in $r$ one obtains
\begin{equation}
	F_{ur}=
	r^{-2}\int_{0}^{r} \zspaceD^{A} F_{A r} \mathrm{d} s
	\,,
\end{equation}
so that
\begin{equation}
	F_{ur}=\frac{\otwo F_{ur}}{r^{2}}
	-\frac{\zspaceD^A \otwo F_{Ar}}{r^3}
	-\frac{\zspaceD^A \othree F_{Ar}}{2 r^4}
	+\ldots
	\,,
	\label{26VI20.t2}
\end{equation}
where
\begin{equation}
	\otwo F_{ur}=
	\int_{0}^{\infty} \zspaceD^{A} F_{A r} \mathrm{d} s
	\,.
\end{equation}
\ptcheck{30VII21}
Integrating (\ref{26VI20.8}) we have
%
\begin{eqnarray}
	F_{AB}&=&
	\ozero F_{AB}
	+
	\frac{\partial_{A} \otwo F_{Br}-\partial_{B} \otwo F_{Ar}}{r}%
	+ \frac{\partial_{A} \othree F_{Br}
		-\partial_{B} \othree F_{Ar}}{2 r^2 }
	+ \ldots
	\, ,
	\phantom{xxx}
	\label{26VI20.t6}
\end{eqnarray}
where
\begin{equation}\label{21V21.5}
	\ozero F_{AB} =
	\int_{0}^{\infty}(\partial_B F_{Ar} - \partial_A F_{Br})\, ds
	\,.
\end{equation}
\ptcheck{30VII21}

Substituting \eqref{26VI20.3}, \eqref{26VI20.t2}, and \eqref{26VI20.t6} into \eqref{22V21.5}, after integration one finds
\ptcheck{30VII21; next order in the file, commented out but unchecked the copy and paste from maple, but checked the formal procedure}
\begin{eqnarray}
	F_{u A}
	&=&
	\ozero F_{u A}
	+
	\frac{
		\alpha^2 \othree F_{A r}
		-\zspaceD_{A} \otwo F_{u r}
		-\zspaceD^{B} \ozero F_{B A}
	}{2 r}
	+ \ldots \, .
	\label{22V21.t2}
\end{eqnarray}
Here the ``integration constant" $\ozero  F_{uA}$ equals
\ptcheck{30VII21; including convergence at the origin}
\begin{equation}
	\ozero  F_{uA}
	=
	\frac12\left[
	\alpha^2 \otwo F_{Ar}
	+
	\int_{0}^{\infty} \left(
	\partial_A F_{u r}
	+ s^{-2}\zspaceD^{B} F_{BA}
	\right) d s
	\right]
	.
\end{equation}

%
Inserting \eqref{26VI20.3}, \eqref{26VI20.t2}, and \eqref{26VI20.t6} into \eqref{22V21.6}, one obtains
\begin{eqnarray}
	\partial_{u} F_{A r}
	&=&
	\frac{\alpha^2 \othree F_{A r}+\zspaceD_{A} \otwo F_{u r}-\zspaceD^{B} \ozero F_{B A} }{2 r^2}
	+ \ldots
	\,.
	\label{30VII21.1}
\end{eqnarray}
Inserting \eqref{26VI20.3} and \eqref{22V21.t2}  into \eqref{26VI20.t8} leads to
\begin{eqnarray}
	\partial_{u} F_{ru } &=&  \frac{\zspaceD^{A}  \ozero  F_{A r}+\alpha^{2} \zspaceD^{A} \otwo F_{A r}}{r^2}
	+ \ldots \,.
\end{eqnarray}
Substituting \eqref{22V21.t2} into \eqref{26VI20.t9}, one finds
\ptcheck{30VII21; whole section now}
\begin{eqnarray}
	\partial_{u} F_{A B} &=& - 2 \zspaceD_{[A} \ozero F_{B] u}
	+
	\frac{\alpha^2 D_{[A} \othree F_{B] r}+\zspaceD_{[A} \zspaceD^{C} \ozero F_{B] C}}{r} + \ldots
	\,.
	\phantom{xxx}
\end{eqnarray}

\section{Noether charges in Maxwell theory}
\label{s30VII21.2}

We are ready to pass to the analysis of Noether-type currents for Maxwell fields in Minkowski, de Sitter and anti-de Sitter spacetimes. In our signature the Lagrangian reads
\begin{equation}\label{23V21.1}
	\mcL (A_{\mu}, \partial A_{\mu}) =- \frac {1}{16 \pi}  \sqrt{|-\det g|}
	g^{\mu\nu} g^{\alpha\beta} F_{\mu \alpha} F_{\nu\beta}
	\,.
\end{equation}

The theory is linear so there is no need to  make a distinction, in the notation of \cite{ChHMS}, between the fields $F_{\mu\nu}$ and $\tilde F_{\mu\nu}$.
Denoting $\partial_\nu A_\mu$ by $A_{\mu,\nu}$, the canonical momentum density reads
\begin{eqnarray}
	\pi^{\alpha \beta}=
	\frac{\partial\mcL}{\partial \big( A_{\alpha,\beta}\big)} =\frac{1}{4 \pi} \TSF^{\alpha \beta} \, ,
	\label{29V21.t2}
\end{eqnarray}
where $\TSF^{\alpha \beta}$ is a density of Maxwell tensor
\begin{equation}
	\TSF^{\alpha \beta}=\sqrt{|-\det g|}F^{\alpha \beta} \, .
	\label{28VI22.t1}
\end{equation}

The standard Noether currents, which we will denote by $\mcH^\mu_c$,   is  defined as
\begin{eqnarray}
	\mcH^\mu_c[X]
	&:= & \frac{\partial \mcL}{\partial A_{\beta,\mu} } \Lie_{X} A_{\beta}
	-
	\mcL
	X^\mu
	\nonumber
	\\
	& = &
	- \frac {1}{4 \pi}  \sqrt{|-\det g|}
	\big(   F^{\mu \beta}\Lie_{X} A_{\beta}
	- \frac 1 4 F^{\alpha \beta} F_{\alpha\beta} X^\mu
	\big)
	\, ,\label{23V21.t1old}
\end{eqnarray}
where $\Lie_X A$ denotes   Lie derivative of a covector field.

It holds that
\begin{equation}\label{18VI21.1}
	\nabla_\mu (\mcH^\mu_c[X]) =0
	\,,
\end{equation}
when $A$ satisfies the field equations and $X$ is a Killing field of the background metric.
This follows of course from a theorem of Noether, but a direct proof can be given starting with the  identity
\ptcheck{21VI21}
\begin{eqnarray}
	\Delta_{\beta \delta}(V,A)&:=&
	[\nabla_{\delta}, \Lie_{V}] A_{\beta}
	\nonumber
	\\
	&=&
	A_{\gamma}\nabla_{\delta} \nabla_{\beta} V^{\gamma}
	-
	V^{\gamma}R^{\sigma}{}_{\beta \delta \gamma} A_{\sigma}
	\, ,
	\label{18VI21.t3}
\end{eqnarray}
where $V$ is an arbitrary vector field  and  $A$ is an arbitrary one-form.
Next, if $V$ is a conformal Killing field of the background metric,
\begin{equation}
	\nabla_{(\alpha} V_{\beta)}=\Keig g_{\alpha \beta} \, ,
\end{equation}
we  have
\ptcheck{21VI21}
\begin{equation}
	\nabla_{\gamma} \nabla_{\alpha} V_{\beta}= R^{\sigma}{}_{\gamma \alpha \beta }V_{\sigma}+ \nabla_{\gamma} \Keig g_{\alpha \beta}+\nabla_{\alpha} \Keig g_{ \beta \gamma}-\nabla_{\beta} \Keig g_{\alpha \gamma}
	\, .
	\label{18VI21.t2}
\end{equation}
Substituting \eqref{18VI21.t2} into \eqref{18VI21.t3}, we obtain for any conformal Killing field V
\begin{equation}
	\Delta _{\beta \delta}(V,A)=\Delta _{(\beta \delta)}(V,A)
	= A_\beta \nabla_\delta \lambda
	+\nabla_\beta \lambda A_\delta - g_{\beta\delta } A^\gamma \nabla_\gamma \lambda
	\, .
\end{equation}
Let
\begin{equation}\label{21VI21.p1}
	j ^\mu := \frac{1}{4\pi}\nabla_\nu F^{\mu\nu}
	\,,
\end{equation}
which of course vanishes when $A$ satisfies the field equations. As is well known, a consequence of the definition \eqref{21VI21.p1} is
\begin{equation}
	\nabla_{\mu} j ^\mu=0 \, .
	\label{22VI20.1}
\end{equation}
We are ready now to calculate, for any vector field $X$, as follows:%
%
\begin{eqnarray}
	\lefteqn{
		-\frac{4 \pi}{\sqrt{|-\det g|}}\nabla_\mu( \mcH^\mu_c[X]) =
		\nabla_{\mu}(F^{\mu \nu} \Lie_{X} A_{\nu}) - \frac{1}{4} \nabla_{\mu}(X^{\mu} F^{\alpha \beta} F_{\alpha \beta})
	}
	& &
	\nonumber
	\\
	&=&
	- 4 \pi j ^\nu \Lie_{X} A_{\nu}
	+
	F^{\mu \nu}( \Lie_{X} \nabla_{\mu}A_{\nu}
	+  \Delta_{\nu \mu}(X,A))
	\nonumber
	\\
	&&- \frac{1}{4} [F^{\alpha \beta} F_{\alpha \beta} \nabla_{\mu}X^{\mu}+X^{\mu}\nabla_{\mu} (F^{\alpha \beta} F_{\alpha \beta})]
	\nonumber
	\\
	&=&
	- 4 \pi j ^\nu \Lie_{X} A_{\nu}
	+
	\frac14 \Lie_{X} \big(F^{\mu \nu} F_{\mu \nu} \big)
	-\frac{1}{2}F_{\mu}{}^{\beta} F_{\nu \beta} \underbrace{
		\Lie_{X} g^{\mu \nu}
	}_{=-2 \nabla^{(\mu} X^{\nu)}}
	\nonumber
	\\
	& &
	+ F^{\mu \nu} \Delta_{\nu \mu}(X,A)
	- \frac{1}{4} [
	F^{\alpha \beta} F_{\alpha \beta} \nabla_{\mu}X^{\mu}
	+
	X^{\mu}\nabla_{\mu} (F^{\alpha \beta} F_{\alpha \beta})
	]
	\nonumber
	\\
	&=&
	- 4 \pi j ^\nu \Lie_{X} A_{\nu}
	+
	F^{\mu \nu}  \Delta_{\nu \mu}(X,A)
	+F_{\mu}{}^{\beta} F_{\nu \beta} \Big(
	\nabla^{(\mu} X^{\nu)}
	-
	\frac{1}{4}g^{\mu \nu} \nabla_{\alpha}X^{\alpha}
	\Big)  \, .
	\phantom{xxxxx}
	\label{18VI21.t5}
\end{eqnarray}
The last line of \eqref{18VI21.t5} vanishes for all sourceless field configurations if
$X$ is a conformal Killing vector field of the background metric.

The problem with the   Hamiltonian \eqref{23V21.t1old} is its  gauge dependence.
This can be fixed by
replacing $\Lie_XA$ by
\begin{equation}
	\label{24V21.t4}
	\myLie_{X}A_{\mu} := X^{\nu}F_{ \nu \mu}
\end{equation}
(which, by the way, is a natural definition for the Lie derivative of a connection one form on a $U(1)$ principal bundle), and defining
\begin{eqnarray}
	\mcH^\mu [X]
	& := &   \frac{\partial \mcL}{\partial A_{\beta,\mu} } \myLie_{X} A_{\beta}
	-
	\mcL
	X^\mu
	\nonumber
	\\
	&= &      -\frac{1}{4 \pi} \sqrt{|-\det g|}
	\Big(F^{\mu \beta} \myLie_{X} A_{\beta}
	-
	\frac {1}{4}
	\big(
	F^{\nu \beta} F_{\nu\beta}
	\big)
	X^\mu
	\Big)
	\nonumber
	\\
	&=&
	-\frac{1}{4 \pi} \sqrt{|-\det g|}
	\Big(F^{\mu \beta} X^{\alpha} F_{\alpha \beta}
	-
	\frac {1}{4}
	\big(
	F^{\nu \beta} F_{\nu\beta}
	\big)
	X^\mu
	\Big)
	\,.
	\phantom{xxxx}
	\label{23V21.t1}
\end{eqnarray}

Let us set
\begin{eqnarray}
	\difH^{\mu} [X]  & := &
	\mcH^\mu [X]-\mcH^\mu_{c} [X]
	\nonumber
	\\
	&= &- \sqrt{|-\det g|}  j ^\mu X^\sigma A_\sigma
	+\frac{1}{4 \pi}\partial_{\beta} \big(
	\TSF^{\mu \beta} X^{\sigma} A_{\sigma}
	\big)
	\, .
	\label{29V21.t3}
\end{eqnarray}
From \eqref{22VI20.1} and \eqref{29V21.t3} we immediately find
\begin{equation}\label{18VI21.2}
	\partial_\mu( \difH^\mu [X]) = - \sqrt{|-\det g|}  j ^\mu \nabla_\mu( X^\sigma A_\sigma)
	\,.
\end{equation}
so that we again have $\partial_\mu \mcH^\mu =0$ when the field equation  $j ^\mu \equiv 0$  is satisfied and when $X^\mu$ is a Killing vector field.

Now, $\mcH^\mu_{c} [X]$ is of the form considered in \cite{ChHMS}. There an alternative form of Hamiltonian density has been derived
\cite[Proposition 1]{ChHMS}, which in our case reads
\begin{eqnarray}
		\mcH^\mu_{c} [X]
		&
		= &
		\frac 12  \omega ^\mu(A, \Lie_X A)
		+   \partial_{{\sigma}} \Big(
		X^{[{{\sigma}}}
		\pi ^{\mu]\nu }   A_\nu
		\Big)
		\,,
		\label{24V21.t5a}
	\end{eqnarray}
	with
	\begin{equation}
		\omega ^\mu(A, \Lie_X A)
		=
		\Lie_X A_{\beta}
		\,
		\pi^{\mu \beta}
		-
		\Lie_X \pi^{\mu \beta}  A_{\beta} \, .
		\label{29V21.2}
	\end{equation}
	and where $\pi^{\mu \beta}$ is given by \eqref{29V21.t2}. This rewriting does not seem to very enlightening in the case of the Maxwell field, with a gauge behaviour  even more cumbersome than that of \eqref{23V21.t1old}.

	In order to determine the flux of energy, we continue by calculating the Lie derivative of the Hamiltonian density in the direction of an arbitrary vector field $Y$.
	Recall the formula for the Lie derivative of a vector density $Z^\mu$:
	\begin{equation}\label{24V21.t3}
		\Lie_X Z^\mu = \partial_\sigma(X^\sigma Z^\mu) - Z^\sigma \partial_\sigma X^\mu
		\equiv \nabla_\sigma(X^\sigma Z^\mu) - Z^\sigma \nabla_\sigma X^\mu
		\,.
	\end{equation}
	In order to calculate  $ \Lie_{Y} \mcH^\mu[X]$ we use this formula to obtain
	\begin{eqnarray}
		\Lie_{Y} \mcH^\mu[X]
		&=&
		\nabla_{\sigma} \Big(Y^{\sigma} \mcH^\mu \Big) - \mcH^\sigma \nabla_{\sigma} Y^{\mu}
		\nonumber
		\\
		&=&
		2\nabla_{\sigma} \Big(Y^{[\sigma} \mcH^{\mu]} \Big)
		+Y^{\mu} \nabla_{\sigma}\mcH^\sigma
		\nonumber
		\\
		&=&
		2\nabla_{\sigma} \Big(Y^{[\sigma} \mcH^{\mu]} \Big)
		+Y^{\mu} \Big[
		\nabla_{\sigma}\mcH^\sigma_{c}
		+
		\nabla_\mu( \difH^\mu)
		\Big]
		\, ,
		\label{17III23.1}
	\end{eqnarray}
	where $\difH^\mu$ has been defined in \eqref{29V21.t3}. Keeping in mind that if $Z^{\alpha}$ is a vector density then $\nabla_{\alpha} Z^{\alpha} =\partial_{\alpha} Z^{\alpha}$, and substituting \eqref{18VI21.t5}, \eqref{23V21.t1} and \eqref{18VI21.2} into \eqref{17III23.1} we find
	\begin{eqnarray}
		\lefteqn{
			\frac{4 \pi}{\sqrt{|-\det g|}} \Lie_{Y} \mcH^{\mu} [X]
			=
			-2\nabla_{\sigma} \left[Y^{[\sigma} F^{\mu] \alpha}
			X^{\kappa}F_{\kappa \alpha} - \frac{1}{4} Y^{[\sigma} X^{\mu]} F^{\alpha \beta} F_{\alpha \beta} \right]
			\nonumber}
		& &
		\\
		& &
		-
		Y^{\mu} \Big\{
		- 4 \pi j ^\nu \big(
		\Lie_{X} A_{\nu}
		-\nabla_\nu( X^\sigma A_\sigma)
		\big)
		+
		F^{\mu \nu}  \Delta_{\nu \mu}(X,A)
		\nonumber
		\phantom{xxxxxxxxxxxxxx}
		\\
		&&
		+F_{\mu}{}^{\beta} F_{\nu \beta} \Big(
		\nabla^{(\mu} X^{\nu)}
		-
		\frac{1}{4}g^{\mu \nu} \nabla_{\alpha}X^{\alpha}
		\Big)
		\Big\}
		\, .
		\label{23VI21.3}
	\end{eqnarray}

\subsection{Noether charges in de Sitter spacetime}
\label{s6VIII21.1}

We wish to determine the Noether charges associated with the Killing fields \eqref{24VI21.t1}-\eqref{24VI21.t2}.
Since the Hamiltonian density \eqref{23V21.t1old} is linear in the Hamiltonian vector field $X$, each charge is given by an integral of a linear combination of the following four functionals
\begin{eqnarray}
	\mcH^{\mu} {[\partial_u]}&=&\HfunaI^{\mu}+\mathcal{T}^{\mu} \Hfunb
	\,,
	\\
	\mcH^{\mu} [\mathcal{R}]&=& \varepsilon^{A B} \zspaceD_{B}(R_i \wtx ^i)
	\HfunA^{\mu} +\mathcal{R}^{\mu} \Hfunb
	\,,
	\\
	\mcH^{\mu} [\LKmom ] &=&e^{\alpha u}\Big[\LKmcon_{i} \wtx ^i \HfunaI^{\mu}
	-\big(\alpha r +1\big)\LKmcon_{i} \wtx ^i \HfunaII^{\mu}
	\nonumber
	\\
	& &
	-\frac{\alpha r +1}{r} \zspaceD^{A}(\LKmcon_{i} \wtx ^i) \HfunA^{\mu} \Big]+ \LKmom^{\mu} \Hfunb
	\,,
	\\
	\mcH^{\mu} [\LKbst] &=&
	e^{-\alpha u}\Big[
	\LKbcon_{i} \wtx ^i \HfunaI^{\mu}
	+\big(\alpha r -1\big)\LKbcon_{i} \wtx ^i \HfunaII^{\mu}
	\nonumber
	\\
	& &
	+\frac{\alpha r -1}{r} \zspaceD^{A}(\LKbcon_{i} \wtx ^i) \HfunA^{\mu} \Big]
	+\LKbst^{\mu} \Hfunb
	\, ,
\end{eqnarray}
where $\varepsilon^{A B}$ is a two-dimensional Levi-Civita tensor (in spherical coordinates $(\theta, \phi)$ we take the sign $\varepsilon^{\theta \phi}=\frac{1}{\sin \theta} \, .$), and
\begin{eqnarray}
	\Hfuna^{\mu}[X] &= &-\frac{1}{4 \pi} \sqrt{|-\det g|}
	F^{\mu \beta} X^{\alpha} F_{\alpha \beta} \, ,
	\label{5VIII21.1C}
	\\
	\HfunaI^{\mu}&=&\Hfuna^{\mu}[\partial_{u}] \, ,
	\label{7VII21.t1}
	\\
	\HfunaII^{\mu}&=&\Hfuna^{\mu}[\partial_r]
	\, ,
	\\
	\HfunA^{\mu}&=&\Hfuna^{\mu}[\partial_A] \, ,
	\\
	\Hfunb &=& \frac{1}{16 \pi} \sqrt{|-\det g|} F^{\nu \beta} F_{\nu\beta} \, .
	\label{7VII21.t2}
\end{eqnarray}
Written-out in detail, the functionals \eqref{7VII21.t1}-\eqref{7VII21.t2}  read
\begin{eqnarray}
	\HfunaI^{u}&=&\frac{1}{4 \pi} \big(
	r^2 F_{u r}^2
	+
	\zh^{AB} F_{u A} F_{r B}
	\big)\sqrt{\det \zzhTBW }
	\, ,
	\\
	\HfunaI^{r}&=&\frac{1}{4 \pi} \big(
	\epsilon N^2 \zh^{AB} F_{r A} F_{u B}
	+
	\zh^{AB} F_{u A} F_{u B}
	\big)\sqrt{\det \zzhTBW }
	\, ,
	\\
	\HfunaII^{u}&=& \frac{1}{4 \pi}
	\zh^{AB} F_{r A} F_{r B}\sqrt{\det \zzhTBW }
	\, ,
	\\
	\HfunaII^{r}&=& \frac{1}{4 \pi} \Big(
	r^2 F_{u r}^2
	+ \zh^{AB} F_{u A} F_{r B}
	+ \epsilon N^2 \zh^{AB} F_{r A} F_{r B}
	\Big)\sqrt{\det \zzhTBW }
	\, ,
	\\
	\HfunA^{u}&=&\frac{1}{4 \pi} \Big(
	r^2 F_{u r} F_{A r}
	+
	\zh^{BC} F_{B r}F_{C A}
	\Big)\sqrt{\det \zzhTBW }
	\, ,
	\\
	\HfunA^{r}&=&\frac{1}{4 \pi} \Big(
	r^2 F_{u r} F_{u A}
	-
	\epsilon N^2 \zh^{BC} F_{r B}F_{C A}
	-\zh^{BC} F_{u B}F_{C A}
	\Big)\sqrt{\det \zzhTBW }
	\, ,
	\phantom{xxxxx}
	\\
	\Hfunb &=&\frac{1}{16 \pi} \Big(
	\frac{1}{r^2} \zh^{AC} \zh^{B D} F_{A B} F_{C D}
	-2 F_{u r}^2
	-2 \epsilon N^2 \zh^{A B} F_{r A} F_{r B}
	\nonumber
	\\
	& &
	-4 \zh^{A B} F_{u A} F_{r B}
	\Big)\sqrt{\det \zzhTBW }
	\, .
\end{eqnarray}

As in \cite{ChHMS} we denote by $\mcC_{u}$ the light cone of constant $u$.
One checks that all charge integrals over $\mcC_{u}$  are convergent. The most interesting charge is  the energy-like integral associated with the motion of the tip of the light cone to the future along the flow of the Killing vector $\mathcal{T} \equiv \partial_u$; recall that $\partial_u$  is timelike at the tip of the light cone so that each subsequent cone so obtained lies to the future of the preceding one.
Letting 
\begin{eqnarray}
	&
	dS_\mu := \partial_\mu \rfloor dx^{\TSred 0}\wedge \cdots \wedge dx^n\,,
	\quad
	dS_{\mu\nu} := \partial_\mu \wedge \partial_\nu \rfloor dx^{\TSred 0}\wedge \cdots \wedge dx^n
	\equiv -\partial_\mu \rfloor dS_\nu
	\,,
	\phantom{xxx}
	&
	\nonumber
\end{eqnarray}
and
\begin{equation}\label{24V21.9}
	d\mu_{\mcC} = \sqrt{\det g_{AB}} \; dr\wedge dx^2\wedge dx^3
	\,, \qquad
	d\mu_{\zzhTBW} = \sqrt{\det \zzhTBW_{AB}} \;  dx^2\wedge dx^3
	\,,
\end{equation}
we find
\begin{eqnarray}
	\nonumber
	{E _\mcH [} \mcC_{u }]
	&  := &
	\int_{\mcC_{u }}  \mcH^\mu [\partial_u] dS_\mu
	=
	\int_{\mcC_{u }}  \mcH^u [\partial_u] dS_u
	=  \int_{\mcC_{u }} \left(\HfunaI^{u}+\Hfunb
	\right)
	\,dr\,dx^2 dx^3
	\nonumber
	\\
	& = &
	\frac{1}{16 \pi}  \int_{\mcC_{u }}
	\Big(
	\frac{1}{r^2} \zh^{AC} \zh^{B D} F_{A B} F_{C D}
	+2 F_{u r}^2%
	-2 \epsilon N^2 \zh^{A B} F_{r A} F_{r B}
	\Big)
	\,dr\,  d\mu_{\zzhTBW}
	\,.
	\nonumber
	\\
	& &
	\label{24V21.t1a}
\end{eqnarray}
Likewise the total angular-momentum is obtained from the following integral:
\begin{eqnarray}
	J[\mathcal{R}]&:=&
	\int_{\mcC_{u }}  \mcH^\mu [\mathcal{R}] dS_\mu
	\equiv  R_i J^i
	\,,
	\label{5VIII21.3GWc}
\end{eqnarray}
where
\begin{eqnarray}
	J^i&:=&
	\int_{\mcC_{u }}
	\varepsilon^{A B} \zspaceD_{B} \wtx ^i
	\HfunA^{u}
	\,dr\,dx^2 dx^3
	\nonumber
	\\
	&  = &
	\frac{1}{4 \pi}  \int_{\mcC_{u }}  \varepsilon^{A B} \zspaceD_{B} \wtx ^i \Big(
	r^2 F_{u r} F_{A r}
	+
	\zh^{BC} F_{B r}F_{C A}
	\Big)
	\,dr\,d\mu_{\zzhTBW}
	\,.
	\phantom{xxxxx}
	\label{5VIII21.3GWbc}
\end{eqnarray}

For completeness we give the formulae for the remaining charges
\begin{eqnarray}
	P  [\LKmom]&:=&
	\int_{\mcC_{u }}  \mcH^\mu [\LKmom] dS_\mu
	\nonumber
	\\
	&
	= &
	\LKmcon_{i}\int_{\mcC_{u }} \left(
	e^{\alpha u}\Big[ \wtx ^i \HfunaI^{u}
	-\big(\alpha r +1\big)  \wtx ^i \HfunaII^{u}
	-\frac{\alpha r +1}{r} \zspaceD^{A}(  \wtx ^i) \HfunA^{u} +  \wtx ^i \Hfunb
	\Big]
	\nonumber
	\right)
	\,dr\,dx^2 dx^3
	\nonumber
	\\
	& = &
	\frac{1}{16 \pi} \LKmcon_{i} \int_{\mcC_{u }}
	e^{\alpha u}
	\left[
	\wtx ^i\left(
	\frac{1}{r^2} \zh^{AC} \zh^{B D} F_{A B} F_{C D}
	+2 F_{u r}^2
	\right.
	\right.
	\nonumber
	\\
	& &
	\left.
	-2 (\alpha r+1)^2 \zh^{A B} F_{A r} F_{B r}
	\right)
	\left.
	-
	4 \frac{\alpha r+1}{r} \zspaceD^{A} \wtx ^i \left(
	r^2 F_{u r} F_{A r}
	+
	\zh^{BC} F_{B r}F_{C A}
	\right)
	\right]
	\,dr\,d\mu_{\zzhTBW}
	\,,
	\nonumber
	\\
	& &
	\label{5VIII21.3GWa}
\end{eqnarray}
and
\begin{eqnarray}
	C  [\LKbst]&:=&
	\int_{\mcC_{u }}  \mcH^\mu [\LKbst] dS_\mu
	\nonumber
	\\
	&
	= &
	\LKbcon_{i}\int_{\mcC_{u }} \left(
	e^{-\alpha u}\Big[ \wtx ^i \HfunaI^{u}
	+\big(\alpha r -1\big)  \wtx ^i \HfunaII^{u}
	+\frac{\alpha r -1}{r} \zspaceD^{A}(  \wtx ^i) \HfunA^{u} +  \wtx ^i \Hfunb
	\Big]
	\right)
	\,dr\,dx^2 dx^3
	\, .
	\phantom{xxxxxx}
	\label{5VIII21.3GWaWI}
\end{eqnarray}
A more detailed formula  for $C  [\LKbst]$ can be obtained from \eqref{5VIII21.3GWa} by
replacing $\alpha$ by $-\alpha$ and $\LKmcon_{i} $ by $ \LKbcon_{i}$.

\subsection{Noether charges in Minkowski spacetime}
\label{s6VIII21.2a}

All the equations in Section~\ref{s6VIII21.1} apply in Minkowski spacetime by taking the limit $\alpha \to 0$.
Indeed, the Killing fields for Minkowski spacetime can be obtained as a limit of those for de Sitter spacetime.  In the notation of Appendix~\ref{A30IX21.1}, the equations \eqref{24VI21.t2a}, \eqref{24VI21.t2} and  \eqref{10VII21.t7a} give
\begin{equation}
	\mathcal{P} = -
	\frac 12
	\lim\limits_{\alpha \to 0} \Big(\LKmom+\LKbst\Big) \, ,
\end{equation}
where in \eqref{10VII21.t7a}-\eqref{10VII21.t7} we set $P_i=\LKmcon_{i}=\LKbcon_{i}$ .
Similarly, \eqref{24VI21.t2a}, \eqref{24VI21.t2} and  \eqref{10VII21.t7} leads to
\begin{equation}
	\mathcal{L} =
	\frac 12 \lim\limits_{\alpha \to 0} \Big(\frac{\LKbst-\LKmom}{\alpha}\Big) \, ,
	\label{7X22.t1}
\end{equation}
where in \eqref{10VII21.t7a}-\eqref{10VII21.t7} we set  $L_i= \LKmcon_{i} = \LKbcon_{i} $. This shows that for Minkowski spacetime, the linear momentum is given by
\begin{equation}
	P_{M}=-
	\frac 12
	\lim\limits_{\alpha \to 0} \big(P  [\LKmom] + C  [\LKbst] \big) \, ,
\end{equation}
while the center of mass
\begin{equation}
	C_{M}=
	\frac 12
	\lim\limits_{\alpha \to 0} \frac{\big(C  [\LKbst] - P  [\LKmom]  \big)}{\alpha} \, .
\end{equation}
Finally the equations for angular momentum and energy are obvious. One checks that all the limits exist.

\subsection{Noether charges in anti-de Sitter spacetime}
\label{s6VIII21.2b}

All the equations in Section~\ref{s6VIII21.1} apply in anti-de Sitter spacetime under the resplacement $\alpha\mapsto \sqrt{-1}\alpha$. We note that under this replacement both the energy and the angular momentum remain real, while $P$ and $C$ become linear combinations of two linearly independent real-valued charges:
\begin{eqnarray}
	P  [\PadS]&:=&
	\int_{\mcC_{u }}  \mcH^\mu [\PadS] dS_\mu
	\nonumber
	\\
	&
	= &
	\PadSc_{i}\int_{\mcC_{u }} \Big[\wtx^{i}  {\cos (\aldS u)}  \HfunaI^{u}+ \wtx^{i}(\aldS r  {\sin (\aldS u)}- {\cos (\aldS u)}) \HfunaII^{u}
	\nonumber
	\\
	& &+\frac{(\aldS r  {\sin (\aldS u)}- {\cos (\aldS u)})}{r}\zspaceD^{A} \wtx^{i}\HfunA^{u} + \wtx^{i}  {\cos (\aldS u)} \Hfunb
	\Big] \,dr\,dx^2 dx^3 \, ,
	\phantom{xxxxx}
\end{eqnarray}
and
\begin{eqnarray}
	C  [\LadS]&:=&
	\int_{\mcC_{u }}  \mcH^\mu [\LadS] dS_\mu
	\nonumber
	\\
	&
	= &
	\LadSc_{i}\int_{\mcC_{u }} \Big[\wtx^{i}  {\sin (\aldS u)} \HfunaI^{u}-\wtx^{i}( {\sin (\aldS u)}+\aldS r  {\cos (\aldS u)})\HfunaII^{u}
	\nonumber
	\\
	& &
	-\frac{( {\sin (\aldS u)}+\aldS r  {\cos (\aldS u)})}{r} \zspaceD^{A} \wtx^{i}\HfunA^{u}
	+\wtx^{i}  {\sin (\aldS u)} \Hfunb
	\Big]
	\,dr\,dx^2 dx^3
	\, .
	\phantom{xxxxxx}
\end{eqnarray}

\subsection{The evolution of Noether charges}
\label{ss10XII22.1}

In this section we address the question of the rate of change of the charge integrals as the tip of the light cone is moved to the future along the flow of the Killing vector $\partial_u \equiv \mathcal{T}$:
\begin{equation}\label{5VIII21.101}
	\frac{dH[X,\mcC_{u}]}{du}   \equiv
	\frac{d}{du } \int_{\mcC_u}  \mcH^{\mu} [X] dS_\mu = \int_{\mcC_u}  {\Lie_{\partial_u}}\mcH^{\mu} [X] dS_\mu
	\,.
\end{equation}
Assuming sourceless Maxwell fields, \eqref{23VI21.3} with two Killing vector fields    $X$ and $Y$  reads
\begin{equation}
	\Lie_{Y} \mcH^{\mu} [X]
	=
	-\frac{\sqrt{|-\det g|}}{2 \pi}\nabla_{\sigma} \left[Y^{[\sigma} F^{\mu] \alpha}
	X^{\kappa}F_{\kappa \alpha} -\frac{1}{4} Y^{[\sigma} X^{\mu]} F^{\alpha \beta} F_{\alpha \beta} \right]
	\,.
\end{equation}
Using the fields \eqref{5VIII21.1C}-\eqref{7VII21.t2} one finds
\begin{eqnarray}
	{\Lie_{\partial_u}} \mcH^{\mu} {[\partial_u]}&=&2 \nabla_{\sigma} \Big[\mathcal{T}^{[\sigma} \HfunaI^{\mu]} \Big] \, ,
	\label{8V23.t1}
	\\
	{\Lie_{\partial_u}}\mcH^{\mu} [\mathcal{R}]&=&2 \nabla_{\sigma} \Big\{ \mathcal{T}^{[\sigma} \Big[
	\varepsilon^{A B} \zspaceD_{B}(R_i \wtx ^i)
	\HfunA^{\mu]} +\mathcal{R}^{\mu]} \Hfunb
	\Big]
	\Big\}
	\, ,
	\phantom{xxx}
	\\
	{\Lie_{\partial_u}}\mcH^{\mu} [\LKmom ]
	&=&
	2 \nabla_{\sigma} \Big\{
	e^{\alpha u} \mathcal{T}^{[\sigma}  \Big[\LKmcon_{i} \wtx ^i \HfunaI^{\mu]}
	-\big(\alpha r +1\big)\LKmcon_{i} \wtx ^i \HfunaII^{\mu]}
	\nonumber
	\\
	& &
	-\frac{\alpha r +1}{r} \zspaceD^{A}(\LKmcon_{i} \wtx ^i) \HfunA^{\mu]}
	+\LKmom^{\mu]} \Hfunb
	\Big]
	\Big \}
	\, ,
	\\
	{\Lie_{\partial_u}}\mcH^{\mu} [\LKbst ] &=&2 \nabla_{\sigma} \Big\{
	e^{-\alpha u} \mathcal{T}^{[\sigma}  \Big[\LKbcon_{i} \wtx ^i \HfunaII^{\mu]}
	+\big(\alpha r -1\big)\LKbcon_{i} \wtx ^i \HfunaII^{\mu]}
	\nonumber
	\\
	& &
	+\frac{\alpha r -1}{r} \zspaceD^{A}(\LKbcon_{i} \wtx ^i) \HfunA^{\mu]}
	+\LKbst^{\mu]} \Hfunb
	\Big]
	\Big\}
	\, .
\end{eqnarray}
In particular we obtain a formula for the flux of energy:
\begin{eqnarray}
	\lefteqn{
		\frac{d {E _\mcH [}\mcC_{u}]}{d u }
		=
		-
		2 \int_{\partial \hyp_\tau}
		\mathcal{T}^{[\sigma}\HscaaI^{\mu]} \, dS_{{{\sigma}} \mu}
	}
	&&
	\nonumber
	\\
	&=&
	- \lim_{R\to \infty}
	\int_{S_{R}} \HscaaI^{r}\Big{|}_{r=R} dx^2 dx^3
	\nonumber
	\\
	&=&
	-\lim_{R\to \infty}
	\frac{1}{4 \pi}\int_{S_{R}}
	\Big[
	r^2 F_{u r}^2
	+ \zh^{AB} F_{u A} F_{r B}
	+ \epsilon N^2 \zh^{AB} F_{r A} F_{r B}
	\Big]_{r=R}
	d\mu_{\zzhTBW}
	\nonumber
	\\
	&=&
	-\frac{1}{4 \pi}\int_{\Sinf}
	\Big[ \zh^{A B}\big(\alpha^2\otwo{F}_{A r} \ozero{F}_{B u}+ \ozero{F}_{A u} \ozero{F}_{B u}\big)
	\Big]
	d\mu_{\zzhTBW}
	\, .
	\label{10XII22.2}
\end{eqnarray}
The
$u$-derivative of angular momentum is given by
\begin{eqnarray}
	\frac{d J[  \mcC_{u,R}]}{d u }
	&=&
	-
	2\int_{\partial \hyp_\tau}
	\Big[\mathcal{T}^{[\sigma}
	\Big( \HscaA^{\mu]}{  \varepsilon^{A B} \zspaceD_{B}(R_i \wtx ^i)}
	+
	\mathcal{R}^{\mu]}\Hfunb
	\Big)
	\Big] \, dS_{{{\sigma}} \mu}
\nonumber
\\
&=&
-
R_i \lim_{R\to \infty}\int_{S_{R}} \Big[
\HscaA^{r}{  \varepsilon^{A B} \zspaceD_{B} \wtx ^i}
\Big]_{r=R} dx^2 dx^3=:R_i \frac{d J^{i}}{d u }
\,,
\phantom{xxx}
\end{eqnarray}
where
\begin{eqnarray}
\frac{d J^{i}}{d u }
&=&
-
\frac{1}{4 \pi}\lim_{R\to \infty}\int_{S_{R}}
{ \varepsilon^{A B} \zspaceD_{B} \wtx ^i} \Big[
r^2 F_{u r} F_{u A}
-
\epsilon N^2 \zh^{BC} F_{r B}F_{C A}
\nonumber
\\
& &
-\zh^{BC} F_{u B}F_{C A}
\Big]_{r=R}d\mu_{\zzhTBW}
\nonumber
\\
&=&
-
\frac{1}{4 \pi}\int_{\Sinf} \Big[
{  \varepsilon^{A B} \zspaceD_{B}( \wtx ^i)} \Big(
\zh^{BC} \big(\alpha^2 \otwo{F}_{B r}+\ozero{F}_{B u}
\big) \ozero{F}_{C A}
\nonumber
\\
& &
-\otwo{F}_{u r} \ozero{F}_{A u}
\Big)
\Big]d\mu_{\zzhTBW}
\, .
\end{eqnarray}
Finally
\begin{eqnarray}
\lefteqn{
	\frac{dP[\LKmom,\mcC_{u}]}{d u }
	=
	-
	2\int_{\partial \hyp_\tau}
	\Big[e^{\alpha u}
	\mathcal{T}^{[\sigma}
	\Big(\LKmcon_{i} \wtx ^i \HfunaI^{\mu]}
	-\big(\alpha r +1\big)\LKmcon_{i} \wtx ^i \HfunaII^{\mu]}
}
&&
\nonumber
\\
& &
-\frac{\alpha r +1}{r} \zspaceD^{A}(\LKmcon_{i} \wtx ^i) \HfunA^{\mu]}
+\LKmom^{\mu]} \Hfunb
\Big)
\Big]\, dS_{{{\sigma}} \mu}
\nonumber
\\
&=&
-
\lim_{R\to \infty}
\int_{S_{R}} e^{\alpha u}\Big[
\Big(\LKmcon_{i} \wtx ^i \HfunaI^{r}
-\big(\alpha r +1\big)\LKmcon_{i} \wtx ^i \HfunaII^{r}
\nonumber
\\
& &
-\frac{\alpha r +1}{r} \zspaceD^{A}(\LKmcon_{i} \wtx ^i) \HfunA^{r}
+\LKmom^{r} \Hfunb
\Big)
\Big]_{r=R} dx^2 dx^3
\nonumber
\\
&=&
-
\frac{1}{16 \pi}\lim_{R\to \infty}
\int_{S_{R}} e^{\alpha u}\Big[
\LKmcon_{i} \wtx^{i} \Big(
4 \epsilon N^2\zh^{AB} F_{u A} F_{r B}
+4 \zh^{AB} F_{u A} F_{u B}
\nonumber
\\
& &
-(\alpha r+1)\big(
\frac{1}{r^2} \zh^{AC} \zh^{B D} F_{A B} F_{C D}
+
2 r^2 F_{u r}^2
+2 \varepsilon N^2 \zh^{AB} F_{r A} F_{r B}
\big)
\Big)
\nonumber
\\
& &
+
4\LKmcon_{i} \zspaceD^{A} \wtx^{i}\Big(
r^2 F_{u r} F_{u A}
-
\epsilon N^2 \zh^{BC} F_{r B}F_{C A}
-\zh^{BC} F_{u B}F_{C A}
\Big)
\Big]_{r=R}
d\mu_{\zzhTBW}
\nonumber
\\
&=&
-\frac{1}{4 \pi}\int_{\Sinf}
\Big\{
e^{\alpha u} \Big[
\LKmcon_{i} \wtx ^i \zh^{A B}\big(
\ozero{F}_{A u}\ozero{F}_{B u}+ \alpha^2 \otwo{F}_{A r}\ozero{F}_{B u}
\big)
\nonumber
\\
& &
+\alpha \zspaceD^{A}(\LKmcon_{i} \wtx ^i) \Big(
\otwo{F}_{u r} \ozero{F}_{A u}
+ \ozero F_{A B} \zh^{B C}\big(
\alpha^2 \otwo{F}_{C r}
+\ozero{F}_{C u}
\big)\Big)
\Big]
\Big\}
d\mu_{\zzhTBW}
\, ,
\phantom{xxxxx}
\label{6VIII21.t1}
\end{eqnarray}
and
\begin{eqnarray}
\lefteqn{
\frac{d C [\LKbst,\mcC_{u}]}{d u }
=
-2
\int_{\partial \hyp_\tau}
\Big\{e^{-\alpha u} \mathcal{T}^{[\sigma}  \Big[\LKbcon_{i} \wtx ^i \HfunaI^{\mu]}
+\big(\alpha r -1\big)\LKbcon_{i} \wtx ^i \HfunaII^{\mu]}
}
&&
\nonumber
\\
& &
+\frac{\alpha r -1}{r} \zspaceD^{A}(\LKbcon_{i} \wtx ^i) \HfunA^{\mu]}
+\LKbst^{\mu]} \Hfunb
\Big]
\Big\}
\, dS_{{{\sigma}} \mu}
\nonumber
\\
&=&
-
\lim_{R\to \infty}
\int_{S_{R}}
e^{-\alpha u}\Big[\LKbcon_{i} \wtx ^i \HfunaI^{r}
+\big(\alpha r -1\big)\LKbcon_{i} \wtx ^i \HfunaII^{r}
\nonumber
\\
& &
+\frac{\alpha r -1}{r} \zspaceD^{A}(\LKbcon_{i} \wtx ^i) \HfunA^{r}  +\big(\alpha r -1\big)\LKbcon_{i} \wtx ^i \Hfunb
\Big]_{r=R} dx^2 dx^3
\phantom{xxx}
\nonumber
\\
&=&
-\frac{1}{4 \pi}\int_{\Sinf}
e^{-\alpha u}
\Big\{
\LKbcon_{i} \wtx ^i \zh^{A B}\big(
\ozero{F}_{A u}\ozero{F}_{B u}+ \alpha^2 \otwo{F}_{A r}\ozero{F}_{B u}
\big)
\nonumber
\\
& &
-\alpha \zspaceD^{A}(\LKbcon_{i} \wtx ^i) \Big[
\otwo{F}_{u r} \ozero{F}_{A u}
+ \ozero F_{A B} \zh^{B C}\Big(
\alpha^2 \otwo{F}_{C r}
+\ozero{F}_{C u}
\Big)\Big]
\Big\}
d\mu_{\zzhTBW}
\, .
\phantom{xxxxx}
\end{eqnarray}

\section{Noether charges for scalar fields}
\label{s27II23.4}

In~\cite{ChHMS} we found that the canonical energy on light cones for a natural class of linear scalar fields in   de Sitter spacetime was generically infinite, and had to be renormalised. The aim of this section is to address this question for the remaining canonical charges.

In our signature the Lagrangian reads
\begin{equation}\label{9VII21.1}
	\mcL =- \frac 12  \sqrt{|\det g|}
	\big(
	g^{\mu\nu}\partial_\mu \SField \, \partial_\nu \SField
	\pmass \SField^2
	\big)
	\,,
\end{equation}
for a constant $m$.

The theory coincides with its linearisation and we will therefore not make a distinction between the fields $\varphi$ and its linearised counterpart  $\tilde \varphi$, as done in~\cite{ChHMS}.

The canonical energy-momentum current $\mcH^\mu$ equals
\begin{equation}\label{9VII21.2}
	\mcH^\mu [X] = -  \sqrt{|\det g|}
	\Big( \nabla^\mu \SField \,\Lie_X \SField -
	\frac 12
	\big(\nabla^\alpha \SField \nabla_\alpha \SField \pmass \SField^2
	\big)
	X^\mu
	\Big)
	\,.
\end{equation}
Analogously to our analysis of the Maxwell field, we start by considering simultaneously the Minkowski space-time and the de Sitter space-time
in  coordinates as in \eqref{8VII20.11}.

The Lie derivative of the Hamiltonian \eqref{9VII21.2} reads:
\begin{eqnarray}
	-\frac{\Lie_{Y}\mcH^\mu [X]}{\sqrt{|\det g|}} &=&
	\nabla_{\sigma} \big(Y^{\sigma}\nabla^\mu \SField \,\Lie_X \SField \big)
	-
	\nabla_{\sigma} Y^{\mu} \nabla^\sigma \SField \,\Lie_X \SField
	\nonumber
	\\
	& &
	-
	\frac 12
	\nabla_{\sigma} \Big[Y^{\sigma}
	\big(\nabla^\alpha \SField \nabla_\alpha \SField \pmass \SField^2
	\big)
	X^\mu
	\Big]
	\nonumber
	\\
	& &
	+
	\frac 12
	\nabla_{\sigma} Y^{\mu} X^{\sigma} \big(\nabla^\alpha \SField \nabla_\alpha \SField \pmass \SField^2
	\big)
	\nonumber
	\\
	&=&
	2\nabla_{\sigma} \big(Y^{[\sigma}\nabla^{\mu]} \SField \,\Lie_X \SField \big)
	+
	Y^{\mu} \nabla_{\sigma} \big(\nabla^\sigma \SField \,\Lie_X \SField \big)
	\nonumber
	\\
	& &
	-
	\frac 12X^\mu
	Y^{\sigma}\nabla_{\sigma}
	\big(\nabla^\alpha \SField \nabla_\alpha \SField \pmass \SField^2
	\big)
	\nonumber
	\\
	& &
	+\frac{1}{2} [X,Y]^{\mu}
	\big(\nabla^\alpha \SField \nabla_\alpha \SField \pmass \SField^2
	\big)
	\nonumber
	\\
	& &
	-
	\frac 12
	\nabla_{\sigma}Y^{\sigma} X^\mu
	\big(\nabla^\alpha \SField \nabla_\alpha \SField \pmass \SField^2
	\big)
	\, .
	\label{10VII21.t1}
\end{eqnarray}
We combine the second  and  third terms with the equation of motion:
\begin{eqnarray}
	\lefteqn{
		Y^{\mu} \nabla_{\sigma} \big(\nabla^\sigma \SField \,\Lie_X \SField \big)
		-
		\frac 12X^\mu
		Y^{\sigma}\nabla_{\sigma}
		\big(\nabla^\alpha \SField \nabla_\alpha \SField \pmass \SField^2
		\big)
	}
	& &
	\nonumber
	\\
	&=&
	\frac 12 Y^\mu
	X^{\sigma}\nabla_{\sigma}
	\big(\nabla^\alpha \SField \nabla_\alpha \SField \pmass \SField^2
	\big)
	+
	Y^\mu \nabla^{\alpha} \SField \nabla_{\alpha}X^{\sigma}\nabla_{\sigma} \SField
	\nonumber
	\\
	& &
	-
	\frac 12X^\mu
	Y^{\sigma}\nabla_{\sigma}
	\big(\nabla^\alpha \SField \nabla_\alpha \SField \pmass \SField^2
	\big)
	\nonumber
	\\
	&=&
	\nabla_{\sigma} \big[Y^{[\mu}
	X^{\sigma]}
	\big(\nabla^\alpha \SField \nabla_\alpha \SField \pmass \SField^2
	\big) \big]
	+
	Y^\mu \nabla^{\alpha} \SField \nabla_{\alpha}X^{\sigma}\nabla_{\sigma} \SField
	\nonumber
	\\
	& &
	-
	\frac{1}{2} \Big(
	[X,Y]^{\mu}
	+ Y^{\mu} \nabla_{\sigma} X^{\sigma}
	-X^{\mu} \nabla_{\sigma} Y^{\sigma}
	\Big)
	\big(\nabla^\alpha \SField \nabla_\alpha \SField \pmass \SField^2
	\big)
	\, .
	\label{10VII21.t2}
\end{eqnarray}
Equations~\eqref{10VII21.t1} and \eqref{10VII21.t2} lead to
\begin{eqnarray}
	-\frac{\Lie_{Y}\mcH^\mu [X]}{\sqrt{|\det g|}} &=&
	2\nabla_{\sigma} \Big(Y^{[\sigma}\nabla^{\mu]} \SField X^{\alpha} \nabla_{\alpha} \SField -\frac{1}{2} Y^{[\sigma}
	X^{\mu]}
	\big(\nabla^\alpha \SField \nabla_\alpha \SField \pmass \SField^2
	\big) \Big)
	\nonumber
	\\
	& &
	+ Y^\mu \nabla_{\alpha}X^{\sigma} \nabla^{\alpha} \SField \nabla_{\sigma} \SField
	-
	\frac 12
	Y^{\mu} \nabla_{\sigma} X^{\sigma}
	\big(\nabla^\alpha \SField \nabla_\alpha \SField \pmass \SField^2
	\big)
	\nonumber
	\\
	&=&
	2\nabla_{\sigma} \Big(Y^{[\sigma}\nabla^{\mu]} \SField X^{\alpha} \nabla_{\alpha} \SField -\frac{1}{2} Y^{[\sigma}
	X^{\mu]}
	\big(\nabla^\alpha \SField \nabla_\alpha \SField \pmass \SField^2
	\big) \Big)
	\nonumber
	\\
	& &
	+Y^\mu \Big(\nabla_{\alpha}X_{\sigma}-\frac 12
	\nabla_{\kappa} X^{\kappa} g_{\alpha \sigma}\Big)  \nabla^{\alpha} \SField \nabla^{\sigma} \SField
	\nonumber
	\\
	& &
	-
	\frac 12
	Y^{\mu} \nabla_{\sigma} X^{\sigma}
	m^2 \SField^2
	\, .
\end{eqnarray}
\ptcheck{2X21; all calculations above}

\subsection{Charges in (anti)-de Sitter spacetime}
\label{ss17X21.1}

We only
consider here a massive scalar field, with the mass chosen so that the equation is conformally covariant,
\begin{equation}\label{5VII20.5}
	\Box_g \phi - \underbrace{\frac{(d-2)R(g)}{4 (d-1)}}_{=:m^2} \phi = 0
	\,,
\end{equation}
where $d$ is the dimension of spacetime and $R(g)$ is the scalar curvature of $g$. In the four-dimensional case, it leads to
\begin{equation}
	m^2 = 2 \alpha ^2
\end{equation}

After a conformal transformation  $g\mapsto \Omega^2 g$ the field $\Omega^{d/2-1} \phi$ satisfies again \eqref{5VII20.5}, with $g$ there replaced by $\Omega^2 g$. This is useful in that solutions of \eqref{5VII20.5} with smooth initial data on a Cauchy surface in de Sitter spacetime extend smoothly, afer the rescaling above, in local coordinates on the conformally completed manifold, across the conformal boundary at infinity. This translates to the following asymptotic behaviour of $\SField$, for large $r$, in spacetime dimension four:
\begin{equation}\label{13VIII21.t1}
	\SField(u,r, x^A) = \frac{\oone \SField(u,x^A)}{r}
	+ \frac{\otwo \SField(u,x^A)}{r^2}
	+ \frac{\othree \SField(u,x^A)}{r^3} + ...
	\, .
\end{equation}
See \cite[Section 2.2.1]{ChHMS} for a discussion.
Here we simply note that the functions $\oone \SField$ and $\otwo \SField$ are freely prescribable, with all remaining expansion coefficients determined uniquely by these two.

We wish to construct the Noether charges associated with the Killing fields \eqref{24VI21.t1}-\eqref{24VI21.t2}.
For this let
\begin{eqnarray}
	\Hscaa^{\mu}[X] &= &- \sqrt{|\det g|}
	\nabla^{\mu} \SField X^{\alpha} \nabla_{\alpha} \SField
	\, ,
	\label{5VIII21.1Cs}
	\\
	\Hscab &=& \frac{1}{2} \sqrt{|\det g|} \big( \nabla^\alpha \SField \nabla_\alpha \SField+ m^2 \SField^2 \big)
	\, ,
\end{eqnarray}
and
\begin{eqnarray}
	\HscaaI^{\mu}&=&\Hfuna^{\mu}[\partial_{u}] \, ,
	\label{7VII21.t1s}
	\\
	\HscaaII^{\mu}&=&\Hfuna^{\mu}[\partial_r]
	\, ,
	\\
	\HscaA^{\mu}&=&\Hfuna^{\mu}[\partial_A]
	\, .
	\label{7VII21.t2s}
\end{eqnarray}
Since the Hamiltonian density \eqref{9VII21.2} is linear in the Hamiltonian vector field $X$, each charge is given by an integral of a linear combination of the following four functionals
\ptcheck{17X21; here and before}
\begin{eqnarray}
	\label{5II21.101}
	\mcH^{\mu} {[\partial_u]}&=&\HscaaI^{\mu}+\mathcal{T}^{\mu} \Hscab
	\,,
	\\
	\mcH^{\mu} [\mathcal{R}]&=&
	\varepsilon^{A B} \zspaceD_{B}(R_i \wtx ^i)
	\HscaA^{\mu} +\mathcal{R}^{\mu} \Hscab
	\label{17X21.31}
	\,,
	\\
	\mcH^{\mu} [\LKmom ] &=&e^{\alpha u}\Big[\LKmcon_{i} \wtx ^i \HscaaI^{\mu}
	-\big(\alpha r +1\big)\LKmcon_{i} \wtx ^i \HscaaII^{\mu}
	\nonumber
	\\
	& &
	-\frac{\alpha r +1}{r} \zspaceD^{A}(\LKmcon_{i} \wtx ^i) \HscaA^{\mu} \Big]+ \LKmom^{\mu} \Hscab
	\,,
	\\
	\mcH^{\mu} [\LKbst] &=&
	e^{-\alpha u}\Big[
	\LKbcon_{i} \wtx ^i \HscaaI^{\mu}
	+\big(\alpha r -1\big)\LKbcon_{i} \wtx ^i \HscaaII^{\mu}
	\nonumber
	\\
	& &
	+\frac{\alpha r -1}{r} \zspaceD^{A}(\LKbcon_{i} \wtx ^i) \HscaA^{\mu} \Big]
	+\LKbst^{\mu} \Hscab
	\, ,
	\label{17X21.33}
\end{eqnarray}
Written-out in detail, the functionals \eqref{7VII21.t1s}-\eqref{7VII21.t2s}  read
\begin{eqnarray}
	\HscaaI^{u}&=&\big(
	r^{2}\partial_{r} \SField \partial_{u} \SField
	\big)\sqrt{\det \zzhTBW }
	\, ,
	\\
	\HscaaI^{r}&=&
	r^{2}\big(\partial_{u} \SField+\big(\alpha^{2} r^{2}-1\big)\partial_{r} \SField \big) \partial_{u} \SField
	\sqrt{\det \zzhTBW }
	\, ,
	\\
	\HscaaII^{u}&=& r^{2}\big(\partial_{r} \SField\big)^{2}\sqrt{\det \zzhTBW }
	\, ,
	\\
	\HscaaII^{r}&=&
	r^{2}\big(\partial_{u} \SField+\big(\alpha^{2} r^{2}-1\big)\partial_{r} \SField\big)\partial_{r} \SField
	\sqrt{\det \zzhTBW }
	\, ,
	\\
	\HscaA^{u}&=&
	r^{2}\partial_{r} \SField \zspaceD_{A} \SField
	\sqrt{\det \zzhTBW }
	\, ,
	\\
	\HfunA^{r}&=&
	r^{2}\big(\partial_{u} \SField+\big(\alpha^{2} r^{2}-1\big)\partial_{r} \SField\big)\zspaceD_{A} \SField
	\sqrt{\det \zzhTBW }
	\, ,
	\phantom{xxxxx}
	\\
	\Hscab &=&\frac{1}{2}\big(\zh^{A B}\zspaceD_{A} \SField \zspaceD_{B} \SField +m^{2} r^{2}\SField^{2}
	\nonumber
	\\
	& &
	-2r^{2} \partial_{r} \SField\partial_{u} \SField+\big(1-\alpha^{2} r^{2}\big)r^{2}\big(\partial_{r} \SField\big)^{2}\big)
	\sqrt{\det \zzhTBW }
	\, .
\end{eqnarray}
\ptcheck{17X21; here and before}


Recall that we denote by $\mcC_{u}$ the light cone of constant $u$, and $\mcC_{u,R}= \mcC_{u}\cap \{r\le R\}$ its truncation to $r=R$.
It turns out that, generically, all charge integrals over $\mcC_{u}$  diverge  as $R$ tends to infinity, and therefore  need to be renormalised.  Therefore we first calculate the charges on $\mcC_{u,R}$ and exhibit their divergent parts, for large $r$. We use the asymptotics \eqref{13VIII21.t1} which applies both to the $\alpha =0$ case with $m=0$ and to the case $\alpha^2= m^2/2$ with $m\ne 0$:
%
\begin{eqnarray}
	\nonumber
	\lefteqn{
		{E _\mcH [} \mcC_{u ,R}]  :=
		\int_{\mcC_{u,R }}  \mcH^\mu [\partial_u] dS_\mu
		=
		\int_{\mcC_{u,R }}  \mcH^u [\partial_u] dS_u
	}
	&&
	\\
	&= &   \int_{\mcC_{u,R }} \big(\HscaaI^{u}+\Hscab
	\big)
	\,dr\,dx^2 dx^3
	\nonumber
	\\
	&=&
	\frac{1}{2}
	\int_{\mcC_{u,R }}
	\Big(
	\zh^{A B}\zspaceD_{A} \SField \zspaceD_{B} \SField
	+
	m^{2} r^{2} \SField^{2}
	+
	\big(r^{2} - \alpha^{2} r^{4}\big)\big(\partial_{r} \SField\big)^{2}
	\Big)
	\,dr \, d\mu_{\zzhTBW}
	\nonumber
	\\
	&=&
	\frac{1}{2}
	\int_{\mcC_{u,R }}
	\Big(
	\zh^{A B}\zspaceD_{A} \SField \zspaceD_{B} \SField
	+
	m^{2} r^{2} \SField^{2}
	\nonumber
	\\
	& &
	+
	\partial_{r}
	\Big[
	\big(r^{2} - \alpha^{2} r^{4}\big) \SField \big(\partial_{r} \SField\big)
	\Big]
	-
	\SField
	\partial_{r}
	\big[
	\big(r^{2} - \alpha^{2} r^{4}\big) \big(\partial_{r} \SField\big)
	\big]
	\Big)
	\,dr \, d\mu_{\zzhTBW}
	\nonumber
	\\
	&=&
	\frac{\alpha^{2} R }{2}  \int_{S_R}
	(\oone \SField)^{2}  \, d\mu_{\zzhTBW}
	+
	\int_{\mcC_{u,R }} O(r^{-2})
	\,dr \, d\mu_{\zzhTBW}
	\,,
	\phantom{xxxxx}
	\label{24V21.t1as}
\end{eqnarray}
where we have used
\begin{eqnarray}
	\big(r^{2} - \alpha^{2} r^{4}\big) \SField \big(\partial_{r} \SField\big)
	=
	\frac{r}{2}\alpha^{2} (\oone \SField)^{2}
	+O(r^{-1})
\end{eqnarray}

As before, the total angular-momentum is obtained from the following integral:
\begin{eqnarray}
	J[  \mcC_{u,R }]&:=&
	\int_{\mcC_{u,R }}  \mcH^\mu [\mathcal{R}] dS_\mu
	\equiv  R_i J^i [\mcC_{u,R }]
	\,,
	\label{5VIII21.3GWs}
\end{eqnarray}
where now
\begin{eqnarray}
	J^i[\mcC_{u,R }]&:=&
	\int_{\mcC_{u,R }}
	\varepsilon^{A B} \zspaceD_{B} \wtx ^i
	\HscaA^{u}
	\,dr\,dx^2 dx^3
	\nonumber
	\\
	&  = &
	\int_{\mcC_{u,R }} r^{2} \varepsilon^{A B} \zspaceD_{B} \wtx ^i \zspaceD_{A} \SField
	\partial_{r} \SField
	\,dr\, d\mu_{\zzhTBW}
	\nonumber
	\\
	&  = &
	\int_{\mcC_{u,R }}
	\Big(
	-\frac{ {\oone \SField}
		\varepsilon^{A B} \zspaceD_{B} \wtx ^i \zspaceD_{A} {\oone \SField} }{r}
	+
	O({r}^{-2})
	\Big)
	\,dr\, d\mu_{\zzhTBW}
	\nonumber
	\\
	&  = &
	\int_{\mcC_{u,R }}
	O({r}^{-2})
	\,dr\, d\mu_{\zzhTBW}
	\, ,
	\label{5VIII21.3GWbs}
\end{eqnarray}
where we have used
\begin{eqnarray}
	\int_{S^2}{\oone \SField}
	\varepsilon^{A B} \zspaceD_{B} \wtx ^i \zspaceD_{A} {\oone \SField}
	d\mu_{\zzhTBW}
	= \frac 12 \int_{S^2}\zspaceD_{A} \left(
	\varepsilon^{A B} \zspaceD_{B} \wtx ^i ({\oone \SField})^2
	\right)
	d\mu_{\zzhTBW}  = 0
	\, .
	\label{17X21.51}
\end{eqnarray}

We further have
\begin{eqnarray}
	P  [\LKmom,\mcC_{u,R }]&:=&
	\int_{\mcC_{u,R }}  \mcH^\mu [\LKmom] dS_\mu
	\nonumber
	\\
	&
	= &
	\LKmcon_{i}\int_{\mcC_{u,R }}
	e^{\alpha u}\Big[ \wtx ^i \HscaaI^{u}
	-\big(\alpha r +1\big)  \wtx ^i \HscaaII^{u}
	-\frac{\alpha r +1}{r} \zspaceD^{A}(  \wtx ^i) \HscaA^{u} +  \wtx ^i \Hscab
	\Big]
	\,dr\,dx^2 dx^3
	\nonumber
	\\
	& = &
	\LKmcon_{i} \int_{\mcC_{u,R }}
	e^{\alpha u}
	\Big[\frac{1}{2}\wtx ^i \Big(
	\zh^{A B}\zspaceD_{A} \SField \zspaceD_{B} \SField
	+m^{2} \SField^{2} r^{2}
	\nonumber
	\\
	& &
	-\big(\alpha^{2} r^{4}+2  \alpha r^{3}+r^{2}\big)\big(\partial_{r} \SField\big)^{2}
	\Big)
	-(\alpha r+1) r \zspaceD^{A}n^{i}\left(\partial_{r} \SField\right)\zspaceD_{A}\SField
	\Big]
	\,dr\, d\mu_{\zzhTBW}
	\nonumber
	\\
	&=&
	\LKmcon_{i} \int_{\mcC_{u,R }}
	\mathrm{e}^{ {\alpha u }} \Big[
	\frac{1}{2} \alpha^{2} {\wtx^i}  {\oone \SField}^{2}
	+
	\frac{1}{r}{\oone \SField}\left(\alpha \zh^{A B}\zspaceD_{A} {n^i}\zspaceD_{B} {\oone \SField}
	- {\wtx^i}  { \alpha  } {\oone \SField}
	\right)
	+ O(r^{-2})
	\Big]
	\,dr\, d\mu_{\zzhTBW}
	\nonumber
	\\
	&=&
	\LKmcon_{i}
	\Big[
	\frac{R \alpha^{2}\mathrm{e}^{ {\alpha u }} }{2} \int_{S_{ R }}
	{\wtx^i}  {\oone \SField}^{2} \, d\mu_{\zzhTBW} +
	\int_{\mcC_{u,R }}
	O(r^{-2})
	\,dr\, d\mu_{\zzhTBW}
	\Big]
	\,,
	\label{5VIII21.3GWas}
\end{eqnarray}
%
and note that the second and third terms in the before-last line integrate out to zero. Finally,
\begin{eqnarray}
	C  [\LKbst,\mcC_{u,R }]&:=&
	\int_{\mcC_{u,R }}  \mcH^\mu [\LKbst] dS_\mu
	\nonumber
	\\
	&
	= &
	\LKbcon_{i}\int_{\mcC_{u,R }} \big(
	e^{-\alpha u}\Big[ \wtx ^i \HscaaI^{u}
	+\big(\alpha r -1\big)  \wtx ^i \HscaaII^{u}
	+\frac{\alpha r -1}{r} \zspaceD^{A}(  \wtx ^i) \HscaA^{u} +  \wtx ^i \Hscab
	\Big]
	\big)
	\,dr\,dx^2 dx^3
	\, .
	\nonumber
	\\
	& &
	\phantom{xxxxxx}
	\label{17X21.34}
\end{eqnarray}
Similarly to the case of the Maxwell field, a more detailed formula version of \eqref{17X21.34} can be obtained from \eqref{5VIII21.3GWas} by
replacing there $\alpha$ by $-\alpha$ and $\LKmcon_{i}$ by $\LKbcon_{i}$.

\subsection{Noether charges in Minkowski spacetime}
\label{s6VIII21.2as}

All the equations in Section~\ref{ss17X21.1} apply to the massless scalar field in Minkowski spacetime by passing to the limit
$m=\alpha=0$. In that case we clearly have a finite energy.
This is also clear for the total momentum, which we denote by $P_M$, using \eqref{5VIII21.3GWas}-\eqref{17X21.34}
with $P_i$ replaced by $p_i$ for consistency of notation with \eqref{10VII21.t3}  (see \eqref{10VII21.t7a}), compare~\eqref{28IV22.1}):
\begin{eqnarray}
	P_{M} [\mathcal{P}, \mcC_{u,R}] &=&-\frac{1}{2}\lim\limits_{\alpha \to 0} \big(P  [\LKmom, \mcC_{u ,r}] + C  [\LKbst, \mcC_{u ,r}]_{\LKbcon_{i}:= P_{i}} \big)
	\, ,
	\nonumber
	\\
	&=&
	P_{i}
	\int_{\mcC_{u,R }}\Big(\wtx ^i \left(\HscaaII^{u}-\HscaaI^{u}
	\right)+\frac{1}{r}\zspaceD^{A} {\wtx^i}  \HscaA^{u}
	-  \wtx ^i \Hscab
	\Big)\Big|_{\alpha=0}
	\,dr\, d\mu_{\zzhTBW}
	\nonumber
	\\
	&  = &
	\int_{\mcC_{u,R }}
	O({r}^{-2})
	\,dr\, d\mu_{\zzhTBW}
	\, .
\end{eqnarray}

Consider, next, the formula for the center of mass, which can be similarly obtained from \eqref{28IV22.2} and \eqref{5VIII21.3GWas}-\eqref{17X21.34}: in the notation of \eqref{10VII21.t4},
\begin{eqnarray}
	C_{M}[\mathcal{L}, \mcC_{u ,R}]&=&\frac{1}{2} \lim\limits_{\alpha \to 0} \frac{\big(C  [\LKbst, \mcC_{u ,r}]-P  [\LKmom, \mcC_{u ,r}]_{\LKmcon_{i}:= L_{i}} \big)}{\alpha}
	\nonumber
	\\
	&=&
	L_{i}
	\int_{\mcC_{u,R }}
	\Big(
	-u \wtx ^i  \HscaaI^{u} +(u+r)  \wtx^ i \HscaaII^{u} +\big(1+\frac{u}{r}\big)\zspaceD^{A} {\wtx^i}  \HscaA^{u} -  u \wtx ^i \Hscab
	\Big)
	\,dr\, d\mu_{\zzhTBW}
	\nonumber
	\\
	&=&
	\int_{\mcC_{u,R }}
	O({r}^{-2})
	\,dr\, d\mu_{\zzhTBW}
	\, ,
\end{eqnarray}

Finally, the total angular momentum is finite,
\begin{eqnarray}
	\hat J^i[\mcC_{u }]& : =&
	\lim_{R\to\infty}
	\bigg[
	\int_{\mcC_{u,R }} r^{2} \varepsilon^{A B} \zspaceD_{B} \wtx ^i \zspaceD_{A} \SField
	\partial_{r} \SField
	\,dr\, d\mu_{\zzhTBW}
	\nonumber
	\\
	&&
	+ \ln R \underbrace{
		\int_{S^2} \varepsilon^{A B} \zspaceD_{B} \wtx ^i {\oone \SField} \zspaceD_{A} {\oone \SField}  d\mu_{\zzhTBW}
	}_{0}
	\bigg]
	\, ,
	\label{17X21.52}
\end{eqnarray}
as the boundary integral in \eqref{17X21.52} is a total divergence.

\subsection{Noether charges in anti-de Sitter spacetime}
\label{s6VIII21.2bs}

All the equations in Section~\ref{ss17X21.1} apply in anti-de Sitter spacetime under the replacement $\alpha\mapsto \sqrt{-1}\alpha$. Then both the energy and the angular momentum remain real, while $P$ and $C$ become linear combinations of two linearly independent real-valued charges. Indeed, using \eqref{28IV22.5}-\eqref{28IV22.6} one finds
\begin{eqnarray}
	\lefteqn{
		P  [\PadS, \mcC_{u ,R}] :=
		\int_{\mcC_{u,R }}  \mcH^\mu [\PadS ] dS_\mu
	} &&
	\nonumber
	\\
	&
	= &
	\PadSc_{i}\int_{\mcC_{u,R }} \Big[\wtx^{i} {\cos (\aldS u)}  \HscaaI^{u}+ \wtx^{i}(\aldS r  {\sin (\aldS u)}- {\cos (\aldS u)}) \HscaaII^{u}
	\nonumber
	\\
	& &+\frac{(\aldS r  {\sin (\aldS u)}- {\cos (\aldS u)})}{r}\zspaceD^{A} \wtx^{i}\HscaA^{u} + \wtx^{i} {\cos (\aldS u)} \Hscab
	\Big] \,dr\,dx^2 dx^3 \, ,
	\phantom{xxxxx}
\end{eqnarray}
and
\begin{eqnarray}
	\lefteqn{
		C  [\LadS, \mcC_{u ,R}] :=
		\int_{\mcC_{u,R }}  \mcH^\mu [\LadS ] dS_\mu
	}
	&&
	\nonumber
	\\
	&
	= &
	\LadSc_{i}\int_{\mcC_{u,R}} \Big[\wtx^{i}  {\sin (\aldS u)} \HscaaI^{u}-\wtx^{i}( {\sin (\aldS u)}+\aldS r {\cos (\aldS u)})\HscaaII^{u}
	\nonumber
	\\
	& &
	-\frac{( {\sin (\aldS u)}+\aldS r {\cos (\aldS u)})}{r} \zspaceD^{A} \wtx^{i}\HscaA^{u}
	+\wtx^{i}  {\sin (\aldS u)} \Hscab
	\Big]
	\,dr\,dx^2 dx^3
	\, .
	\phantom{xxxxxx}
\end{eqnarray}

\subsection{The time-evolution of Noether charges}

A question of interest is the rate of change of the charge integrals as the tip of the light cone is moved to the future along the flow of the Killing vector $\partial_u \equiv \mathcal{T}$:
\begin{equation}\label{5VIII21.101s}
	\frac{dH[X,\mcC_{u,R}]}{du}   \equiv
	\frac{d}{du } \int_{\mcC_{u,R}}  \mcH^{\mu} [X] dS_\mu = \int_{\mcC_{u,R}}  {\Lie_{\partial_u}}\mcH^{\mu} [X] dS_\mu
	\,.
\end{equation}
Assuming two Killing vector fields    $X$ and $Y$, we have
\begin{eqnarray}
	\Lie_{Y} \mcH^{\mu} [X]
	&=&
	-
	2\sqrt{|\det g|}\nabla_{\sigma} \Big(Y^{[\sigma}\nabla^{\mu]} \SField X^{\alpha} \nabla_{\alpha} \SField
	\nonumber
	\\
	& &
	-\frac{1}{2} Y^{[\sigma}
	X^{\mu]}
	\big(\nabla^\alpha \SField \nabla_\alpha \SField \pmass \SField^2
	\big) \Big)
	\,.
\end{eqnarray}
Using the fields \eqref{5VIII21.1Cs}-\eqref{7VII21.t2s} one finds 
\begin{eqnarray}
	\label{5X22.t1}
	{\Lie_{\partial_u}} \mcH^{\mu} {[\partial_u]}&=&2 \nabla_{\sigma} \Big[\mathcal{T}^{[\sigma} \HscaaI^{\mu]} \Big] \, ,
	\\
	\label{5X22.t2}
	{\Lie_{\partial_u}}\mcH^{\mu} [\mathcal{R}]&=&2 \nabla_{\sigma} \Big\{ \mathcal{T}^{[\sigma} \Big[
	{ \varepsilon^{A B} \zspaceD_{B}(R_i \wtx ^i)}\HscaA^{\mu]} +\mathcal{R}^{\mu]} \Hscab
	\Big]
	\Big\}
	\, ,
	\phantom{xxx}
	\\
	{\Lie_{\partial_u}}\mcH^{\mu} [\LKmom ]
	&=&
	2 \nabla_{\sigma} \Big\{
	e^{\alpha u} \mathcal{T}^{[\sigma}  \Big[\LKmcon_{i} \wtx ^i \HscaaI^{\mu]}
	-\big(\alpha r +1\big)\LKmcon_{i} \wtx ^i \HscaaII^{\mu]}
	\nonumber
	\\
	& &
	-\frac{\alpha r +1}{r} \zspaceD^{A}(\LKmcon_{i} \wtx ^i) \HscaA^{\mu]}
	+\LKmom^{\mu]} \Hscab
	\Big]
	\Big \}
	\, ,
	\\
	{\Lie_{\partial_u}}\mcH^{\mu} [\LKbst ] &=&2 \nabla_{\sigma} \Big\{
	e^{-\alpha u} \mathcal{T}^{[\sigma}  \Big[\LKbcon_{i} \wtx ^i \HscaaI^{\mu]}
	+\big(\alpha r -1\big)\LKbcon_{i} \wtx ^i \HscaaII^{\mu]}
	\nonumber
	\\
	& &
	+\frac{\alpha r -1}{r} \zspaceD^{A}(\LKbcon_{i} \wtx ^i) \HscaA^{\mu]}
	+\LKbst^{\mu]} \Hscab
	\Big]
	\Big\}
	\, .
	\label{5X22.t3}
\end{eqnarray}

In particular we obtain the energy-flux:
\footnote{We take this opportunity to correct a misprint in \cite[Equation~(2.68)]{ChHMS}, where the terms involving $\otwo \SField$ are missing.}
\ptcheck{28IV22}
\begin{eqnarray}
	\frac{d {E _\mcH [} \mcC_{u ,R}]}{d u }
	& = &
	-
	\int_{\partial \mcC_{u ,R}}
	\mathcal{T}^{[\sigma}\HscaaI^{\mu]} \, dS_{{{\sigma}} \mu}
	\nonumber
	\\
	&=&
	-\int_{S_{R}} \HscaaI^{r}\Big{|}_{r=R} dx^2 dx^3
	\nonumber
	\\
	&=&
	-
	\int_{S_{R}}
	\Big[
	r^{2}\left(\partial_{u} \SField+\left(\alpha^{2} r^{2}-1\right)\partial_{r} \SField\right)\partial_{u} \SField
	\Big]_{r=R}
	d\mu_{\zzhTBW}
	\nonumber
	\\
	&=&
	\int_{S_{R}}
	\Big[
	\alpha^{2} {\oone \SField}\partial_{u} {\oone \SField} R
	\nonumber
	\\
	& &
	+\alpha^{2} {\oone \SField}\partial_{u} {\otwo \SField}+\left(2 \alpha^{2} {\otwo \SField}-\partial_{u} {\oone \SField}\right)\partial_{u} {\oone \SField}
	\Big]
	d\mu_{\zzhTBW}
	+ o(1)
	\, ,
	\phantom{xxxxx}
	\label{8X22.1}
\end{eqnarray}
where $o(1)$ tends to zero as $R$ tends to infinity.

We turn now our attention to the rate of change of  the functional  $E_\omega[ \hyp]$ of \eqref{S6IX18.5+a}, when the tip of the light cone is moved to the future along the flow of the Killing vector $\partial_u$:
\begin{equation}
	\frac{d  E_\omega[\mcC_{u,R}]}{du}   \equiv
	\frac 12 \frac{d}{du } \int_{\mcC_{u,R}} \omega^\mu({\phi}, {\Lie_{\partial_u}} {\phi}) dS_\mu
	= \frac 12\int_{\mcC_{u,R}}  {\Lie_{\partial_u}} \omega^\mu({\phi}, {\Lie_{\partial_u}} {\phi}) dS_\mu
	\,.
	\label{1XI22.t1}
\end{equation}
We find
\begin{eqnarray}
	\Lie_{Y} \omega^\mu({\phi}, {\Lie_{\partial_u}} {\phi})
	&=&
	\partial_\sigma \big(Y^\sigma \omega^\mu \big)
	-\omega^\sigma  \partial_\sigma Y^{\mu}
	\nonumber
	\\
	&=&
	2 \partial_\sigma \big(Y^{[\sigma} \omega^{\mu]}\big)
	+Y^\mu \partial_\sigma \omega^\sigma
	\, .
\end{eqnarray}
Assuming that the field equations hold, we have $\partial_{\sigma}
\big(\omega^\sigma({\phi}, {\Lie_{\partial_u}} {\phi})
\big)
=0$. 
The flux of  $E_\omega[\mcC_{u,R}]$ reads
\begin{eqnarray}
	\frac{d  E_\omega[ \mcC_{u,R}]}{du}
	&=&
	\frac 12\int_{\mcC_{u,R}} 2 \partial_\sigma \big(\mathcal{T}^{[\sigma} \omega^{\mu]}\big) dS_\mu
	\nonumber
	\\
	&=&
	-\frac12\int_{S_{R}} \mathcal{T}^{[\sigma} \omega^{\mu]}  d S_{\sigma \mu}
	\nonumber
	\\
	&=&\frac12\int_{S_{R}}
	r^2\Big(
	\phi(\alpha^2 r^2-1) \partial_{u} \partial_{r} \phi+\phi \partial^{2}_{u} \phi
	\nonumber
	\\
	& & -\big(\partial_{u} \phi+(\alpha^2 r^2-1 )\partial_{r} \phi\big)\partial_{u} \phi
	\Big)_{r=R}
	\, d\mu_{\zzhTBW}
	\,,
\end{eqnarray}
with asymptotic expansion
\begin{eqnarray}
	\frac{d  E_\omega[ \mcC_{u,R}]}{du}
	& = &
	\frac12\int_{S_{R}}
	\Big(
	-\alpha^2 \oone \phi \partial_u \otwo \phi +\alpha^2 \otwo \phi \partial_{u} \oone \phi +\oone \phi \partial^{2}_{u}  \oone \phi-(\partial_u \oone \phi )^2
	\Big)
	\, d\mu_{\zzhTBW}
	\nonumber
	\\
	&&
	+O(\frac{1}{R})
	\,.
	\label{2XI22.t1}
\end{eqnarray}
Note that when $\alpha=0$ the limit of $dE_\omega/du$  as $R$ tends to infinity  is \emph{not} negative, which makes questionable the interpretation of $E_\omega$ as the right functional for a physically significant definition of energy.

We continue with the
$u$-derivative of angular momentum, given by
\begin{eqnarray}
	\frac{d J[  \mcC_{u,R}]}{d u }
	&=&
	-
	\int_{\mcC_{u,R}}
	\Big[\mathcal{T}^{[\sigma}
	\Big( \HscaA^{\mu]}{ \varepsilon^{A B} \zspaceD_{B}(R_i \wtx ^i)}
	+
	\mathcal{R}^{\mu]}\Hscab
	\Big)
	\Big] \, dS_{{{\sigma}} \mu}
\nonumber
\\
&=&
-
R_i \int_{S_{R}} \Big[
\HscaA^{r}{ \varepsilon^{A B} \zspaceD_{B} \wtx ^i}
\Big]_{r=R} dx^2 dx^3=:R_i \frac{d J^{i}}{d u }
\,,
\phantom{xxx}
\end{eqnarray}
with
\begin{eqnarray}
\frac{d J^{i}}{d u }
&=&
-\int_{S_{R}}
\sqrt{\det \zzhTBW }
{ \varepsilon^{A B} \zspaceD_{B} \wtx ^i} \Big[
r^{2}\big(\partial_{u} \SField+\left(\alpha^{2} r^{2}-1\right)\partial_{r} \SField\big)
\zspaceD_{A} \SField
\Big]_{r=R}d\mu_{\zzhTBW}
\nonumber
\\
&=&
\int_{S_{R}} \Big[
{ \varepsilon^{A B} \zspaceD_{B} \wtx ^i } \Big(
\alpha^{2} {\oone \SField} \zspaceD_{A} {\oone \SField} R
\nonumber
\\
& &
+\alpha^{2} {\oone \SField} \zspaceD_{A} {\otwo \SField}
+\left(2 \alpha^{2} {\otwo \SField}-\partial_{u} {\oone \SField}\right)\zspaceD_{A} {\oone \SField}
\Big)
\Big]d\mu_{\zzhTBW}
+ o(1)
\nonumber
\\
&=&
-
\int_{S_{R}} \Big[
\alpha^{2} {\oone \SField} \zspaceD_{A} {\otwo \SField}
+
\partial_{u} {\oone \SField} \zspaceD_{A} {\oone \SField}
\Big)
\Big]d\mu_{\zzhTBW}
+ o(1)
\, ,
\phantom{xxxxx}
\label{6X22.h2}
\end{eqnarray}
\ptcheck{28IV22}
where the terms proportional to $R$ integrated-out to zero.
Finally
\begin{eqnarray}
\frac{dP[\LKmom,\mcC_{u,R}]}{d u }
&=&
-
2  \int_{\mcC_{u,R}}
\Big[e^{\alpha u}
\mathcal{T}^{[\sigma}
\Big(\LKmcon_{i} \wtx ^i \HscaaI^{\mu]}
-\big(\alpha r +1\big)\LKmcon_{i} \wtx ^i \HscaaII^{\mu]}
\nonumber
\\
& &
-\frac{\alpha r +1}{r} \zspaceD^{A}(\LKmcon_{i} \wtx ^i) \HscaA^{\mu]}
+\LKmom^{\mu]} \Hscab
\Big)
\Big]\, dS_{{{\sigma}} \mu}
\nonumber
\\
&=&
-
\int_{\mcC_{u,R}}
\Big[e^{\alpha u}
\Big(\LKmcon_{i} \wtx ^i \HscaaI^{r}
-\big(\alpha r +1\big)\LKmcon_{i} \wtx ^i \HscaaII^{r}
\nonumber
\\
& &
-\frac{\alpha r +1}{r} \zspaceD^{A}(\LKmcon_{i} \wtx ^i) \HscaA^{r}
-\big(\alpha r +1\big) \Hscab
\Big)
\Big]\, dx^2 dx^3
\nonumber
\\
&=&
-
\int_{S_{R}} e^{\alpha u} \LKmcon_{i}\Big\{
\wtx ^i\Big[r^{2}\Big(\partial_{u} \SField+\left(\alpha^2 r^{2}-1\right)\partial_{r} \SField\Big)\partial_{u} \SField
\nonumber
\\
& &
-(\alpha r+1) r^{2}\left(\partial_{u} \SField+\left(\alpha^{2} r^{2}-1\right)\partial_{r} \SField\right)\partial_{r} \SField
\nonumber
\\
& &
-\frac{1}{2}(\alpha r+1)\Big(
\zh^{A B}\zspaceD_{A} \SField \zspaceD_{B} \SField
-2r^{2}\left(\partial_{r} \SField\right)\left(\partial_{u} \SField\right)
\nonumber
\\
& &
+m^{2} r^{2} \SField^{2}
-\left(\alpha^{2} r^{2}-1\right)r^{2}\left(\partial_{r} \SField\right)^{2}\Big)\Big]
\nonumber
\\
& &
-(\alpha r+1) r  \zspaceD^{A} \wtx ^i \partial_{r} \SField \zspaceD_{A} \SField
\Big\}_{r=R}
d\mu_{\zzhTBW}
\nonumber
\\
&=&
\int_{S_{R}}
e^{\alpha u} \LKmcon_{i} \wtx ^i \Big\{
\alpha  {\oone \SField} \Big[
\alpha^{2} {\oone \SField}+ m^{2} {\oone \SField} +2 \alpha \partial_{u} {\oone \SField}
\Big] R
\nonumber
\\
& &
+ \Big[
4 \alpha^{3} {\otwo \SField}
+2 \alpha m^{2} {\otwo \SField}
+2 \alpha^{2} \partial_{u} {\otwo \SField}
+\alpha^{2} {\oone \SField}
+m^{2} {\oone \SField}
\Big] {\oone \SField}
\nonumber
\\
& &
+4 \alpha^{2}\partial_{u} {\oone \SField}  {\otwo \SField}
-2\left(\partial_{u} {\oone \SField}\right)^{2}
\Big\}
d\mu_{\zzhTBW}
+ o(1)
\, .
\phantom{xxxxx}
\label{6VIII21.t1s}
\end{eqnarray}
\ptcheck{28IV22; sprawdzone poza ostatnia rownoscia; stopping here in this section}

Comparing \eqref{24VI21.t2a} with \eqref{24VI21.t2}, we see that an analogous formula for $ {d C [\LKbst,\mcC_{u,R}]}/{d u }$ can be obtained from \eqref{6VIII21.t1s} by
replacing  $\alpha$ by $-\alpha$ and $\LKmcon_{i}$   by $\LKbcon_{i}$:
\begin{eqnarray}
\frac{d C [\LKbst,\mcC_{u,R}]}{d u }
&=&
- 2
\int_{\mcC_{u,R}}
\Big\{e^{-\alpha u} \mathcal{T}^{[\sigma}  \Big[\LKbcon_{i} \wtx ^i \HscaaI^{\mu]}
+\big(\alpha r -1\big)\LKbcon_{i} \wtx ^i \HscaaII^{\mu]}
\nonumber
\\
& &
+\frac{\alpha r -1}{r} \zspaceD^{A}(\LKbcon_{i} \wtx ^i) \HscaA^{\mu]}
+\LKbst^{\mu]} \Hscab
\Big]
\Big\}
\, dS_{{{\sigma}} \mu}
\nonumber
\\
&=&
\int_{S_{R}}
e^{-\alpha u} \LKbcon_{i} \wtx ^i \Big\{
-\alpha  {\oone \SField} \Big[
\alpha^{2} {\oone \SField}+ m^{2} {\oone \SField} -2 \alpha \partial_{u} {\oone \SField}
\Big] R
\nonumber
\\
& &
+ \Big[
2 \alpha^{2} \partial_{u} {\otwo \SField}
+\alpha^{2} {\oone \SField}
+m^{2} {\oone \SField}
-4 \alpha^{3} {\otwo \SField}
-2 \alpha m^{2} {\otwo \SField}
\Big] {\oone \SField}
\nonumber
\\
& &
+4 \alpha^{2}\partial_{u} {\oone \SField}  {\otwo \SField}
-2\left(\partial_{u} {\oone \SField}\right)^{2}
\Big\}
d\mu_{\zzhTBW}
+ o(1)
\, .
\phantom{xxxxx}
\label{6X22.h1}
\end{eqnarray}

\section{An alternative Lagrangian for the scalar field}
\label{s27II23.2}

The Lagrangian for a conformally-covariant scalar field theory on the de Sitter background reads
\begin{equation}\label{28I2022.t1}
	\mcL =- \frac 12  \sqrt{|\det g|}
	\big(
	g^{\mu\nu}\nabla_\mu \SField \, \nabla_\nu \SField+ 2 \alpha^2 \SField^2
	\big)
	\,,
\end{equation}
which coincides with \eqref{9VII21.1} with   $m^2=2 \alpha^2$. With some work this can be rewritten as
\begin{equation}\label{28I2022.t2}
	\frac{\mcL}{\sqrt{|\det g|}} =- \frac{1}{2 r^2}
	\Big(
	g^{\mu\nu}\nabla_\mu \big( r \SField \big) \, \nabla_\nu \big( r \SField \big) \Big)
	-
	\frac{1}{4} \nabla_{\nu} \Big(\big(r \SField \big)^2 \nabla^{\nu}r^{-2}  \Big)
	\,.
\end{equation}

Since boundary terms in a Lagrangian do not change the Euler--Lagrange equations,  after
neglecting the boundary term in \eqref{28I2022.t2}  we obtain a Lagrangian which leads us to an equivalent theory
\begin{equation}
	\label{28I2022.t3}
	\amcL =- \frac{1}{2 r^2} \sqrt{|\det g|}
	g^{\mu\nu}\nabla_\mu   \aSField   \, \nabla_\nu   \aSField
	\, ,
\end{equation}
where
$$
\aSField=r \SField
\,.
$$
As already announced, all Noether charges turn out to be finite, no renormalisation is required. The price is that the time-derivatives of some charges are not boundary integrals anymore, because both the Lagrangian and the Hamiltonian depend explicitly on the coordinate $r$ now.

The canonical momentum for \eqref{28I2022.t3} reads
\begin{equation}
	\api^{\alpha}=\frac{\partial\amcL}{\partial (\nabla_{\alpha} \aSField) }=-\sqrt{|\det g|} \frac{1}{r^2}\nabla^{\alpha}  \aSField  \, .
\end{equation}

The canonical energy-momentum current equals
\begin{equation}\label{28I2022.t4}
	\amcH^\mu [X] = -  \sqrt{|\det g|}
	\Big( \frac{1}{r^2} \nabla^\mu  \aSField   \,\Lie_X \aSField -
	\frac{1}{2 r^2}
	\big(\nabla^\alpha   \aSField   \, \nabla_\alpha   \aSField
	\big)
	X^\mu
	\Big)
	\,.
\end{equation}

The large-r asymptotic behaviour of $\SField$ of \eqref{13VIII21.t1} translates into the following asymptotics for $\aSField$:
\begin{equation}\label{30I22.t1}
	\aSField(u,r, x^A) =
	\ozero \aSField(u,x^A)
	+ \frac{\oone \aSField(u,x^A)}{r}
	+ \frac{\otwo \aSField(u,x^A)}{r^2} + ...
	\, .
\end{equation}

\subsection{Noether charges}
\label{s27II23.1}

The analysis of  the Noether charges associated with the Killing fields \eqref{24VI21.t1}-\eqref{24VI21.t2} proceeds now in a way completely analogous to that for $\SField$.
The charges are again of the form \eqref{5II21.101}-\eqref{17X21.33}, where now
%
%
%
\begin{eqnarray}
	\HscaaI^{u}&=&\big(
	\partial_{r}   \aSField   \partial_{u} \aSField
	\big)\sqrt{\det \zzhTBW }
	\, ,
	\\
	\label{7III22.t1}
	\HscaaI^{r}&=&
	\big(\partial_{u}  \aSField+\big(\alpha^{2} r^{2}-1\big)\partial_{r} \aSField \big) \partial_{u} \aSField
	\sqrt{\det \zzhTBW }
	\, ,
	\\
	\HscaaII^{u}&=&  \big(\partial_{r} \aSField\big)^2\sqrt{\det \zzhTBW }
	\, ,
	\\
	\HscaaII^{r}&=&
	\big(\partial_{u} \aSField+\big(\alpha^{2} r^{2}-1\big)\partial_{r}  \aSField  \big)\partial_{r} \aSField
	\sqrt{\det \zzhTBW }
	\, ,
	\\
	\HscaA^{u}&=&
	\partial_{r} \aSField \zspaceD_{A} \aSField
	\sqrt{\det \zzhTBW }
	\, ,
	\\
	\label{7III22.t2}
	\HscaA^{r}&=&
	\big(\partial_{u} \aSField+\big(\alpha^{2} r^{2}-1\big)\partial_{r} \aSField\big)\zspaceD_{A} \aSField
	\sqrt{\det \zzhTBW }
	\, ,
	\phantom{xxxxx}
	\\
	\Hscab &=&\frac{1}{2}\Big(\frac{1}{r^2}\zh^{A B}\zspaceD_{A} \aSField \zspaceD_{B} \aSField
	-2 \partial_{r} \aSField\partial_{u} \aSField
	\nonumber
	\\
	& &
	+\big(1-\alpha^{2} r^{2}\big)\big(\partial_{r} \aSField\big)^{2}\Big)
	\sqrt{\det \zzhTBW }
	\, .
	\label{7III22.t2b}
\end{eqnarray}

The asymptotic behaviour \eqref{30I22.t1} leads to
\begin{eqnarray}
	\nonumber
	\lefteqn{
		{\amcE  _\mcH [} \mcC_{u ,R}]  :=
		\int_{\mcC_{u,R }}  \amcH^\mu [\partial_u] dS_\mu
		=
		\int_{\mcC_{u,R }}  \amcH^u [\partial_u] dS_u
	}
	&&
	\\
	&= &   \int_{\mcC_{u,R }} \big(\HscaaI^{u}+\Hscab
	\big)
	\,dr\,dx^2 dx^3
	\nonumber
	\\
	&=&
	\frac{1}{2}
	\int_{\mcC_{u,R }}
	\underbrace{
		\Big( \frac{1}{r^2}
		\zh^{A B}\zspaceD_{A} \aSField \zspaceD_{B} \aSField
		+
		\big(1 - \alpha^{2} r^{2}\big)\big(\partial_{r} \aSField\big)^{2}
		\Big)
	}_ { O(r^{-2})}
	\,dr \, d\mu_{\zzhTBW}
	\,,
	\phantom{xxxxx}
	\label{31I22.t1}
\end{eqnarray}
hence the volume integral has a finite limit as $R$ tends to infinity, resulting in a finite total energy.

As before, the total angular-momentum is obtained from the following integral:
\begin{eqnarray}
	\amcJ[  \mcC_{u,R }]&:=&
	\int_{\mcC_{u,R }}  \amcH^\mu [\mathcal{R}] dS_\mu
	\equiv  R_i \amcJ^i [\mcC_{u,R }]
	\,,
	\label{31I22.t2}
\end{eqnarray}
where now
\begin{eqnarray}
	\amcJ^i[\mcC_{u,R }]&:=&
	\int_{\mcC_{u,R }}
	\varepsilon^{A B} \zspaceD_{B} \wtx ^i
	\HscaA^{u}
	\,dr\,dx^2 dx^3
	\nonumber
	\\
	&  = &
	\int_{\mcC_{u,R }}
	\underbrace{
		\varepsilon^{A B} \zspaceD_{B} \wtx ^i \zspaceD_{A} \aSField
		\partial_{r} \aSField
	}_{
		O({r}^{-2}) }
	\,dr\, d\mu_{\zzhTBW}
	\, ,
	\label{31I22.t3}
\end{eqnarray}
again an integral which converges to a finite value as $R$ tends to infinity.

We further have
\begin{eqnarray}
	\amcP  [\LKmom,\mcC_{u,R }]&:=&
	\int_{\mcC_{u,R }}  \amcH^\mu [\LKmom] dS_\mu
	\nonumber
	\\
	&
	= &
	\LKmcon_{i}\int_{\mcC_{u,R }}
	e^{\alpha u}\Big[ \wtx ^i \HscaaI^{u}
	-\big(\alpha r +1\big)  \wtx ^i \HscaaII^{u}
	-\frac{\alpha r +1}{r} \zspaceD^{A}(  \wtx ^i) \HscaA^{u} +  \wtx ^i \Hscab
	\Big]
	\,dr\,dx^2 dx^3
	\nonumber
	\\
	& = &
	\LKmcon_{i} \int_{\mcC_{u,R }}
	e^{\alpha u}
	\Big[\frac{1}{2}\wtx ^i \Big(
	\frac{1}{r^2}\zh^{A B}\zspaceD_{A} \aSField \zspaceD_{B} \aSField
	-\big(\alpha^{2} r^{2}+2  \alpha+1\big)\big(\partial_{r} \aSField\big)^{2}
	\Big)
	\nonumber
	\\
	& &
	-\frac{(\alpha r+1)}{r} \zspaceD^{A}n^{i}\left(\partial_{r} \aSField\right)\zspaceD_{A}\aSField
	\Big]
	\,dr\, d\mu_{\zzhTBW}
	\nonumber
	\\
	&=&
	\int_{\mcC_{u,R }}
	O(r^{-2})
	\,dr\, d\mu_{\zzhTBW}
	\,.
	\label{31I22.t4}
\end{eqnarray}
Finally,
\begin{eqnarray}
	\amcC  [\LKbst,\mcC_{u,R }]&:=&
	\int_{\mcC_{u,R }}  \amcH^\mu [\LKbst] dS_\mu
	\nonumber
	\\
	&
	= &
	\LKbcon_{i}\int_{\mcC_{u,R }} \big(
	e^{-\alpha u}\Big[ \wtx ^i \HscaaI^{u}
	+\big(\alpha r -1\big)  \wtx ^i \HscaaII^{u}
	+\frac{\alpha r -1}{r} \zspaceD^{A}(  \wtx ^i) \HscaA^{u} +  \wtx ^i \Hscab
	\Big]
	\big)
	\,dr\,dx^2 dx^3
	\, .
	\nonumber
	\\
	& &
	\phantom{xxxxxx}
	\label{31I22.t5}
\end{eqnarray}
A more detailed  version of the integral \eqref{31I22.t5}, which is again finite in the limit $R\to\infty$, can be obtained from \eqref{31I22.t4} by
replacing there $\alpha$ by $-\alpha$ and $\LKmcon_{i}$ by $\LKbcon_{i}$.

\subsection{Time derivatives}
\label{ss3XI22.5}

Recall that the Lie derivatives of the {Noether current}  read
\begin{eqnarray}
	\Lie_{Y} \amcH^{\mu}
	&=&
	\nabla_{\alpha}\left(Y^{\alpha} \amcH^{\mu}\right)
	-\amcH^{\alpha} \nabla_{\alpha} Y^{\mu}
	\nonumber
	\\
	&=&
	2 \nabla_{\alpha}\left(Y^{[\alpha} \amcH^{\mu]}\right)
	+Y^{\mu} \nabla_{\alpha} \amcH^{\alpha}
	\, .
	\label{6III22.t1}
\end{eqnarray}
We associate Hamiltonian density with the canonical energy-momentum tensor through the formula
\begin{equation}
	\amcH^{\mu}[X]=\amcT^{\mu}{}_{\alpha} X^{\alpha} \, ,
\end{equation}
where
\begin{equation}
	\amcT^{\mu}{}_{\alpha}=\tpi^{\mu} \nabla_{\alpha} \aSField-\delta^{\mu}_{\alpha} \amcL
	\, .
\end{equation}
Since the alternative Lagrangian \eqref{28I2022.t2} depends explicitly upon the coordinate $r$, for solutions of the field equations we find
\ptcheck{4X22}
\begin{equation}
	\nabla_{\mu} \amcH^{\mu}
	= \nabla_{\mu} \Big(\amcT^{\mu}{}_{\alpha} X^{\alpha}\Big)
	=
	\amcT^{\mu}{}_{\alpha} \nabla_\mu X^\alpha
	-\frac{\partial \amcL  }{\partial r} X^{\alpha} \partial_{\alpha} r
	\, .
	\label{6III22.t2}
\end{equation}
Using \eqref{6III22.t1}, \eqref{6III22.t2}, and assuming that $X$ is Killing vector field, the Lie derivative of the {Noether current} reads
\begin{equation}
	\Lie_{Y} \amcH^{\mu}
	=
	2
	\nabla_{\alpha}\left[Y^{[\alpha} \amcH^{\mu]}\right]
	\nonumber
	\\
	-
	Y^{\mu} \frac{\partial \amcL  }{\partial r}
	X^{\alpha} \partial_{\alpha} r
	\, .
	\label{6III22.t3}
\end{equation}
Using \eqref{28I2022.t3} and \eqref{28I2022.t4} one obtains
\ptcheck{4X22}
\begin{eqnarray}
	\Lie_{Y} \amcH^{\mu}
	&=&
	-  2 \sqrt{|\det g|} \nabla_{\alpha}\left[Y^{[\alpha}  \Big( \frac{1}{r^2} \nabla^{\mu]}  \aSField   \,\Lie_X \aSField -
	\frac{1}{2 r^2}
	\nabla^\beta   \aSField   \, \nabla_\beta   \aSField
	X^{\mu]}
	\Big)\right]
	\nonumber
	\\
	& &
	-
	Y^{\mu}
	\frac{1}{r^3} \sqrt{|\det g|}
	g^{ \nu\rho }\nabla_\nu   \aSField   \, \nabla_\rho   \aSField
	X^{\alpha} \partial_{\alpha} r
	\, .
	\label{6III22.t4}
\end{eqnarray}

The ``non-divergence term'' $(...) Y^\mu X^\alpha \partial_\alpha r$ in this equation   implies that some charges might have volume terms in their evolution formulae. No such terms will certainly occur when either $Y$ is tangent to $\hyp$ (which will be the case for rotations), or when $r$ is invariant under the flow of $X$ (which will be the case for $u$-translations and rotations).

For instance, consider \eqref{6III22.t4} with  $X=Y=\partial_u \equiv \mathcal{T}$.
In this case  \eqref{5X22.t1} applies,   with the relevant component given by \eqref{7III22.t1}.
Passing to the limit $R\to\infty$ in  \eqref{5VIII21.101s} with \eqref{5X22.t1} and \eqref{7III22.t1}   one obtains
\ptcheck{5X22}
\begin{eqnarray}
	\frac{d {\amcE  _\mcH [} \mcC_{u }]}{d u }
	& = &
	-
	2 \int_{\partial \hyp_\tau}
	\mathcal{T}^{[\sigma}\HscaaI^{\mu]} \, dS_{{{\sigma}} \mu}
	\nonumber
	\\
	&=&
	-\lim\limits_{R \to \infty}\int_{S_{R}} \HscaaI^{r}\Big{|}_{r=R} dx^2 dx^3
	\nonumber
	\\
	&=&
	-
	\lim\limits_{R \to \infty}\int_{S_{R}}
	\Big[
	\left(\partial_{u} \aSField+\left(\alpha^{2} r^{2}-1\right)\partial_{r} \aSField\right)\partial_{u} \aSField
	\Big]_{r=R}
	d\mu_{\zzhTBW}
	\nonumber
	\\
	&=&
	\int_{S}
	\big(\alpha^{2} {\oone \aSField}-\partial_{u} {\ozero \aSField}\big)\partial_{u} {\ozero \aSField}
	d\mu_{\zzhTBW}
	\, .
	\label{8X22.5}
\end{eqnarray}

As another example, the $u$-derivative of angular momentum is obtained from \eqref{5VIII21.101s}, 
and  \eqref{5X22.t2} with  $X= \mathcal{R}$ and $Y=\partial_u \equiv \mathcal{T}$:
\begin{eqnarray}
	\frac{d \amcJ[  \mcC_{u}]}{d u }
	&=&
	-
	2\int_{\partial \hyp_\tau}
	\mathcal{T}^{[\sigma}
	\Big( \HscaA^{\mu]}{ \varepsilon^{A B} \zspaceD_{B}(R_i \wtx ^i)}
	+
	\mathcal{R}^{\mu]}\Hscab
	\Big)
	\, dS_{{{\sigma}} \mu}
\nonumber
\\
&=&
-
R_i \lim\limits_{R \to \infty} \int_{S_{R}} \Big[
\HscaA^{r}{ \varepsilon^{A B} \zspaceD_{B} \wtx ^i}
\Big]_{r=R} dx^2 dx^3=:R_i \frac{d \amcJ^{i}}{d u }
\,.
\phantom{xxx}
\end{eqnarray}
Using \eqref{7III22.t2} we find
\ptcheck{6X22}
\begin{eqnarray}
\frac{d \amcJ^{i}}{d u }
&=&
-\lim\limits_{R \to \infty}\int_{S_{R}}
\sqrt{\det \zzhTBW }
{ \varepsilon^{A B} \zspaceD_{B} \wtx ^i} \Big[
\big(\partial_{u} \aSField+\left(\alpha^{2} r^{2}-1\right)\partial_{r} \aSField\big)
\zspaceD_{A} \aSField
\Big]_{r=R}d\mu_{\zzhTBW}
\nonumber
\\
&=&
\int_{S}
{ \varepsilon^{A B} \zspaceD_{B} \wtx ^i }
\Big(2 \alpha^{2} {\oone \aSField}-\partial_{u} {\ozero \aSField}\Big)\zspaceD_{A} {\ozero \aSField}
d\mu_{\zzhTBW}
\, ,
\phantom{xxxxx}
\end{eqnarray}
which does not coincide with  \eqref{6X22.h2}.

The remaining $u$-derivatives have both fluxes and volume integrals. For instance,
calculating similarly to \eqref{6VIII21.t1s}
and taking into account the volume term in \eqref{6III22.t4},
\begin{eqnarray}
\frac{d\amcP[\LKmom,\mcC_{u}]}{d u }
&=&
-
2\int_{\partial \hyp_\tau}
\Big[e^{\alpha u}
\mathcal{T}^{[\sigma}
\Big(\LKmcon_{i} \wtx ^i \HscaaI^{\mu]}
-\big(\alpha r +1\big)\LKmcon_{i} \wtx ^i \HscaaII^{\mu]}
\nonumber
\\
& &
-\frac{\alpha r +1}{r} \zspaceD^{A}(\LKmcon_{i} \wtx ^i) \HscaA^{\mu]}
+\LKmom^{\mu]} \Hscab
\Big)
\Big]\, dS_{{{\sigma}} \mu}
\nonumber
\\
& &
-\int_{ \hyp_\tau} \frac{2}{r}\Big[\mathcal{T}^{\mu} \Hscab \LKmom^{r} \Big] dS_\mu
\nonumber
\\
&=&
-
\int_{\partial \hyp_\tau}
\Big[e^{\alpha u}
\Big(\LKmcon_{i} \wtx ^i \HscaaI^{r}
-\big(\alpha r +1\big)\LKmcon_{i} \wtx ^i \HscaaII^{r}
\nonumber
\\
& &
-\frac{\alpha r +1}{r} \zspaceD^{A}(\LKmcon_{i} \wtx ^i) \HscaA^{r}
-\big(\alpha r +1\big) \Hscab
\Big)
\Big]\, dS_{{{\sigma}} \mu}
\nonumber
\\
& &
+\int_{ \hyp_\tau} \frac{2}{r}\Big[ \Hscab \big(\alpha r +1\big)\LKmcon_{i} \wtx ^i \Big]
dr \, dx^2 dx^3
\nonumber
\\
&=&
-
\int_{S_{R}} e^{\alpha u} \LKmcon_{i}\Big\{
\wtx ^i\Big[\Big(\partial_{u} \aSField+\left(\alpha^2 r^{2}-1\right)\partial_{r} \aSField\Big)\partial_{u} \aSField
\nonumber
\\
& &
-(\alpha r+1) \left(\partial_{u} \aSField+\left(\alpha^{2} r^{2}-1\right)\partial_{r} \aSField\right)\partial_{r} \aSField
\nonumber
\\
& &
-\frac{1}{2}(\alpha r+1)\Big(
\frac{1}{r^2}\zh^{A B}\zspaceD_{A} \aSField \zspaceD_{B} \aSField
-2\left(\partial_{r} \aSField\right)\left(\partial_{u} \aSField\right)
\nonumber
\\
& &
-\left(\alpha^{2} r^{2}-1\right)\left(\partial_{r} \aSField\right)^{2}\Big)\Big]
\nonumber
\\
& &
-\frac{(\alpha r+1)}{r}   \zspaceD^{A} \wtx ^i \partial_{r} \aSField \zspaceD_{A} \aSField
\Big\}_{r=R}
d\mu_{\zzhTBW}
\nonumber
\\
& &
+\int_{ \hyp_\tau} \frac{1}{r}\Big[ \Big(\frac{1}{r^2}\zh^{A B}\zspaceD_{A} \aSField \zspaceD_{B} \aSField
-2 \partial_{r} \aSField\partial_{u} \aSField
\nonumber
\\
& &
+\big(1-\alpha^{2} r^{2}\big)\big(\partial_{r} \aSField\big)^{2}\Big)
\sqrt{\det \zzhTBW } \big(\alpha r +1\big)\LKmcon_{i} \wtx ^i \Big] dS_\mu
\nonumber
\\
&=&
\int_{\partial \hyp_\tau}
e^{\alpha u} \LKmcon_{i} \wtx ^i \Big[
\alpha^2 {\oone \aSField}
-\partial_{u} {\ozero \aSField}
\Big]\partial_{u} {\ozero \aSField}
d\mu_{\zzhTBW}
\nonumber
\\
& &+\int_{ \hyp_\tau} O(r^{-2})
\,dr\, d\mu_{\zzhTBW}
\, .
\phantom{xxxxx}
\label{8III22.t1}
\end{eqnarray}
It is not clear whether a meaningful comparison to \eqref{6VIII21.t1s} is possible because of the volume term appearing here.

A formula for $\frac{d C [\LKbst,\mcC_{u,R}]}{d u }$ can be obtained from \eqref{8III22.t1} by
replacing there  $\alpha$ by $-\alpha$ and $\LKmcon_{i}$   by $\LKbcon_{i}$.

\section{Poisson algebras}
 \label{s6II21.1}

Having obtained a set of  global charges, either directly or after renormalisation, the question arises whether the charges satisfy a well-defined Poisson algebra. As we will see, the question is far from clear, because of the boundary integrals arising when varying the charges.

Quite generally, we consider two Hamiltonian functionals,  $H[\hyp,X]$ and $H[\hyp, Y]$, defined as integrals on a hypersurface $\hyp$ with boundary $\partial \hyp$, with two  vector field $X$ and $Y$. Here the boundary might be at finite distance,
before a limit to infinity is taken, or it can be a boundary at infinity in the conformally compactified spacetime.
We take an approach similar to that of~\cite{BrownHenneaux} to define the Poisson algebra of charges  through the Poisson algebra of fields on $\hyp$. When there are no constraints, as is the case of the scalar field, \emph{and} when there are no boundary terms in the variations, \emph{and} when  $\hyp$ is spacelike, the algebra is straightforward. When the hypersurface is null the algebra of the fields is more demanding.
We avoid the work associated with the last problem by  deforming  $\hyp$ to a   hypersurface which is spacelike, and calculating the Poisson brackets on the deformed  hypersurface. We expect this to give a correct answer in situations where the charges are independent of the hypersurface, within the family of hypersurfaces sharing the same boundary.

The problem of boundary integrals that remain after a variation of the charges has been carried-out, which arises in the situations of interest in this work, will be addressed in Section~\ref{s23IV22.1}.

Let us pass now to an analysis of the Poisson algebra of Noether charges associated with diffeomorphisms generated by two  vector fields $X$ and $Y$. We consider   first order Lagrangian densities depending upon the fields, the metric, and possibly upon coordinates:   $\mcL=\mcL ({\SField^A}, \partial_{\mu} {\SField^A},  g_{\alpha \beta}, x^{\sigma}) $.
The key assumption in this section is that there are no Hamiltonian constraints; thus some of the calculations that follow do not apply to Maxwell fields, which will be discussed elsewhere.

As elsewhere in this work, the Noether current associated with a vector field $X$ reads
\begin{equation}
	\mcH^{\mu}=\frac{\partial \mcL}{\partial (\partial_\mu \SField^A)} \Lie_{X} {\SField^A} - X^{\mu} \mcL \, .
\end{equation}
Given a hypersurface $\hyp =\{x^0=0\}$, one thus obtains
a charge integral
\begin{equation}
	H[\hyp,X]=\int_{\hyp} \mcH^{0} d S_{0} \, .
	\label{19II22.1}
\end{equation}
The canonical momentum on $\hyp$ is defined as
\begin{equation}
	\label{19II22.1a}
	{\pi_A}\equiv
	{\pi_A}^{0}:=
	\frac{\partial \mcL}{\partial\left(\partial_{0} {\SField^A}\right)} \, .
\end{equation}
%
%

\subsection{Hamilton equations}

To avoid ambiguities, variations of fields are defined as follows: given a  one parameter family of fields $\lambda \mapsto \phi^A(\lambda)$ one sets
$$
\delta \phi^A := \frac{d\phi^A}{d\lambda}
\,,
\qquad
\delta \pi_A{}^ {k} := \frac{d\pi_A{}^k}{d\lambda}
\,,
$$
etc.

The calculations that follow are standard. We carry them out in detail in order to keep track of  the boundary terms that arise in the process. We assume that $
\delta X $ vanishes, in particular $[\Lie_{X},\delta]{\SField^A}=0$.
The variation of the functional \eqref{19II22.1} is defined as
\begin{equation}
	\delta H:=
	\frac{d}{d\lambda}\int_{\hyp} \mcH^{0} d S_{0}
	=\int_{\hyp} \frac{d\mcH^{0}}{d\lambda} d S_{0}
	\equiv \int_{\hyp} \delta \mcH^{0} d S_{0}
	\, ,
\end{equation}
assuming that differentiation under the integral is justified,
with
\begin{eqnarray}
	\delta \mcH^{0}&=& \delta \left(
	{\pi_A} \Lie_{X} {\SField^A} - X^{0} \mcL
	\right)
	\nonumber
	\\
	&=&\Lie_{X} {\SField^A} \delta {\pi_A}
	+{\pi_A} \Lie_{X} \delta {\SField^A}
	-X^{0}\left(
	\frac{\partial \mcL}{\partial {\SField^A}} \delta {\SField^A}
	+\frac{\partial \mcL}{\partial\left(\partial_{\mu} {\SField^A}\right)} \partial_{\mu} \delta {\SField^A}
	\right)
	\nonumber
	\\
	&=&\Lie_{X} {\SField^A} \delta {\pi_A}
	-\Lie_{X} {\pi_A} \delta {\SField^A}
	+\Lie_{X}\left( {\pi_A} \delta {\SField^A} \right)
	\nonumber
	\\
	& & -X^{0}\Big\{
	\underbrace{
		\Big[
		\frac{\partial \mcL}{\partial {\SField^A}}
		-\partial_{\mu}\Big(\frac{\partial \mcL}{\partial (\partial_{\mu} {\SField^A} )}\Big)
		\Big]
	}_{\mathcal{E}_{A}}
	\delta {\SField^A}
	+\partial_{\mu}\Big(
	\frac{\partial \mcL}{\partial (\partial_{\mu}{\SField^A} )}
	\delta {\SField^A}\Big)
	\Big\}
	\,,
	\phantom{xxxxx}
	\label{7III22.t3}
\end{eqnarray}
where the
vanishing of $\mathcal{E}_{A}$ is the contents of the Euler--Lagrange equations.
If we assume additionally that ${\pi_A} \delta {\SField^A}\equiv {\pi_A}^{0} \delta {\SField^A}$ is the $0$-component of a vector density, using the definition of the Lie derivative of a vector density we find
\begin{equation}
	\Lie_{X}({\pi_A} \delta {\SField^A})
	=\Lie_{X}({\pi_A}^{0} \delta {\SField^A})
	=\partial_{\alpha} (X^{\alpha}{\pi_A}^{0} \delta {\SField^A})- \delta {\SField^A} {\pi_A}^{\alpha} \partial_{\alpha} X^{0}
	\, .
	\label{7III22.t4}
\end{equation}
Inserting \eqref{7III22.t4} into \eqref{7III22.t3} we obtain
\begin{eqnarray}
	\delta \mcH^{0}&=&\Lie_{X} {\SField^A} \delta {\pi_A}
	-\Lie_{X} {\pi_A} \delta {\SField^A}
	+\partial_{\alpha} (X^{\alpha}{\pi_A}^{0} \delta {\SField^A})- \delta {\SField^A} {\pi_A}^{\alpha} \partial_{\alpha} X^{0}
	\nonumber
	\\
	& & -X^{0}\Big\{
	{{\mathcal{E}_{A}}} \delta {\SField^A}
	+\partial_{\mu}\Big({\pi_A}^{\mu} \delta {\SField^A}\Big)
	\Big\}
	\phantom{xxxxx}
	\nonumber
	\\
	&=& \Lie_{X} {\SField^A} \delta {\pi_A}
	-\left[
	\Lie_{X} {\pi_A}
	+X^{0} {\mathcal{E}_{A}}
	\right] \delta {\SField^A}
	\nonumber
	\\
	& &
	+\partial_{\mu}\Big[
	(X^{\mu} {\pi_A}
	-X^{0} {\pi_A}^{\mu} ) \delta {\SField^A}
	\Big]
	\, .
	\label{14II2022.t2}
\end{eqnarray}
Here $\Lie_X \pi_A$ is understood as the $\alpha=0$-component of the field
$\Lie_X \pi_{A}{}^{ \alpha}$.

Recall  that by assumption the Lagrangian, and therefore also $\mcH$, is a functional of the fields and their first derivatives. \emph{Let us suppose} that the equation defining $\pi_A$ can be inverted to express the $x^0$-derivative  of $\SField^A$ as a function of $\pi_A$, $\SField^A$, and  of the  derivatives of fields $\SField^A$ in directions tangential to $\hyp$ (which we denote by  $\partial_k\SField^A$); we emphasise that \eqref{14II2022.t2} holds regardless of whether or not this assumption is true.
Reexpressing $\mcH^0$ as a functional of $\pi_A$ and $\SField^A$ we can then calculate as
\begin{eqnarray}
	\delta \mcF
	&=&
	\underbrace{
		\frac{\partial \mcF}{\partial {\pi_A}} }_{=:
		\frac{\delta \mcF}{\delta {\pi_A}}
	} \delta {\pi_A}
	+\frac{\partial \mcF}{\partial {\SField^A}} \delta {\SField^A}
	+\frac{\partial \mcF}{\partial\left(\partial_{k} {\SField^A}\right)} \delta \partial_{k} {\SField^A}
	\nonumber
	\\
	&=&
	\frac{\delta \mcF}{\delta {\pi_A}} \delta {\pi_A}
	+\underbrace{\left(
		\frac{\partial \mcF}{\partial {\SField^A}}
		-\partial_{k}\left(
		\frac{\partial \mcF}{\partial\left(\partial_{k} {\SField^A}\right)}
		\right)\right)}_{=:\frac{\delta \mcF}{\delta {\SField^A}}} \delta {\SField^A}
	+\partial_{k}\left(
	\frac{\partial \mcF}{\partial \left(\partial_{k} {\SField^A}\right)}
	\delta {\SField^A}\right)
	\, .
	\phantom{xxxxx}
	\label{14II2022.t1}
\end{eqnarray}
In this equation the notation $ \frac{\delta \mcF}{\delta {\pi_A}}$ is somewhat of an overkill: by assumption $\mcL$ depends only on the first derivatives of $\SField^A$, thus $\mcF$ does \emph{not} depend on the derivatives of $\partial_0 \SField^A$, and $ \frac{\delta \mcF}{\delta {\pi_A}}$ is  simply a partial derivative with respect to $\pi_A$.

Comparing \eqref{7III22.t3} with \eqref{14II2022.t1}, for  variations $\delta \pi_A$ and $\delta \SField^A$ of compact support, and supported away from the boundaries of $\hyp$ if any, we find
\begin{equation}
	\int_{\hyp}
	\bigg(
	(\Lie_{X} {\SField^A} - \frac{\delta \mcF}{\delta {\pi_A}})  \delta {\pi_A}
	+
	\big(
	\frac{\delta \mcF}{\delta {\SField^A}}
	-[
	\Lie_{X} {\pi_A}
	+X^{0} \mathcal{E}_{A}
	] \big)
	\delta {\SField^A}
	\bigg)
	d S_{0} \, .
\end{equation}
\emph{If} all the variations $\delta \pi_A$ and $\delta \SField^A$ are independent and arbitrary we can conclude that
\begin{eqnarray}
	\frac{\delta \mcH^{0}}{\delta {\pi_A}}&=&
	\Lie_{X} {\SField^A}
	\, ,
	\label{14II2022.t3}
	\\
	\frac{\delta \mcH^{0}}{\delta {\SField^A}}&=&
	-\Lie_{X} {\pi_A}
	-X^{0} \mathcal{E}_{A}
	\, .
	\label{14II2022.t4}
\end{eqnarray}
We emphasise that the assumptions above are satisfied for a scalar field, but \emph{are not} for a Maxwell field.

It further holds that, for variations that do not necessarily vanish on $\partial \hyp$,
\begin{equation}
	\int_{\partial\hyp}
	\Big(
	\frac{\partial \mcH^{0}}{ \partial  {\SField^A{}_{,k}} }
	-\big(X^{k} {\pi_A}-X^{0} {\pi_A}^{k} \big)
	\Big) \delta {\SField^A} dS_{0k} =0
	\, ,
\end{equation}
where the integration over $\partial \hyp$, when not compact, is understood by exhausting $\hyp$ with a family of compact domains with smooth boundary, and passing to the limit.
In situations where both $\partial \hyp$ and  the variations of the fields on $\partial\hyp$ are arbitrary, we conclude that
\begin{equation}
	\frac{\partial \mcH^{0}}{ \partial  {\SField^A{}_{,k}} }
	=
	X^{k} {\pi_A}-X^{0} {\pi_A}^{k}
	\, .
\end{equation}

\subsection{Algebra of charges}
\label{ss24II23.1}

Consider two functionals $F$ and $G$ depending upon the fields $\pi_A$, $\SField^A$  and the tangential derivatives $\partial_k \SField^A$, which in the adapted coordinates as above take the form
\begin{equation}\label{4IV23.1}
	F = \int_\hyp f(\phi^A,\partial_i \phi^A, \pi^A) \, dS_0
	\,,
	\quad
	G = \int_\hyp g(\phi^A,\partial_i \phi^A, \pi^A) \, dS_0
	\,.
\end{equation}
Following \cite{BrownHenneaux} we set
\begin{equation}
	\{F, G \}_\hyp: =
	\int_{\hyp}\left(
	\frac{\delta f}{\delta {\SField^A}} \frac{\delta g}{\delta {\pi_A}}
	-\frac{\delta f}{\delta {\pi_A}} \frac{\delta g}{\delta {\SField^A}}
	\right) d S_{0}  \, ,
	\label{16II22.2}
\end{equation}
with
$$
\frac{\delta f}{\delta  \SField^A} := \frac{\partial f}{\partial {\SField^A}}
- \partial_i  \left(\frac{\partial f}{\partial  \SField^A{}_{,i}}\right)
\,,
\quad \frac{\delta f}{\delta  \pi_A} \equiv  \frac{\partial f}{\partial \pi_A}
\,,
$$
similarly for $g$.

We note that there is no reason for $\{F,G\}_\hyp$ to be independent of $\hyp$, e.g.\ when the original functionals $F$ or $G$ depend upon $\hyp$. We will, however, see shortly that $\{H_X,H_Y\}_\hyp$ will be independent of $\hyp$, within its homology class, in situations of interest.

We note that the question of boundary terms in the variations of Noether charges arising from flows in spacetime can be shuffled under the carpet by defining instead
\begin{equation}\label{24II23.1}
	\left\{H_{X}, H_{Y}\right\} _{\hyp}:=
	-\Omega_\hyp(\mcL_X\phi,\mcL_Y\phi)
	\,,
\end{equation}
where
\begin{equation}\label{24II23.1b}
	\Omega_\hyp :=
	\int_\hyp \omega^\mu dS_\mu
	\equiv
	\int_\hyp\delta \pi_A\wedge \delta \SField ^A\, dS_\mu
	\,.
\end{equation}
Equation~\ref{24II23.1}
is a special case of \eqref{16II22.2} \emph{whenever} no boundary terms arise in the variations of $H_X$ and $H_Y$. We will, however, use the more fundamental equation \eqref{16II22.2} in our calculation of the left-hand side of \eqref{24II23.1}.

Using \eqref{14II2022.t3} and \eqref{14II2022.t4}, the Poisson bracket of two Hamiltonian functionals $H_X$ and $H_Y$ thus equals
\ptcheck{29 III 22, rechecked TS 10I23 }
\begin{eqnarray}
	\lefteqn{
		\left\{H_{X}, H_{Y}\right\} _{\hyp}=
		\int_{\hyp} \left [
		\frac{\delta \mcH^{0}_{X}}{\delta {\SField^A}} \frac{\delta \mcH^{0}_{Y}}{\delta {\pi_A}}
		-\frac{\delta \mcH^{0}_{X}}{\delta {\pi_A}} \frac{\delta \mcH^{0}_{Y}}{\delta {\SField^A}}
		\right] d S_{0}
	}
	&&
	\nonumber
	\\
	&=&
	-
	\int_{\hyp} \Big[
	\Lie_{Y} {\SField^A} \left(\Lie_{X} {\pi_A}
	+X^{0} \mathcal{E}_{A}
	\right)
	-
	\Lie_{X} {\SField^A} \left(\Lie_{Y} {\pi_A}
	+Y^{0} \mathcal{E}_{A}
	\right)
	\Big ]
	d S_{0}
	\nonumber
	\\
	&=&
	-
	\int_{\hyp} \Big \{\Lie_{X} \Big({\pi_A} \Lie_{Y} {\SField^A} \Big)
	-\Lie_{Y} \Big({\pi_A} \Lie_{X} {\SField^A} \Big) +
	{\pi_A}\left(
	\Lie_{Y} \Lie_{X}  {\SField^A}
	-\Lie_{X} \Lie_{Y}  {\SField^A}
	\right)
	\nonumber
	\\
	& &
	+\mathcal{E}_{A}\left(
	X^{0} \Lie_{Y} {\SField^A}-Y^{0} \Lie_{X} {\SField^A}
	\right)
	\Big \}
	d S_{0}
	\nonumber
	\\
	&=&
	-
	\int_{\hyp} \big \{\Lie_{X} \mcH^{0}_{Y}
	-\Lie_{Y} \mcH^{0}_{X}
	+\Lie_{X}\Big(Y^{0} \mcL\Big)
	-
	\Lie_{Y}\Big(X^{0} \mcL\Big)
	+
	{\pi_A}
	\Lie_{[Y,X]}  {\SField^A}
	\nonumber
	\\
	& &
	+\mathcal{E}_{A}\left(
	X^{0} \Lie_{Y} {\SField^A}-Y^{0} \Lie_{X} {\SField^A}
	\right)
	\big \}
	d S_{0}
	\, .
	\phantom{xxx}
	\label{14II2022.t5}
\end{eqnarray}
The following relations hold
\ptcheck{29III22}
\begin{eqnarray}
	\label{7III22.t6}
	\Lie_{X} \mcH^{ \beta}_{Y}
	-\Lie_{Y} \mcH^{ \beta}_{X}
	&=&2 \partial_{\alpha} \Big(X^{[\alpha} \mcH^{\beta]}_{Y}-Y^{[\alpha} \mcH^{\beta]}_{X}\Big)
	\nonumber
	\\
	& &
	+X^{ \beta} \partial_{\alpha} \mcH^{\alpha}_{Y}-Y^{ \beta} \partial_{\alpha} \mcH^{\alpha}_{X}
	\, ,
	\\
	\label{7III22.t5}
	\Lie_{X}\Big(Y^{ \beta} \mcL\Big)
	-
	\Lie_{Y}\Big(X^{ \beta} \mcL\Big)
	&=&
	2 \partial_{\alpha} \Big(X^{[ \alpha} Y^{\beta]} \mcL\Big)
	+[X,Y]^{ \beta} \mcL
	\,.
\end{eqnarray}
Inserting \eqref{7III22.t5} and \eqref{7III22.t6} into \eqref{14II2022.t5} results in
\begin{eqnarray}
		\left\{H_{X}, H_{Y}\right\}_{\hyp}
		& =
		&
		\int_{\hyp} \Big \{
		\mcH^{\beta}_{[X,Y]}
		-\mathcal{E}_{A}\left(
		X^{\beta} \Lie_{Y} {\SField^A}-Y^{\beta} \Lie_{X} {\SField^A}
		\right)
		\nonumber
		\\
		&&
		+
		Y^{\beta} \partial_{\alpha} \mcH^{\alpha}_{X}
		-X^{\beta} \partial_{\alpha} \mcH^{\alpha}_{Y}
	\nonumber
	\\
	& &
	-2 \partial_{\alpha} \Big(X^{[\alpha} \mcH^{\beta]}_{Y}-Y^{[\alpha} \mcH^{\beta]}_{X}
	+X^{[ \alpha} Y^{\beta]} \mcL \Big)
	\Big \}
	d S_{\beta}
	\, .
	\phantom{xxxxxxx}
	\label{19IX22.1}
\end{eqnarray}
We conclude that:
\begin{proposition}
	\label{P10XII22.1}
	If $ \partial_\alpha \mcH^{\alpha}_{Y}=\partial_\alpha \mcH^{\alpha}_{Y}=\partial_\alpha \mcH^{\alpha}_{[X,Y]}
	=0$, if the field equations are satisfied, and if $\partial \hyp_1 = \partial \hyp_2$, then
	\begin{equation}
		\left\{H_{X}, H_{Y}\right\}_{\hyp_1}=
		\left\{H_{X}, H_{Y}\right\}_{\hyp_2}
		\,.
	\end{equation}
\end{proposition}

\proof
Under the conditions listed the right-hand side of \eqref{19IX22.1} does not depend upon $\hyp$.
\qedskip

Another immediate consequence of  \eqref{19IX22.1}  is:

\begin{proposition}
	\label{P10XII22.2}
	If
	\begin{enumerate}
		\item
		\begin{eqnarray}
			\int_{\partial \hyp} \Big( X^{[ \alpha} Y^{\beta]} \mcL
			+
			\big(X^{[\alpha} \Lie_{Y} {\SField^A}-Y^{[\alpha} \Lie_{X} {\SField^A}\big){\pi_A}^{\beta]}
			\Big)
			d S_{\alpha\beta} =0
			\, ,
		\end{eqnarray}
		\cbk
		and if
		\item  $X$ is tangent to $\hyp$ \emph{or} the field equations are satisfied and $ \partial_\alpha \mcH^{\alpha}_{Y}$ vanishes,
		and if
		\item   $Y$ is tangent to $\hyp$ \emph{or} the field equations are satisfied and $ \partial_\alpha \mcH^{\alpha}_{X}$ vanishes,
	\end{enumerate} then it holds that
	\begin{eqnarray}
		\left\{H_{X}, H_{Y}\right\}_{\hyp}
		= H_{[X,Y]}
		\, .
	\end{eqnarray}
	\qed
\end{proposition}

A comment on the vanishing of  $ \partial _\alpha\mcH^{\alpha} $ is in order. For this, we recall that in~\cite{ChHMS} theories satisfying the following were considered:

\begin{itemize}
	\item[H1.]
	\label{p6III23.1} $\mcL$ is a scalar density.
	\item[H2.] There exists a notion of derivation with respect to a family of vector fields $X$, which we will denote by $\Lie_X$, which coincides with the usual Lie derivative on vector densities, and which  we will  call  \emph{Lie derivative} regardless of whether or not this is the usual Lie derivative on the remaining fields, such that the following holds:
	\begin{enumerate}
		\item[a)] $\Lie_X$ preserves the type of a field, thus $\Lie_X$ of a scalar density is a scalar density, etc.;
		\item[b)] the field $\pi_A{}^\mu \Lie_X \phi^A$ is a vector density;
		\item[c)] in a coordinate system in which $X={\partial_0}$ we have $\Lie_X= {\partial_0}$;
		\item[d)] $\Lie_X$ satisfies the Leibniz rule.
	\end{enumerate}
	\label{H18XI19.1}
\end{itemize}

In our case the Lagrangian also depends on a background structure, namely the background metric.
Let us collectively denote  background fields by $\psi^I$, with the understanding that if the Lagrangian depends upon both a background field $\chi $ and its derivatives, then these derivatives  appear as a separate entry in $\psi^I = (\chi, \partial_\mu\chi, \ldots)$. As will be seen shortly, under H1-H2 one then has the identity
\begin{equation}\label{30III22.1}
	\partial_\mu \mcH^\mu_X =  \mcE_A   \Lie_X \phi^A - \frac{\partial \mcL}{\partial \psi^I} \Lie_X \psi^I
	\,.
\end{equation}
Hence the divergence of $\mcH^\mu$ vanishes when the field equations are satisfied and the background quantities are invariant under $\mcL$. For the scalar field, or for linearised gravity, this requires $X$ to be a Killing vector field of the background. In the Maxwell case,
the divergence of $\mcH$ also  vanishes for conformal Killing vector fields of the background metric.

The proof of \eqref{30III22.1} is simplest in adapted coordinates as in~\cite{ChHMS}.
An ``explicitly covariant'' proof can be given for  tensor fields, in which case
we can write
$$
\nabla_{\mu} \phi^{A}= \partial_{\mu} \phi^{A} +\Gamma^A{}_{B\mu} \phi^B
\,,
$$
where $\nabla$ is the covariant derivative operator of $g_{\mu\nu}$.
Let us assume, for simplicity, that the Lagrangian  $\mcL \big(\phi^{A}, \partial_{\mu} \phi^{A}  , g_{\mu \nu} , \partial_\sigma g_{\mu \nu}\big)$ depends upon the derivatives of the metric through the connection coefficients only:
$$
\mcL \big(\phi^{A}, \partial_{\mu} \phi^{A}  , g_{\mu \nu} , \partial_\sigma g_{\mu \nu}\big)
= \omcL\big(\phi^{A}, \nabla_{\mu} \phi^{A}  , g_{\mu \nu}\big)
\,.
$$
%
Then
\begin{eqnarray}
	\frac{\partial \overline{\mcL}}{\partial( \nabla_{\mu} \phi^{A})}
	&=&
	\frac{\partial  \mcL }{\partial( \partial_{\mu} \phi^{A})} \equiv \pi_A{}^\mu
	\,,
	\label{12XII22.2}
	\\
	\frac{\partial \overline{\mcL}}{\partial  \phi^{A} }
	&=&
	\frac{\partial \mcL }{\partial \phi^{A} } -
	\frac{\partial  \mcL }{\partial( \partial_{\mu} \phi^{B})}\Gamma ^B{}_{A \mu}  = \frac{\partial  \mcL }{\partial \phi^{A} } - \pi_B{} ^\mu \Gamma ^B{}_{A \mu}
	\,,
	\label{12XII22.3}
	\\
	\frac{\partial \overline{\mcL}}{\partial( \partial_{\sigma} g_{\alpha\beta})}
	&=&
	\frac{\partial  \mcL }{\partial( \partial_{\mu} \phi^{A})}
	\frac{\partial\Gamma^A{}_{B\mu}}{\partial( \partial_{\sigma} g_{\alpha\beta})} \phi^B
	\equiv
	\pi_A{}^\mu
	\frac{\partial\Gamma^A{}_{B\mu}}{\partial( \partial_{\sigma} g_{\alpha\beta})} \phi^B
	\,.
	\label{12XII22.2n}
\end{eqnarray}
Assuming that $\mcL$ is a scalar density,
it follows from \eqref{12XII22.3} that the Euler-Lagrange equations
$$
\partial_\mu \pi_A{}^\mu
= \frac{\partial  \mcL }{\partial \phi^{A} } = \frac{\partial  \omcL }{\partial \phi^{A} } +\pi_B{} ^\mu \Gamma ^B{}_{A \mu}
$$
can be equivalently  written as
$$
\nabla_\mu \pi_A{}^\mu = \frac{\partial {\omcL} }{\partial \phi^{A} }
\,,
$$
since   $\pi_{A}{}^{\alpha} \Lie_{X} \phi^{A} \partial_{\alpha}$ is a vector density.
Further:
\begin{eqnarray}
	\lefteqn{
		\Lie_{X} {\omcL} \big(\phi^{A}, \nabla_{\mu} \phi^{A}, g_{\mu \nu},\partial_\sigma g_{\mu \nu}\big)
	}
	&&
	\nonumber
	\\
	&=&
	\frac{\partial {\omcL}}{\partial \phi^{A}} \Lie_{X} \phi^{A}
	+ \pi_{A}{}^{\alpha} \Lie_{X} \nabla_{\alpha} \phi^{A}
	+\frac{\partial {\omcL}}{\partial g_{\mu \nu}} \Lie_{X} g_{\mu \nu}
	+
	\pi_A{}^\mu \phi^B
	\Lie_X \Gamma^A{}_{B\mu}
	\, ;
	\phantom{xxxxx}
	\label{31III22.t3}
\end{eqnarray}
recall that\ptcheck{10I23 }
\begin{equation}\label{14XII22.p1}
	\Lie_X \Gamma^\alpha_{\beta\gamma}
	= \nabla _\beta \nabla_\gamma X^\alpha + R^\alpha{}_{\beta\sigma\gamma}X^\sigma
	\,,
\end{equation}
which carries over to  $\Lie_X \Gamma^A{}_{B\gamma}$  according to the rank of the tensor field $\phi^A$.

The divergence of the Noether currents reads
\begin{eqnarray}
	\partial_{\alpha} \mcH^{\alpha}
	&=&
	\partial_{\alpha}
	\big(
	\pi_{A}{}^{\alpha} \Lie_{X} \phi^{A}-X^{\alpha} {\mcL} \big)
	=
	\partial_{\alpha}
	\big(
	\pi_{A}{}^{\alpha} \Lie_{X} \phi^{A}-X^{\alpha} {\omcL} \big)
	\nonumber
	\\
	&=&\partial_{\alpha}\big(\pi_{A}{}^{\alpha} \Lie_{X} \phi^{A} \big)
	-\Lie_{X} {\omcL}
	\, ,
	\label{31III22.t2}
\end{eqnarray}
where we used the fact that  the Lagrangian is a scalar density:
\begin{equation}
	\Lie_{X} {\omcL}=\partial_{\alpha}\big(X^{\alpha} {\omcL}\big)
	\, .
\end{equation}
Using again the fact that the vector field $\pi_{A}{}^{\alpha} \Lie_{X} \phi^{A} \partial_{\alpha}$ is a vector density, we note that
$$
\partial_{\alpha}\big(\pi_{A}{}^{\alpha} \Lie_{X} \phi^{A} \big)=\nabla_{\alpha}\big(\pi_{A}{}^{\alpha} \Lie_{X} \phi^{A} \big)
\,.
$$
From  \eqref{31III22.t2} and \eqref{31III22.t3}  we conclude that
\begin{eqnarray}
	\lefteqn{
		\partial_{\alpha} \mcH^{\alpha}
		=
		\nabla_{\alpha}\big(\pi_{A}{}^{\alpha} \Lie_{X} \phi^{A} \big)
	}
	&&
	\nonumber
	\\
	&&
	-
	\Big(
	\frac{\partial {\omcL}}{\partial \phi^{A}} \Lie_{X} \phi^{A}
	+ \pi_{A}{}^{\alpha} \Lie_{X} \nabla_{\alpha} \phi^{A}
	+\frac{\partial {\omcL}}{\partial g_{\mu \nu}} \Lie_{X} g_{\mu \nu}
	+
	\frac{\partial {\omcL}}{\partial \phi^{A}{}_{,\mu}} \phi^B
	\Lie_X \Gamma^A{}_{B\mu}
	\Big)
	\nonumber
	\\
	&=&
	\mcE_A   \Lie_X \phi^A
	- \pi_{A}{}^{\alpha} [\Lie_{X}, \nabla_{\alpha}] \phi^{A}
	- \frac{\partial {\omcL}}{\partial g_{\mu \nu}} \Lie_{X} g_{\mu \nu}
	-
	\frac{\partial {\omcL}}{\partial \phi^{A}{}_{,\mu}} \phi^B
	\Lie_X \Gamma^A{}_{B\mu}
	\, ,
	\nonumber
	\\
	&&
	\label{8V23.t2}
\end{eqnarray}
which provides an explicitly covariant derivation of  \eqref{30III22.1}.

The above treatment applies to any theories of tensor fields with a coordinate-invariant Lagrangian and without constraints, e.g.\ for a scalar field. This does, however, fail for theories where constraints are present, which require further considerations.

\subsection{The Maxwell field}
\label{ss29III22.1}

We turn now our attention to Maxwell fields. 
Unless explicitly indicated otherwise we consider a general Lagrangian
\begin{equation}\label{5V23.41}
	\mcL(A_\mu, \partial_\alpha A_\beta, g_{\rho\sigma})
	\equiv
	\mcL( \partial_{[\alpha} A_{\beta]}, g_{\rho\sigma})
	\,,
\end{equation}
thus $\mcL$ neither involves the undifferentiated potential $A_\mu$ nor derivatives of the metric, and the canonical momentum is antisymmetric:
$$
\pi^{\mu\nu}=\pi^{[\mu\nu]}
\,.
$$

We start by noting that there are several ways to proceed:

\begin{enumerate}
	\item We view $A_\mu$ as a covector field on spacetime, with the Noether currents $\mcH_c$ of \eqref{23V21.t1old}  providing a starting point of   further analysis; or
	\item we view $A_\mu$ as a $U(1)$-gauge field, using the Noether currents $\mcH $ instead; and
	\item in either case we may, or we may not, gauge fix, to address issues arising from the vanishing of the momentum conjugate to $A_0$.
	\item
	Yet another approach is presented in \cite[Chapter~3]{IBBBook}.
\end{enumerate}

We continue by noting that the Lagrangian  \eqref{23V21.1} is a scalar density, so that the condition  H1, p.~\pageref{p6III23.1}  is satisfied in all cases.

Next,  while the replacement of $\Lie_{X}A$ in the Noether currents by $\myLie_{X}A_{\mu}$ as given by \eqref{24V21.t4}  renders the current $\mcH^\mu$ given by \eqref{23V21.t1} manifestly gauge invariant, it  leads to problems with  point c) of H2.
For instance,
if $X=\partial_{0}$ the partial derivative $  \partial_{0} A_{\beta}\equiv \Lie_{\partial_0} A_\beta$ will be equal to
\begin{equation}
	\myLie_{\partial_0} A_\beta = F_{0 \beta}
	=\partial_{0} A_{\beta}-\partial_{\beta} A_{0}
	\label{7XI22.98}
\end{equation}
only in a gauge where
\begin{equation}
	\partial_{\beta} A_{0}\equiv 0 \, .
\end{equation}
But we do not wish to gauge-fix, and therefore we need to revisit the scheme.

\subsubsection{Hamilton's equations}
\label{ss3V23.1}

For future reference we calculate on a general hypersurface $\hyp$, in a general metric, for a general vector field $X$, 
but using adapted coordinates in which
$\hyp =\{x^0=0\}$. 

Choosing the covector-field  approach leads to
the Noether charge integral
\begin{equation}
	H_c[\hyp,X]=\int_{\hyp} \mcH^{0}_c d S_{0} \,,
	\label{31III22.t1}
\end{equation}
cf.\ \eqref{23V21.t1old}, while the  $U(1)$-gauge field approach leads instead to
\begin{equation}
	H[\hyp,X]=\int_{\hyp} \mcH^{0} d S_{0} \, ,
	\label{31III22.t1b}
\end{equation}
where $\mcH^{0}$ is defined by \eqref{24V21.t4}.

The variation of $ H_c[\hyp,X]$ is obtained immediately by setting
$(\SField^A)=(A_\alpha)$ in \eqref{14II2022.t2}:
\begin{eqnarray}
	\delta \mcH^{0}_c&=&
	\Lie_{X} {A_\alpha} \delta {\pi^{\alpha 0}}
	-\left[
	\Lie_{X} {\pi^{\alpha 0}}
	+X^{0} {\mathcal{E}^{\alpha}}
	\right] \delta {A_\alpha}
	\nonumber
	\\
	& &
	+\partial_{\mu}\Big[
	(X^{\mu} {\pi^{\alpha 0}}
	-X^{0}  \pi^{\alpha  \mu} ) \delta {A_\alpha}
	\Big]
	\, ,
	\label{14II2022.t2b}
\end{eqnarray}
where $\Lie_X \pi^{\alpha 0}$ is understood as the $\alpha=0$-component of the field
$\Lie_X \pi^{\alpha \mu}$, and
where $\mcE^\alpha$ denotes the field equations operator,
\ptcheck{6III, convention weird but consistent with \eqref{7III22.t3}}
\begin{equation}\label{11XII22.93c}
	\mathcal{E}^\alpha
	\equiv
	\frac{\delta \mcL}{\delta {A_\alpha}}:=
	\frac{\partial \mcL}{\partial {A_\alpha}}
	-\partial_{\mu}\Big(\frac{\partial \mcL}{\partial (\partial_{\mu} {A_\alpha} )}\Big)
	\equiv
	-
	\partial_\mu {\pi}^{\alpha \mu }
	\,.
\end{equation}
Integration gives
\begin{eqnarray}
	\int_\hyp \delta \mcH^{0}_c
	& = &
	\int_{\hyp}
	\Big[\Lie_{X}A_i
	\delta {\pi^{i 0}}
	-\left[
	\Lie_{X} {\pi^{i 0}}
	+X^{0} {\mathcal{E}^i}
	\right] \delta {A_i }
	- X^{0} {\mathcal{E}^0}
	\delta {A_0 }
	\Big] dS_0
	\nonumber
	\\
	& &
	+ \int_{\partial \hyp}  \Big[ (X^{i} {\pi^{ j 0}}
	-X^{0} \pi^{j i} ) \delta A_{j}
	- X^{0} \pi^{0i}  \delta {A_0}
	\Big] dS_{0i}
	\, .
	\label{13XII22.41intb}
\end{eqnarray}

For the variations of $ H [\hyp,X]$, one can   recycle the calculations leading to \eqref{14II2022.t2b} by  rewriting $ \mcH^\mu [X]$ as
\ptcheck{6III23}
\begin{eqnarray}
	\mcH^\mu [X]
	& = &   \pi^{\beta \mu}
	\myLie_X A_\beta
	-\mcL
	X^\mu
	\nonumber
	\\
	& = &   \pi^{\beta \mu }
	\big( \Lie_X A_\beta +
	\underbrace{X^{\alpha} F_{\alpha \beta} - \Lie_X A_\beta
	}_{
		- \partial_\beta (X^\alpha A_\alpha)
	}
	\big)
	-\mcL
	X^\mu
	\nonumber
	\\
	& = &
	\underbrace{
		\pi^{\beta \mu}  \Lie_X A_\beta
		-\mcL
		X^\mu
	} _{{
			\mcH^\mu_c [X]}}
	- \partial_\beta \Big (\pi^{\beta \mu}  X^\alpha A_\alpha
	\Big)
	+ X^\alpha A_\alpha  \,
	\underbrace{
		\partial_\beta
		\pi^{\beta \mu}
	}_{
		\mcE^\mu  }
	\,.
	\phantom{xxxxx}
	\label{11XII22.96}
\end{eqnarray}
(Setting  $\mu=0$, we observe the well known fact that $\mcH^0$ and $\mcH^0_c$ differ by a divergence when the constraint equation $\mcE^{ 0}=0$ holds.)
We can apply   \eqref{14II2022.t2b} to  the first two terms in the right-hand side of \eqref{11XII22.96}, obtaining thus
\ptcheck{6III23, by TS}
\begin{eqnarray}
	\delta \mcH^{0}
	&=&
	\underbrace{
		\Lie_{X}A_\nomu
	}_{
		\myLie_{X}A_\nomu + \partial_\nomu (X^\nonealpha A_\nonealpha)
	}
	\delta {\pi^{\nomu 0}}
	-\left[
	\Lie_{X} {\pi^{\nomu 0}}
	+X^{0} {\mathcal{E}^\nomu}
	\right] \delta {A_\nomu}
	\nonumber
	\\
	& &
	+\partial_{\nomu}\Big[
	(X^{\nomu} {\pi^{\nonealpha0}}
	-X^{0} \pi^{\nonealpha \nomu} ) \delta {A_\nonealpha}
	\Big]
	- \delta\partial_\alpha \Big (\pi^{\alpha 0}  X^\beta  A_\beta
	\Big)
	-  \delta\big(X^\alpha A_\alpha  \, \mcE^0
	\big)
	\nonumber
	\\
	&=&
	\myLie_{X}A_\nomu
	\delta {\pi^{\nomu 0}}
	-\left[
	\Lie_{X} {\pi^{\nomu 0}}
	+X^{0} {\mathcal{E}^\nomu}
	- X^{\nomu} {\mathcal{E}^0}
	\right] \delta {A_\nomu}
	\nonumber
	\\
	& &
	+ \partial_{\nomu}\Big[ (X^{\nomu} {\pi^{\nonealpha0}}
	-X^{0} \pi^{\nonealpha \nomu}
	- X^{\nonealpha} \pi^{\nomu 0} ) \delta {A_\nonealpha}
	\Big]
	\nonumber
	\\
	&=&
	\myLie_{X}A_i
	\delta {\pi^{i 0}}
	-\left[
	\Lie_{X} {\pi^{i 0}}
	+X^{0} {\mathcal{E}^i}
	- X^{i} {\mathcal{E}^0}
	\right] \delta {A_i }
	\nonumber
	\\
	& &
	+ \partial_{i}\Big[ (X^{i} {\pi^{ j 0}}
	-X^{0} \pi^{j i}
	- X^{j} \pi^{i 0} ) \delta {A_j}
	\Big]
	\, ,
	\label{13XII22.41}
\end{eqnarray}
where we use lower-case latin indices for coordinates on  $\hyp$. Note that   $\delta A_0$ vanished from this formula.
Equation~\eqref{13XII22.41} is the  $U(1)$-gauge-field-equivalent of \eqref{14II2022.t2b}, and leads to
\begin{eqnarray}
	\int_\hyp \delta \mcH^{0}
	&=&
	\int_{\hyp}
	\Big[\myLie_{X}A_i
	\delta {\pi^{i 0}}
	-\left[
	\Lie_{X} {\pi^{i 0}}
	+X^{0} {\mathcal{E}^i}
	- X^{i} {\mathcal{E}^0}
	\right] \delta {A_i }
	\Big] dS_0
	\nonumber
	\\
	& &
	+ \int_{\partial \hyp}
	\big(X^{i} {\pi^{ j 0}}
	-X^{0} \pi^{j i}
	- X^{j} \pi^{i 0} \big) \delta {A_j}
	\, dS_{0i}
	\, ,
	\label{13XII22.41int}
\end{eqnarray}
where the integration over $\partial \hyp$ might  be understood by exhausting $\hyp$ with a family of compact domains with smooth boundary, and passing to the limit.

To continue we set
\begin{equation}
	\label{31III22.t4a}
	{\pi^{\mu}}:=
	{\pi}^{\mu 0}
\end{equation}
Anti-symmetry of $\pi^{\mu\nu}$ leads to
\begin{equation}\label{7XI22.99}
	{\pi}^{0 }\equiv{\pi}^{0 0}=0
	\,,
\end{equation}
and so the variations of the momenta are \emph{not} arbitrary.
Further, the field equations give in particular
\begin{equation}
	\label{13V23.1}
	\partial_k \pi^k=0
\end{equation}
which gives a constraint on the $\pi^k$'s when the field equations are assumed.

In view of \eqref{7XI22.99}, and keeping in mind that $\delta A_0$ does not appear in \eqref{13XII22.41},  one could be tempted to drop the field $A_0$ from the Hamiltonian formalism altogether. But then treating  $A_\mu$ as a vector field on spacetime, or a gauge-field on a $U(1)$-principal bundle over spacetime, will not make sense. Likewise neither \eqref{7XI22.98} would not make sense, nor the usual expression for the Lie derivative
$$
X^\mu \partial_\mu A_\nu + \partial_\nu X^\mu A_\mu
\,,
$$
should one wish to use this expression instead of \eqref{7XI22.98} in the definition of the Noether charge.
So it is natural to keep the field $A_0$ as part of the variables, even though it does not appear in some equations below.

To continue, recall that the variation $\delta H$ is defined as follows: given any one parameter family of fields $\lambda \mapsto A_{\mu}(\lambda)$ one sets
$$
\delta A_{k} := \frac{dA_k}{d\lambda}
\,,
\qquad
\delta \pi^ {k} := \frac{d\pi^k}{d\lambda}
\,,
$$
etc.
Now,
in the $U(1)$-bundle Hamiltonian picture both the time derivatives of $A_k$ and the space derivatives of $A_0$ are eliminated in terms of $\pi^k$.  We can therefore calculate as follows
\begin{eqnarray}
	\lefteqn{
		\delta H[\hyp,X]
		:=
		\int_{\hyp} \frac{d \mcH^{0}}{d\lambda}  d S_{0}
		\equiv
		\int_{\hyp} \delta \mcH^{0} d S_{0}
	}
	&&
	\nonumber
	\\
	&=&
	\int_{\hyp}
	\Big[\frac{\partial \mcH^{0}}{\partial {\pi^{k}}}   \delta \pi^{k}
	+
	\frac{\partial \mcH^{0}}{\partial {A_{k}}}
	\delta A_{k}
	+
	\frac{\partial \mcH^{0}}{\partial {A_{k,\ell}}}
	\delta \partial_\ell A_{k}
	\Big] dS_0
	\nonumber
	\\
	&=&
	\int_{\hyp}
	\Big[\frac{\partial \mcH^{0}}{\partial {\pi^{k}}}   \delta \pi^{k}
	+
	\Big(\frac{\partial \mcH^{0}}{\partial {A_{k}}}
	- \partial_\ell \big(
	\frac{\partial \mcH^{0}}{\partial {A_{k,\ell}}}
	\big)
	\Big)
	\delta A_{k}
	\Big] dS_0
	\nonumber
	\\
	&&
	+
	\int_{\partial\hyp}
	\frac{\partial \mcH^{0}}{\partial {A_{k,\ell}}}
	\delta   A_{k} \,  dS_{0k}
	\nonumber
	\\
	&\equiv &
	\int_{\hyp}
	\Big[\frac{\delta \mcH^{0}}{\delta {\pi^{k}}}   \delta \pi^{k}
	+
	\frac{\delta \mcH^{0}}{\delta {A_{k}}}
	\delta A_{k}
	\Big] dS_0
	+
	\int_{\partial\hyp}
	\frac{\partial \mcH^{0}}{\partial {A_{k,\ell}}}
	\delta   A_{k} \,  dS_{0\ell}
	\, .
	\phantom{xxx}
	\label{14XII22.51}
\end{eqnarray}
	On the other hand, in the covector-field Hamiltonian approach we find
	\begin{eqnarray}
		\lefteqn{
			\delta H_c[\hyp,X]
			:=
			\int_{\hyp} \frac{d \mcH^{0}_c}{d\lambda}  d S_{0}
			\equiv
			\int_{\hyp} \delta \mcH^{0}_c d S_{0}
		}
		&&
		\nonumber
		\\
		&=&
		\int_{\hyp}
		\Big[\frac{\delta \mcH^{0}_c}{\delta {\pi^{k}}}   \delta \pi^{k}
		+
		\frac{\delta \mcH^{0}_c}{\delta {A_{\mu}}}
		\delta A_{\mu}
		\Big] dS_0
		+
		\int_{\partial\hyp}
		\frac{\partial \mcH^{0}_c}{\partial {A_{\mu,\ell}}}
		\delta   A_{\mu} \,  dS_{0\ell}
		\, .
		\phantom{xxx}
		\label{14XII22.51b}
	\end{eqnarray}

	Hamilton's equations of motion will be obtained after comparing \eqref{13XII22.41int} with \eqref{14XII22.51}, or \eqref{13XII22.41intb} with \eq{14XII22.51b}.
	
	Now,
	comparison of \eqref{13XII22.41int} with \eqref{14XII22.51} leads to
	\ptcheck{3VI23}
	\begin{eqnarray}
		0 & = & \int_{\hyp}
		\Big[
		\Big(\frac{\delta \mcH^{0}}{\delta {\pi^{k}}} -\myLie_{X}A_k
		\Big) \delta \pi^{k}
		+
		\Big(
		\frac{\delta \mcH^{0}}{\delta {A_{k}}}
		+
		\Lie_{X} {\pi^{k }}
		+X^{0} {\mathcal{E}^k}
		-X^{k} {\mathcal{E}^0}
		\Big)
		\delta A_{k}
		\Big] dS_0
		\nonumber
		\\
		&&
		+
		\int_{\partial\hyp} \Big[
		\frac{\partial \mcH^{0}}{\partial {A_{k,\ell}}}
		- (X^{\ell} {\pi^{k}}
		-X^{0} \pi^{k \ell}
		- X^{k} \pi^{\ell } )\Big]
		\delta   A_{k}\,  dS_{0\ell}
		\, .
		\phantom{xxx}
		\label{14XII22.51comp}
	\end{eqnarray}
	The question then arises  whether or not, and if so how, to take into account  the 
	Gauss constraint  \eqref{13V23.1}. (Strictly speaking, this is the Gauss constraint when the Lagrangian for the standard Maxwell electrodynamics is considered, but we will keep using this terminology for the more general theories considered here.)
	We emphasise that  \eqref{14XII22.51comp}, as well as  \eqref{14XII22.51compb} below, are identities which hold for all variations, whether or not the constraints are satisfied.
	%
	So we have now at least two options:
	
	\begin{enumerate}
		\item  We allow any variations, perhaps but not necessarily assuming that the Gauss constraint is satisfied at the field configuration at which the variation is carried-out; or
		\item we assume that  the Gauss constraint is satisfied, and we restrict ourselves to variations which satisfy this constraint.
		
	\end{enumerate}
	
	Consider, then, Equation~\eqref{14XII22.51comp}. In the first case,   both $\delta A_i$ and $\delta \pi^k$ are arbitrary. Restricting to variations which vanish at $\partial \hyp$ we obtain
	\begin{eqnarray}
		\myLie_{X} A_k & = &  \frac{\delta \mcH^{0}}{\delta {\pi^{k}}}
		\,, \label{15XII22.613}
		\\
		\Lie_{X} {\pi^{k }}
		& = &
		- \frac{\delta \mcH^{0}}{\delta {A_{k}}}
		-X^{0} {\mathcal{E}^k}
		+ X^{k} {\mathcal{E}^0}
		\,.
		\label{15XII22.61}
	\end{eqnarray}
	It then also follows that
	\begin{eqnarray}
		\int_{\partial\hyp} \Big[
		\frac{\partial \mcH^{0}}{\partial {A_{k,\ell}}}
		- (X^{\ell} {\pi^{k}}
		-X^{0} \pi^{k \ell}
		- X^{k} \pi^{\ell } )\Big]
		\delta   A_{k}\,  dS_{0\ell} =0
		\, ,
		\label{12IV23.5}
	\end{eqnarray}
	(If $\delta A_k$ is arbitrary on $\partial \hyp$ we can further conclude that
	$$
	\big(
	\frac{\partial \mcH^{0}}{\partial {A_{k,\ell}}}
	- (X^{\ell} {\pi^{k}}
	-X^{0} \pi^{k \ell}
	- X^{k} \pi^{\ell } )
	\big)\big|_{\partial \hyp} n_\ell= 0
	\,,
	$$
	where $n_\ell$ is the field of conormals to $\partial \hyp$,
	but in some situations it might be appropriate to restrict the   class of field variations allowed at $\partial \hyp$. If moreover  $\delta A_k$ is arbitrary on $\partial \hyp$ and
	we allow $\partial \hyp$ to vary we further find
	\begin{eqnarray}
		\frac{\partial \mcH^{0}}{\partial {A_{k,\ell}}}
		=  X^{\ell} {\pi^{k}}
		-X^{0} \pi^{k \ell}
		- X^{k} \pi^{\ell }
		\, .)
		\label{15XII22.612}
	\end{eqnarray}

	Next, we return to~\eqref{14XII22.51comp} in
	the second case  where the variations of $A_k$ remain arbitrary but those of $\pi^k$  are subject to the constraint
	\begin{equation}\label{12IV23.1}
		\delta \mcE^0 \equiv \partial_k \delta \pi^k=0
		\,.
	\end{equation}
	Now, the vanishing of the divergence of $\delta \pi^k$ implies that for any function  $\lambda$ we have
	\begin{equation}
		\int_\hyp \delta  \pi^i \partial_i \lambda dS_0 =
		-
		\underbrace{ \int_\hyp  \delta \partial_i \pi^i \lambda dS_0 }_0
		+
		\int_{\partial \hyp}
		\delta \pi^i \lambda \, dS_{0i}
		\,.
	\end{equation}
	The right-hand side vanishes when $\lambda$ or when the normal component of $\pi^i$ vanish on $\partial \hyp$. We expect therefore that  \eqref{15XII22.613} should be replaced by
	\begin{eqnarray}
		&
		\displaystyle
		X^\mu (A_{k,\mu} - A_{\mu,k}) \equiv   \myLie_{X} A_k   =    \frac{\delta \mcH^{0}}{\delta {\pi^{k}}} + \partial_k \lambda
		\,,
		&
		\label{15XII22.615}
	\end{eqnarray}
	and it is conceivable that this equation can be justified for classes of fields with restricted boundary conditions, but we have not attempted to do this.

	The apparent discrepancy between the last equation and \eqref{15XII22.613} is easiest to understand in Minkowski spacetime, on the standard slices $t=\const$, with $X=\partial_t$. Then \eqref{15XII22.615}  reads
	\begin{eqnarray}
		\partial_t A_k - \partial_k A_0 & = &  \pi_{k} + \partial_k \lambda
		\,,
		\label{26}
	\end{eqnarray}
	so that in this case the function $\lambda$ can be absorbed in a redefinition of $A_0$.
	
	More generally, \eqref{15XII22.615} can be rewritten as
	\begin{eqnarray}
		&
		\displaystyle
		\Lie_{X} A_k  -\partial_k (X^\mu A_\mu)  =    \frac{\delta \mcH^{0}}{\delta {\pi^{k}}} + \partial_k \lambda
		\,,
		&
		\label{13IV23.1}
	\end{eqnarray}
	which makes it clear that the freedom in the choice of $\lambda$ is closely  related to the gauge freedom of the theory.

	Regardless of whether or not \eqref{15XII22.615} provides the correct  way to proceed in whole generality, one can take into account the  constraint \eqref{12IV23.1} by using variations of the form
	\begin{equation}\label{12IV23.2}
		\delta \pi^k = \epsilon^{k\ell m }D_\ell \Yvec_m
		\,,
	\end{equation}
	which have vanishing divergence for all vector fields $\Yvec$. For such variations, and after taking into account \eqref{15XII22.613}-\eqref{12IV23.5}, Equation~\eqref{14XII22.51comp} becomes
	\begin{eqnarray}
		0 & = & \int_{\hyp}
		\Big(\frac{\delta \mcH^{0}}{\delta {\pi^{k}}} -\myLie_{X}A_k
		\Big) \epsilon^{k\ell m }D_\ell \Yvec_m
		dS_0
		\nonumber
		\\
		& = &
		\int_{\hyp}
		\epsilon^{k\ell m }D_\ell \Big(\frac{\delta \mcH^{0}}{\delta {\pi^{k}}} -\myLie_{X}A_k
		\Big) \Yvec_m
		dS_0
		\nonumber
		\\
		&&          +
		\int_{\partial \hyp}
		\Big(\frac{\delta \mcH^{0}}{\delta {\pi^{k}}} -\myLie_{X}A_k
		\Big) \epsilon^{k\ell m }  \Yvec_m  dS_{0\ell}
		\,.
		\label{12IV23.3}
	\end{eqnarray}
	As this holds in particular for all vector fields $\Yvec$ vanishing at $\partial \hyp$, we conclude that
	we must have
	\begin{equation}\label{12IV23.6}
		\epsilon^{k\ell m }D_\ell \Big(\frac{\delta \mcH^{0}}{\delta {\pi^{k}}} -\myLie_{X}A_k
		\Big) = 0
		\,.
	\end{equation}
	In the Minkowskian case as in \eqref{26}, this is the usual Maxwell equation
	\begin{equation}\label{12IV23.7}
		\partial_t \vec B =  - \rot \, \vec E
		\,.
	\end{equation}

	We finish this section by comparing  \eqref{13XII22.41intb} with \eq{14XII22.51b}:
	\begin{eqnarray}
		0 & = & \int_{\hyp}
		\Big[
		\Big(\frac{\delta \mcH^{0}_c}{\delta {\pi^{k}}} -\Lie_{X}A_k
		\Big) \delta \pi^{k}
		+
		\Big(
		\frac{\delta \mcH^{0}_c}{\delta {A_{k}}}
		+
		\Lie_{X} {\pi^{k }}
		+X^{0} {\mathcal{E}^k}
		\Big)
		\delta A_{k}
		\nonumber
		\\
		&& \phantom{ \int_{\hyp}
			\Big[}
		+  \Big(
		\frac{\delta \mcH^{0}_c}{\delta {A_{0}}}
		+
		X^{0}{\mathcal{E}^0}
		\Big)
		\delta A_{0}
		\Big] dS_0
		\nonumber
		\\
		&&
		+
		\int_{\partial\hyp}
		\Big\{\Big[
		\frac{\partial \mcH^{0}_c}{\partial {A_{k,\ell}}}
		-  (X^{\ell} {\pi^{k}}
		-X^{0} \pi^{k \ell} ) \Big] \delta {A_k}
		\nonumber
		\\
		&&   \phantom{ +  \int_{\partial \hyp}
			\Big\{}
		+
		\Big(\frac{\partial \mcH^{0}_c}{\partial {A_{0,\ell}}} +
		X^{0} \pi^{\ell }
		\Big) \delta {A_0}
		\Big\}
		\,
		dS_{0\ell}
		\, .
		\phantom{xxx}
		\label{14XII22.51compb}
	\end{eqnarray}
	A discussion similar to the one for $\mcH^0$ applies, we leave the details to the reader.

\subsubsection{Noether charge algebra}
\label{ss3V23.51}

In this section, unless explicitly indicated otherwise we consider a theory with a general Lagrangian density of the form \eqref{5V23.41}.
As before we set
\begin{equation}\label{5V23.61}
	\mcH^\mu_X = \pi^{\nu\mu}\myLie_X A_\nu - X^\mu \mcL
	\,.
\end{equation}

Given two functionals $F$ and $G$ of the  form
\begin{equation}\label{4IV23.1-ii}
	F = \int_\hyp f(A_k,\partial_i A_k, \pi^\ell) \, dS_0
	\,,
	\quad
	G = \int_\hyp g(A_k,\partial_i A_k, \pi^\ell) \, dS_0
	\,,
\end{equation}
following~\cite{BrownHenneaux} we set
\begin{equation}
	\{F, G \}_\hyp : =
	\int_{\hyp}\left(
	\frac{\delta f}{\delta A_{l}} \frac{\delta g}{\delta \pi^{l}}
	-\frac{\delta f}{\delta \pi^{l}} \frac{\delta g}{\delta A_{l}}
	\right) d S_{0}  \, .
	\label{10IV22.t1}
\end{equation}
\emph{In this formula the operator $\delta/\delta \pi^i$ is defined by ignoring the fact that $\delta \pi^i$ should satisfy the Gauss constraint,} so that \eqref{15XII22.613}-\eqref{15XII22.61} apply.

The Poisson bracket of two Hamiltonian functionals $H_X$ and $H_Y$ with integrands $\mcH^\mu_X$ and  $\mcH^\mu_Y$   thus equals
\begin{equation}
	\left\{H_{X}, H_{Y}\right\}_\hyp =
	\int_{\hyp} \left (
	\frac{\delta \mcH^{0}_{X}}{\delta A_{l}}
	\frac{\delta \mcH^{0}_{Y}}{\delta \pi^{l}}
	-
	\frac{\delta \mcH^{0}_{X}}{\delta \pi^{l}}
	\frac{\delta \mcH^{0}_{Y}}{\delta A_{l}}
	\right) d S_{0}
	\, ,
\end{equation}
Let
$$
E_X^k:= X^k \mcE^0 - X^0 \mcE^k
\,,
\quad
E_Y^k:= Y^k \mcE^0 - Y^0 \mcE^k
\,.
$$
Recall that $\pi^0 =0$ and that $\Lie_{X} \pi^0:=\Lie_{X} \pi^{00}=0$ for any vector field.
Using \eqref{15XII22.613}-\eqref{15XII22.61}  we find
\ptcheck{5V and 13 V}
\begin{eqnarray}
	\lefteqn{
		\frac{\delta \mcH^{0}_{X}}{\delta A_{l}}
		\frac{\delta \mcH^{0}_{Y}}{\delta \pi^{l}}
		-
		\frac{\delta \mcH^{0}_{X}}{\delta \pi^{l}}
		\frac{\delta \mcH^{0}_{Y}}{\delta A_{l}}
		=
	}
	&&
	\nonumber
	\\
	&=&
	\big(
	-\Lie_{X} \pi^{k}+E_{X}^{k}
	\big)
	\myLie_{Y} A_{k}
	-
	\big(
	-\Lie_{Y}\pi^{k}+E_{Y}^{k}
	\big)
	\myLie_{X} A_{k}
	\nonumber
	\\
	&=&
	-\Lie_{X} \big(\pi^{\nu} \myLie_{Y} A_{\nu}\big)
	+ \pi^{\nu} \Lie_{X} \big(\myLie_{Y} A_{\nu}\big)
	+\Lie_{Y} \big(\pi^{\nu} \myLie_{X} A_{\nu}\big)
	\nonumber
	\\
	& & - \pi^{\nu} \Lie_{Y} \big( \myLie_{X} A_{\nu}\big)
	+\big(E_{X}^{k} Y^{\mu} -E_{Y}^{k} X^{\mu}  \big)F_{\mu k}
	\nonumber
	\\
	&=&
	\pi^{\nu} \big(\Lie_{X} \myLie_{Y} A_{\nu}-\Lie_{Y} \myLie_{X} A_{\nu}\big)
	- \Lie_{X} \big(\pi^{\nu} \myLie_{Y} A_{\nu}-Y^{0} \mcL\big)
	\nonumber
	\\
	& &
	+ \Lie_{Y} \big(\pi^{\nu} \myLie_{X} A_{\nu}-X^{0} \mcL\big)
	-\Lie_{X} \big(Y^{0} \mcL\big)+\Lie_{Y} \big(X^{0} \mcL\big)
	\nonumber
	\\
	& &
	+\big(E_{X}^{k} Y^{\mu} -E_{Y}^{k} X^{\mu}  \big)F_{\mu k}
	\, .
	\label{10IV22.t1a}
\end{eqnarray}
Additionally, we have
\ptcheck{5V}
\begin{eqnarray}
	\lefteqn{
		\big(\Lie_{X} \myLie_{Y} A_{\nu}-\Lie_{Y} \myLie_{X} A_{\nu}\big) =} & &
	\nonumber
	\\
	&=&
	\Lie_{X} \big( Y^{\mu} F_{\mu \nu}\big)
	-\Lie_{Y} \big( X^{\mu} F_{\mu \nu}\big)
	\nonumber
	\\
	&=&
	X^{\alpha} \partial_{\alpha} \big(
	Y^{\mu} F_{\mu \nu}
	\big)
	+Y^{\mu} F_{\mu \alpha} \partial_{\nu} X^{\alpha}
	-Y^{\alpha} \partial_{\alpha} \big(X^{\mu} F_{\mu \nu}\big)
	-X^{\mu} F_{\mu \alpha} \partial_{\nu} Y^{\alpha}
	\nonumber
	\\
	&=& \underbrace{
		\big(
		X^{\alpha} \partial_{\alpha}Y^{\mu}-Y^{\alpha} \partial_{\alpha}X^{\mu}
		\big)F_{\mu \nu}
	}_{\myLie_{[X,Y]} A_{\nu}}
	+ \big(
	X^{\alpha} Y^{\mu}-Y^{\alpha} X^{\mu}
	\big) \partial_{\alpha} F_{\mu \nu}
	\nonumber
	\\
	& &
	+\big(
	Y^{\mu} \partial_{\nu} X^{\alpha}-X^{\mu} \partial_{\nu} Y^{\alpha}
	\big) F_{\mu \alpha}
	\nonumber
	\\
	&=& \myLie_{[X,Y]} A_{\nu}
	+ X^{\alpha} Y^{\mu} \big(
	\partial_{\alpha} F_{\mu \nu}-\partial_{\mu} F_{\alpha \nu}
	\big)
	+F_{\mu \alpha}\partial_{\nu} \big(Y^{\mu} X^{\alpha}\big)
\end{eqnarray}
Using the identity
\ptcheck{5V}
\begin{equation}
	\partial_{\alpha} F_{\mu \nu}-\partial_{\mu} F_{\alpha \nu}
	=\partial_{\nu} F_{\mu \alpha}
\end{equation}
we continue as follows:
\ptcheck{5V}
\begin{eqnarray}
	\lefteqn{
		\big(\Lie_{X} \myLie_{Y} A_{\nu}-\Lie_{Y} \myLie_{X} A_{\nu}\big) =} & &
	\nonumber
	\\
	&=&
	\myLie_{[X,Y]} A_{\nu}
	+ X^{\alpha} Y^{\mu} \partial_{\nu} F_{\mu \alpha}
	+F_{\mu \alpha}\partial_{\nu} \big(Y^{\mu} X^{\alpha}\big)
	\nonumber
	\\
	&=&
	\myLie_{[X,Y]} A_{\nu}
	+\partial_{\nu} \big(F_{\mu \alpha} Y^{\mu} X^{\alpha}\big)
	\,.
	\label{10IV22.t2}
\end{eqnarray}
Recall the identities \eqref{7III22.t6}-\eqref{7III22.t5}
for vector densities,
\ptcheck{5V}
\begin{eqnarray}
	\Lie_{X} \mcH^{  \beta}_{Y}
	-\Lie_{Y} \mcH^{ \beta}_{X}
	&=&2 \partial_{\alpha} \Big(X^{[\alpha} \mcH^{\beta]}_{Y}-Y^{[\alpha} \mcH^{\beta]}_{X}\Big)
	\nonumber
	\\
	& &
	+X^{ \beta} \partial_{\alpha} \mcH^{\alpha}_{Y}-Y^{ \beta} \partial_{\alpha} \mcH^{\alpha}_{X}
	\, ,
	\label{10IV22.t3}
	\\
	\Lie_{X}\Big(Y^{ \beta} \mcL\Big)
	-
	\Lie_{Y}\Big(X^{ \beta} \mcL\Big)
	&=&
	2 \partial_{\alpha} \Big(X^{[ \alpha} Y^{\beta]} \mcL\Big)
	+[X,Y]^{ \beta} \mcL
	\,.
	\label{10IV22.t4}
\end{eqnarray}
Inserting \eqref{10IV22.t2}-\eqref{10IV22.t4} into \eqref{10IV22.t1a} results in
\ptcheck{5V}
\begin{eqnarray}
	\left\{H_{X}, H_{Y}\right\}
	& = &
	\int_{\hyp} \Big \{
	\mcH^{\beta}_{[X,Y]}
	+ Y^{\beta} \partial_{\alpha} \mcH^{\alpha}_{X}
	-X^{\beta} \partial_{\alpha} \mcH^{\alpha}_{Y}
	\nonumber
	\\
	&&
	-2 \partial_{\alpha} \Big(X^{[\alpha} \mcH^{\beta]}_{Y}-Y^{[\alpha} \mcH^{\beta]}_{X} +
	X^{[ \alpha} Y^{\beta]} \mcL \Big)
	\nonumber
	\\
	& &
	+\big(E_{X}^{k} Y^{\mu} -E_{Y}^{k} X^{\mu}  \big)F_{\mu k}
	\Big \}
	d S_{\beta}
	\, .
	\phantom{xxxx}
	\label{10IV22.t4a}
\end{eqnarray}

To continue, we wish to show that the divergence of the Noether currents, which appear above,
vanishes when the field equations hold and when the Lagrangian does not depend upon $A_\mu$. For this, recall that
we have assumed that the Lagrangian is a scalar density, so that
\begin{equation}
	\Lie_{X} \mcL=\partial_{\alpha}\big(X^{\alpha} \mcL\big)
	\, .
\end{equation}
Thus
\begin{eqnarray}
	\partial_{\alpha} \mcH^{\alpha}_X
	&=&
	\partial_{\alpha}
	\big(
	\pi^{\nu \alpha} \myLie_{X} A_{\nu}-X^{\alpha} \mcL \big)
	\nonumber
	\\
	&=&\partial_{\alpha}\big(\pi^{\nu \alpha} \myLie_{X} A_{\nu} \big)
	-\Lie_{X} \mcL
	\, ,
	\label{5V23.32}
\end{eqnarray}
which can be rearranged as
\begin{eqnarray}
	\partial_{\alpha} \mcH^{\alpha}_X
	&=&\partial_{\alpha}\big(\pi^{\nu \alpha} (\myLie_{X} A_{\nu}-\Lie_{X} A_{\nu} \big)
	+ \partial_{\alpha} \mcH^{\alpha}_{c}[X]
	\, .
	\label{8V23.t3}
\end{eqnarray}
For $\partial_{\alpha} \mcH^{\alpha}_{c}[X]$ the formulae \eqref{8V23.t2} holds with $ {\omcL}=\mcL$. Note that $\mcL$ does not depend on connection coefficients. Equation \eqref{8V23.t3} becomes
\begin{eqnarray}
	\partial_{\alpha} \mcH^{\alpha}_X&=&
	\partial_{\alpha}\big(\pi^{\nu \alpha} (\myLie_{X} A_{\nu}-\Lie_{X} A_{\nu} \big)
	\nonumber
	\\
	& &
	+\mcE^{\kappa}   \Lie_X A_{\kappa}
	- \pi^{\lambda \kappa} [\Lie_{X}, \nabla_{\kappa}] A_{\lambda}
	- \frac{\partial \mcL}{\partial g_{\kappa \lambda}} \Lie_{X} g_{\kappa \lambda}
	\, .
	\label{9V23.t1}
\end{eqnarray}%
Now, the divergence term in \eqref{9V23.t1} reads
\cbk
\ptcheck{5V23}
\begin{eqnarray}
	\lefteqn{\partial_{\alpha}\big[\pi^{\nu \alpha} \big( \myLie_{X} A_{\nu}- \Lie_{X} A_{\nu} \big) \big] =}
	& &
	\nonumber
	\\
	&=& \mcE^{\nu} \big( \myLie_{X} A_{\nu}- \Lie_{X} A_{\nu} \big)+ \pi^{\nu \alpha} \partial_{\alpha}\big( \myLie_{X} A_{\nu}- \Lie_{X} A_{\nu} \big)
	\nonumber
	\\
	&=&\mcE^{\nu} \Big( \myLie_{X} A_{\nu}- \Lie_{X} A_{\nu} \big)
	\nonumber
	\\
	& &
	+ \pi^{\nu \alpha} \partial_{\alpha}\big(
	\underbrace{
		X^{\mu} F_{\mu \nu}
		-X^{\mu} \partial_{\mu} A_{\nu}+ X^{\mu} \partial_{\nu} A_{\mu}
	}_{=0}
	-\partial_{\nu} \big( X^{\mu} A_{\mu} \big)
	\Big)
	\nonumber
	\\
	&=&
	\mcE^{\nu} \Big( \myLie_{X} A_{\nu}- \Lie_{X} A_{\nu} \big)
	\, .
	\label{25VII22.t2}
\end{eqnarray}
\cbk
Summarising, we have shown that
\begin{eqnarray}
	\partial_{\alpha} \mcH^{\alpha}_X
	&=&
	\mcE^{\nu}  \myLie_{X} A_{\nu}
	- \pi^{\lambda \kappa} [\Lie_{X}, \nabla_{\kappa}] A_{\lambda}
	- \frac{\partial \mcL}{\partial g_{\kappa \lambda}} \Lie_{X} g_{\kappa \lambda}
	\, .
	\label{25VII22.t1ax}
\end{eqnarray}
Inserting this into  \eqref{10IV22.t4a} we conclude that
\begin{eqnarray}
	\left\{H_{X}, H_{Y}\right\}
	&=&
	\int_{\hyp} \Big \{
	\mcH^{\beta}_{[X,Y]}
	-2 \partial_{\alpha} \Big(X^{[\alpha} \mcH^{\beta]}_{Y}-Y^{[\alpha} \mcH^{\beta]}_{X}+X^{[ \alpha} Y^{\beta]} \mcL\Big)
	\nonumber
	\\
	& &
	+Y^{\beta} \Big(
	\mcE^{\kappa}   \myLie_X A_{\kappa}
	- \pi^{\lambda \kappa} [\Lie_{X}, \nabla_{\kappa}] A_{\lambda}
	- \frac{\partial \mcL}{\partial g_{\kappa \lambda}} \Lie_{X} g_{\kappa \lambda}\Big)
	\nonumber
	\\
	& &
	-X^{\beta} \Big(
	\mcE^{\kappa}   \myLie_Y A_{\kappa}
	- \pi^{\lambda \kappa} [\Lie_{Y}, \nabla_{\kappa}] A_{\lambda}
	- \frac{\partial \mcL}{\partial g_{\kappa \lambda}} \Lie_{Y} g_{\kappa \lambda}\Big)
	\nonumber
	\\
	& &
	+\big(E_{X}^{k} Y^{\mu} -E_{Y}^{k} X^{\mu}  \big)F_{\mu k}
	\Big \}
	d S_{\beta}
	\, .
	\phantom{xxxx}
	\label{25VII22.t3}
\end{eqnarray}
where \eqref{9V23.t1}-\eqref{25VII22.t2} have been used.

\subsubsection{$\mcH^0$: $3+1$ decomposition}
\label{s3V23.21}

We continue by deriving the formula for $\mcH^0$, and its variation, assuming the standard Maxwell Lagrangean. It is convenient to introduce some notation.  In the remainder of this section we will assume that $\hyp$ is spacelike. We will use the ADM parametrisation of the metric,
\begin{eqnarray}
	&
	\threeg_{ij}:=g_{ij}
	\,,
	\quad
	N :=\frac{1}{\sqrt{-g^{00}}} \, ,
	\quad
	N_{k}
	:=
	g_{0k} \, ,
	\quad
	N^{k}
	=
	\threeg ^{k i}N_{i} \, ,
\end{eqnarray}
where $\threeg ^{ij}$ is an inverse of the three-dimensional metric
$\threeg_{ij} $ induced by $g_{\mu\nu}$ on $\hyp$:
\begin{eqnarray}
	g_{\mu \nu}&=&
	\left[\begin{array}{cc}
		g_{00}&g_{0j} \\
		g_{i0}&g_{ij}
	\end{array} \right]=
	\left[\begin{array}{cc}
		-N^2 +N^{k} N_{k}&N_{j} \\
		N_{i}&\threeg_{i j}
	\end{array} \right] \, ,
	\nonumber
	\\
	g^{\mu \nu}&=&\left[\begin{array}{cc}
		g^{00}&g^{0j} \\
		g^{i0}&g^{ij}
	\end{array} \right]=
	\left[\begin{array}{cc}
		-\frac{1}{N^2}&\frac{N^{j}}{N^2} \\
		\frac{N^{i}}{N^2}&\threeg ^{i j}-\frac{N^{i}N^{j}}{N^{2}}
	\end{array} \right] \, .
\end{eqnarray}
It holds that
\begin{equation}
	\sqrt{|\det g_{\mu\nu}|}=N \sqrt{\det  \threeg_{ij} } \, .
\end{equation}

Let $T^\mu$ denote the field of unit future-directed normals to $\hyp$, thus $T_\mu = - N\,  dt$. We define the electric field $E^k$ as
\begin{equation}\label{15XII22.1}
	E^k := N  F^{0k}  = - F^{\mu k} T_\mu
	\,.
\end{equation}
The canonical momentum $\pi^k$ is related to the electric field
as
\begin{equation}\label{3XI22.1}
	\pi^k =
	-
	\frac{1}{4\pi}   \sqrt{\det  \threeg_{ij} } E^k
	\,,
\end{equation}
and we note that
\begin{equation}\label{3XI22.2}
	D_k E^k \equiv
	\frac{1}{\sqrt{\det  \threeg_{ij} }  }\partial_k \big(
	\sqrt{\det  \threeg_{ij} } E^k
	\big) =
	{-}
	\frac{1}{\sqrt{\det  \threeg_{ij} }  }\partial_\mu  \big(
	\sqrt{\det  g_{\alpha\beta} } F^{0\mu}
	\big) =0
	\,,
\end{equation}
where $D$ is the covariant derivative operator of the metric $\threeg$.

The decomposition of the Maxwell tensor density \eqref{28VI22.t1} associated with the (3+1)-decomposition of the metric reads
\begin{eqnarray}
	\TSF_{0k}&=&N^{l}\TSF_{lk}-N^{2}\TSF^{0l}\threeg_{lk}
	\, ,
	\label{21VII22.t1}
	\\
	\TSF^{kl}&=&
	(N^{l}\TSF^{0k}-N^{k}\TSF^{0l})
	+\threeg ^{km}\threeg ^{ln}\TSF_{mn}
	\, ,
	\label{21VII22.t2}
\end{eqnarray}
where
\begin{equation}
	\TSF^{\alpha \beta}=\sqrt{|-\det g|}F^{\alpha \beta} \, .
	\label{21VII22.t3}
\end{equation}
For \eqref{21VII22.t2} the following calculation is useful:
\begin{eqnarray}
	\TSF_{k l} \threeg ^{k p} \threeg ^{l q}&=& \TSF^{\mu \nu} g_{\mu k} g_{\nu l} \threeg ^{k p} \threeg ^{l q}  \nonumber \\
	&=&\Big(\TSF^{0 m} g_{0 k} g_{m l} +\TSF^{m 0} g_{m k} g_{0 l} +\TSF^{m n} g_{ m k} g_{nl} \Big)\threeg ^{k p} \threeg ^{l q} \nonumber \\
	&=& \TSF^{0 q} N^{p}+\TSF^{p 0} N^{q}+\TSF^{p q}
	\,,
\end{eqnarray}
and this last equation is also useful as an intermediate step for  \eqref{21VII22.t1}.
\ptcheck{19IX22} The field equations operator $\mcE^k$ reads
\begin{eqnarray}
	4 \pi \mcE^{k}&=&
	\partial_{l}(N^{l}\TSF^{0k}-N^{k}\TSF^{0l})
	+ \partial_{l}(\threeg ^{km}\threeg ^{ln}\TSF_{mn})
	-\partial_{0}\TSF^{0k}
	\, ,
	\label{21VII22.t2a}
\end{eqnarray}
which has to be supplemented with the constraint equation
$$
4 \pi \mcE^0 =- \partial_i \TSF^{i0}
\,.
$$
\ptcheck{19IX22 }

Using \eqref{21VII22.t1}-\eqref{21VII22.t2} the Noether current \eqref{23V21.t1}  can be rewritten as
\ptcheck{23IX22, and rechecked together 3XI}
\begin{eqnarray}
	\mcH^{0}
	&=&
	\frac12 \pi^{k} X^{0} F_{0 k}
	+
	\pi^{k} X^{\l} F_{l k}
	+
	\frac{1}{16 \pi}
	\sqrt{|\det g|}
	X^0
	F^{k l} F_{k l}
	\nonumber
	\\
	&=&
	\frac12 \pi^{k} X^{0} \big(N^{l} F_{lk}-N^{2} F^{0l}\threeg_{lk}  \big)
	+
	\pi^{k} X^{\l} F_{l k}
	\nonumber
	\\
	& &
	+
	\frac{1}{16 \pi}
	\sqrt{|\det g|}
	X^0
	\big(
	(N^{l}F^{0k}-N^{k}F^{0l})
	+\threeg ^{km}\threeg ^{ln} F_{mn}
	\big)
	F_{k l}
	\nonumber
	\\
	&=&
	- \frac12  X^{0} N^{2} \pi^{k} F^{0l}
	\threeg_{lk}
	+
	\frac{1}{16 \pi}
	\sqrt{|\det g|}
	X^0 \threeg ^{km}\threeg ^{ln} F_{mn} F_{k l}
	\nonumber
	\\
	& &
	+
	X^{0} \pi^{k}N^{l} F_{lk}
	+
	\pi^{k} X^{\l} F_{l k}
	\nonumber
	\\
	&=&
	\frac{\sqrt{\det \threeg} }{8 \pi}
	\Big[ N X^{0}
	\big(  E^{k} E^{l}\threeg_{lk}
	+
	\frac 12  \threeg ^{km}\threeg ^{ln} F_{mn} F_{k l}
	+  2  N^{-1}  E^{k}N^{l} F_{lk}
	\big)
	\nonumber
	\\
	& &
	+ 2 E^{k} X^{\l} F_{l k}
	\Big]
	\, .
	\label{10XII22.11}
\end{eqnarray}

We can therefore write $\mcH^{0}$  in terms of the $\pi^k$'s as
\begin{eqnarray}
	\mcH^{0}&=&
	\frac{2 \pi  X^{0} N}{\sqrt{\det \threeg}} \pi^{k} \pi^{l}
	\threeg_{lk}
	+
	\frac{1}{16 \pi}
	N \sqrt{\det \threeg}
	X^0 \threeg^{km}\threeg^{ln} F_{mn} F_{k l}
	\nonumber
	\\
	& &
	+
	X^{0} \pi^{k}N^{l} F_{lk}
	+
	\pi^{k} X^{\l} F_{l k}
	\,.
	\label{20IX22.t11}
\end{eqnarray}

For completeness we calculate the variation of $\mcH^0$ as given by \eqref{10XII22.11}:
\begin{eqnarray}
	\delta \mcH^{0}&=&
	\frac{\sqrt{\det \threeg} }{8 \pi}
	\Big[ N X^{0}
	\big(  2 \threeg_{lk} E^{k} \delta E^{l}
	+
	\threeg ^{km}\threeg ^{ln} F_{mn} \delta F_{k l}
	\nonumber
	\\
	& &
	+  2  N^{-1} N^{l} (E^{k} \delta F_{lk}+ F_{lk} \delta E^{k})
	\big)
	+ 2X^{\l} \big(E^{k}  \delta F_{l k}+ F_{l k} \delta E^{k} \big)
	\Big]
	\nonumber
	\\
	&=&
	\frac{\sqrt{\det \threeg} }{8 \pi}
	\Big\{
	2 \big[ N X^{0} \big(
	\threeg_{lk} E^{k} \delta E^{l}
	+ N^{-1} N^{k} F_{kl}
	\big)
	+
	X^{k} F_{k l}
	\big]\delta E^{l}
	\nonumber
	\\
	& &-2 \partial_{k}\big[
	N X^{0}
	\big(
	\threeg ^{km}\threeg ^{ln} F_{mn}
	-  2  N^{-1} N^{[l} E^{k]}
	\big)
	- 2X^{[l} E^{k]}
	\big]
	\delta  A_{l}
	\nonumber
	\\
	& &-2 \partial_{k}\big[ \big(
	N X^{0}
	\big(
	\threeg ^{km}\threeg ^{ln} F_{mn}
	-  2  N^{-1} N^{[l} E^{k]}
	\big)
	- 2X^{[l} E^{k]}
	\big)
	\delta  A_{l}
	\big]
	\Big\}
	\,.
	\nonumber
	\\
	\phantom{xxxxx}
	\label{2V23.t1}
\end{eqnarray}

\section{Plumbing the leakage}
\label{s23IV22.1}

The variational identities discussed so far suffer from the existence of ``leaky boundary terms'', i.e., non-zero boundary terms in the variational formulae. These create problems when attempting to define Poisson brackets. In this section we show how this can be avoided by suitably extending the phase spaces.

\subsection{De Sitter background}
\label{ss29VII22.1}

In what follows we will need explicit formulae for Fefferman-Graham coordinates in de Sitter spacetime; the aim of this section is to address this.

In addition to the form \eqref{8VII20.11}  of the de Sitter metric, let us recall the more standard form
\begin{equation}\label{23IV22.1}
	g =   -(1-\alpha^2 r^2)dt^2 + \frac{dr^2}{1-\alpha^2 r^2}
	+ r^2
	(d\theta^2+\sin^2 \theta d\phi^2)
	\,.
\end{equation}

The Bondi form \eqref{8VII20.11} of $g$ can be obtained from \eqref{23IV22.1} with $\alpha r >1$   by introducing a coordinate $u$ through the formula
\begin{equation}\label{16I20.3}
	du : = dt - \frac{dr}{1-\alpha^2 r^2} \equiv d
	\Big(
	t + \frac{1}{2 \alpha} \ln \big(\frac{\alpha r -1}{\alpha r + 1} \big)
	\Big)
	\,;\end{equation}
cf., e.g., \cite{FischerDeSitter}.

Instead of either form of the metric above, for the purpose of global Hamiltonian analysis it seems best to use a globally defined, manifestly regular  representation of the metric on the cylinder $\mathbb{R} \times S^{3}$ for de Sitter spacetime. For instance, the apparent singularity of the metric \eqref{23IV22.1} at $r=\alpha$ is due to a poor choice of coordinates, as can be seen by setting,
\begin{eqnarray}
	\label{27VII22.t1b}
	r=\alpha^{-1} \sin \psi \cosh (\alpha \tau)
	\,,
	\quad t=
		\alpha^{-1} \atanh \left(\frac{\cos\psi}{\tanh(\alpha \tau)} \right)
		\,.
	\end{eqnarray}
	Using \eqref{16I20.3}, we can obtain a relation between Bondi coordinates and the coordinates on the cylinder $\mathbb{R} \times S^{3}$,
	\ptcheck{23VIII, mathematica file Checking alpha u}
	\begin{equation}
		u=\frac{1}{\alpha} \atanh \left(\frac{\sinh (\alpha \tau) \cos (\psi)-\sin (\psi)}{\cosh (\alpha \tau)}\right) \, ,
		\label{19VIII22.t1}
	\end{equation}
	toghether with the first equation in \eqref{27VII22.t1b}.
	
	After the coordinate transformation \eqref{27VII22.t1b} the metric \eqref{23IV22.1} becomes
	\ptcheck{29VII22, mathematica file dScoordinates.nb, old version, and 17VIII22 new version }
	\begin{equation}
		g=-d \tau^{2}+\frac{\cosh^{2} (\alpha \tau)}{\alpha^2}\left(d \psi^{2}
		+\sin ^{2}\psi \left(d \theta^{2}
		+\sin ^{2}\theta d \varphi^{2}\right)\right)
		\, ,
		\label{30VII12.1}
	\end{equation}
	with $\sqrt{ \det|g|}=\alpha^{-3}\cosh (\alpha \tau)^{3} \sin^{2}\psi \sin \theta$. The Killing vector field
	\begin{equation}
		\mathcal{T}=\partial_{t} \equiv \partial_u\, ,
	\end{equation}
	defining the Hamiltonian energy,
	in the coordinates \eqref{27VII22.t1b} reads
	\ptcheck{29VII22, mathematica file dScoordinates.nb }
	\begin{equation}
		\mathcal{T}=
		\cos \psi \partial_{\tau} - \alpha \sin \psi \tanh (\alpha \tau) \partial_{\psi}   \, .
		\label{30VII22.4}
	\end{equation}
	For future reference we note
	\ptcheck{29VII22, mathematica file dScoordinates.nb}
	\begin{equation}
		\partial_{r}=\frac{1}{1-\sin^{2} \psi \cosh^{2} (\alpha \tau) }\Big[ -\sin \psi \sinh (\alpha \tau) \partial_{\tau}+ \frac{\alpha \cos \psi}{\cosh (\alpha \tau)} \partial_{\psi} \Big]
		\, .
	\end{equation}

	The metric \eqref{30VII12.1} can be rewritten in a manifestly conformally smooth form
	\begin{eqnarray}
		g
		&= &
		\underbrace{
			\cosh^{2}(\alpha \tau)
		}_{=:\xdS^{-2}}
		\big(
		-{\cosh^{-2}(\alpha \tau)} d \tau^{2}
		+  \alpha^{-2}
		\underbrace{
			\left(d \psi^{2}+\sin ^{2}\psi \left(d \theta^{2}+\sin ^{2}\theta d \varphi^{2}\right)\right)
		}_{=:\spherem}
		\big)
		\nonumber
		\\
		&= &
		(\alpha\xdS)^{-2}
		\big(
		-  \frac { d \xdS^{2}} {1-\xdS^2}
		+
		\spherem
		\big)
		\, ,
		\label{30VII12.2}
	\end{eqnarray}
	so that the  coordinate $\xdS$ (not to be confused with the coordinate $x$ of~\eqref{26VI20.1}) is a time coordinate for $|\xdS |<1$, with spacelike level sets there.
	
	The conformal boundary is obtained by attaching the hypersurface
	$$ \Scri^{+}:=\{\xdS =0\}
	$$
	to the physical spacetime. The Killing vector \eqref{30VII22.4} becomes
	\begin{eqnarray}
		\mathcal{T}
		& = &  -\alpha\tanh (\alpha \tau)
		\big[ \sech(\alpha\tau)
		\cos \psi \partial_{\xdS} + \sin \psi \partial_{\psi}\big]
		\nonumber
		\\
		& = &  -\alpha \sqrt{ {1-\xdS ^2}}
		\big[ \xdS
		\cos \psi \partial_{\xdS} + \sin \psi \partial_{\psi}\big]   \, ,
		\label{30VII22.5}
	\end{eqnarray}
	and extends smoothly to
	$\scrip$.
	For further reference we note
	\ptcheck{23VIII}
	\begin{equation}
		\partial_{r}=
		-\frac{\alpha \xdS^{2}}{\sin^2 \psi -\xdS^{2}}\Big(
		(1-\xdS^2)\sin \psi \partial_{\xdS}
		+\xdS \cos \psi \partial_{\psi}
		\Big)
		\,.
	\end{equation}

	Let $\tilde g:= \xdS^{2} g$, the $\tilde g$-Lorentzian norm-squared of $\mathcal{T}$ is
	\begin{eqnarray}
		\tilde g(\mathcal{T},\mathcal{T})
		& \equiv &
		\xdS ^{2}g(\mathcal{T},\mathcal{T}) = \xdS ^{2} (\alpha^2 r^2 - 1)
		=  \xdS ^{2}(\sin ^2(\psi ) \cosh ^2(\alpha  \tau )-1)
		\nonumber
		\\
		& = &
		\sin ^2(\psi ) -\xdS ^2
		\,,
		\label{30VII22.7}
	\end{eqnarray}
	thus $\mathcal{T}$ is spacelike througout $\Scri^+$.

\subsection{Conformally-covariant scalar field}
\label{ss23II23.1}

Using the coordinates as in \eqref{30VII12.2}, the phase space of Cauchy data on three-dimensional spheres of constant $x$ consists of smooth fields $(\phi,\partial_x \phi)$ with  symplectic form
\begin{equation}\label{12II23.1}
	\Omega = - \int_{x=\const} \delta \pi^x \wedge \delta \phi
	\,
	dS_x
	\,,
\end{equation}
where the minus sign in front of the integral comes from the fact that $\partial_x$ is past pointing. 
Writing $\pi  $ for $-\pi^x$ 
(where  again the negative sign is motivated by the time orientation of $\partial_x$),
and $\hyp_c:=\{x=c\}$, there is an associated Poisson bracket,
without problems with  boundary terms since $S^3$ has no boundary:
\begin{equation}
	\{F, G \}_{\hyp_c}: =
	\int_{x=c}\left(
	\frac{\delta F}{\delta {\SField }} \frac{\delta G}{\delta \pi}
	-\frac{\delta F}{\delta \pi} \frac{\delta G}{\delta {\SField }}
	\right) dS_x
	\, .
	\label{5III23.1a}
\end{equation}

It follows from conformal invariance of the equation satisfied by $\SField$ that the field
\begin{equation}\label{1VIII22.1a}
	\cphi:= \frac{\SField}{\xdS}
\end{equation}
extends smoothly to the boundary $\{\xdS=0\}$.
We have the expansions, for small $x$,
\begin{equation}\label{30VII22.9a}
	\SField =   {\omone \SField} \xdS +  {\omtwo \SField} \xdS^2  + \ldots
	\qquad
	\Longleftrightarrow
	\qquad
	\cphi=   {\ozero \cphi}   +  {\omone \cphi} \xdS   + \ldots
	\,,
\end{equation}
with coefficients which are functions on $S^3$, where
\begin{equation}\label{18II23.1}
	{\omone \cphi} \equiv \partial_x \cphi|_{x=0}
	\,.
\end{equation}
Since $\Omega$ is $x$-independent when applied to variations satisfying the field equations,  it is tempting to pass  with $x$ to zero. Using \eqref{30VII12.2} and \eqref{1VIII22.1a} one obtains
\begin{equation}\label{12II23.2}
	\Omega = -
	\frac{1}{\alpha^2}
	\int_{x=0} \delta \partial_x \chi \wedge \delta \chi
	\, d\mu_{\spherem}
	\,,
\end{equation}
with Poisson bracket
\begin{equation}
	\{F, G \}_{\hyp_0}: =
	\int_{S^3}\left(
	\frac{\delta f}{\delta {\chi }} \frac{\delta g}{\delta \pi}
	-\frac{\delta f}{\delta \pi} \frac{\delta g}{\delta {\chi }}
	\right) dS_x
	\, .
	\label{5III23.1}
\end{equation}
where now $\pi= -
\alpha^{-2} \partial_x\chi \sqrt{\det \spherem}$.

\begin{figure}
	\centering
	\begin{tikzpicture}
		\draw (0,0) node [] {\tiny \textbullet} ;
		\draw (0,0) node [anchor=north] {$r=0$} ;
		\draw[-, line width = \widl mm] (0,0) -- (2,3) ;
		\draw[-, line width = \widl mm] (0,0) -- (-2,3) ;
		\draw[-, line width = \widl mm,dashed] (2+0.05,3+3/2*0.05) -- (3-0.05,4.5-3/2*0.05) ;
		\draw[-, line width = \widl mm,dashed] (-2-0.05,3+3/2*0.05) -- (-3+0.05,4.5-3/2*0.05) ;
		\draw[-,line width = \widl mm] (-3.5,3) -- (-2,3) ;
		\draw[-,line width = \widl mm] (2,3) -- (3.5,3) ;
		\draw (3.5,3) node [anchor=west] {$x=x(R) \Leftrightarrow R=R(x)$} ;
		\draw[-,line width = \widl mm] (-3.5,4.5) -- (3.5,4.5) ;
		\draw (3.5,4.5) node [anchor=west] {$x=0$} ;
		\draw (1,3/2) node [anchor=north west] {$\mcC_{u,R }$} ;
		\draw[line width = \widl mm] (0,0) .. controls (-0.3,0.5) and (0.5,1) .. (0.3,1.5);
		\draw (0.3,1.5) node [] {\tiny \textbullet} ;
		\draw (0.2,0.7) node [anchor=south east] {$\gamma$} ;
		\draw[-, line width = \widl mm,dashed] (0.3,1.5) -- (1.3,3) ;
		\draw[-, line width = \widl mm,dashed] (0.3,1.5) -- (1.3,3) ;
		\draw[-, line width = \widl mm,dashed] (0.3,1.5) -- (-0.7,3) ;
		\draw [decorate,
		decoration = {calligraphic brace, amplitude=10 pt},line width = \widl+0.05 mm] (-3,4.5) --  (3,4.5)
		node[pos=0.5,above=7 pt,black]{$\overline{I^+(\mcC_u)\cap \{x=0\} }$};
	\end{tikzpicture}
	\caption{ The integral curve of $\partial_u$ passing through  $r=0$ is denoted by $\gamma$. 
	}
	\label{F23IV22.1}
\end{figure}
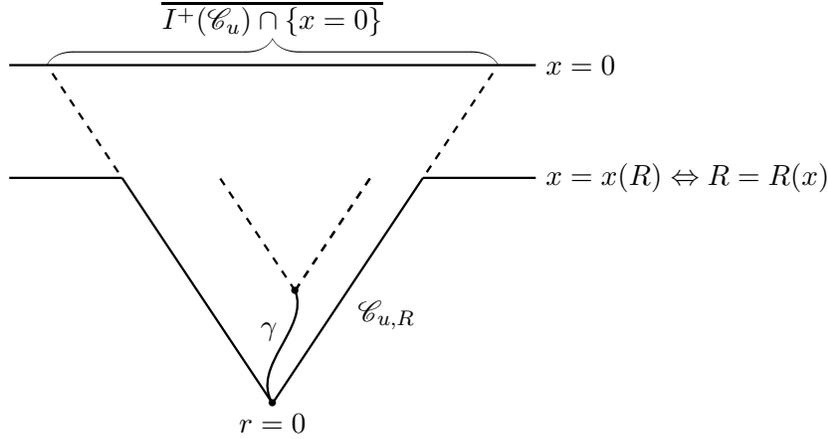

In order to avoid leakage for fields on light-cones $\mcC_u$, for each $u$ we can consider the phase space consisting of the field $\phi$ on $\mcC_u$  and of the fields $(\chi,\partial_x \chi)$ on
\begin{equation}
	\label{23II23.1}
	\scrip_u:= \{x=0\}\setminus I^+(\mcC_u)
\end{equation}
(compare Figure~\ref{F23IV22.1}),
equipped with the symplectic form
\begin{equation}\label{12II23.21}
	\int_{\mcC_u}
	\partial_r \delta \phi \wedge \delta \phi
	\,   r^2   dr \,  d\mu_{\zzhTBW}
	-
	\frac{1}{\alpha^2}  \int_{\scrip_u} \delta \partial_x \chi \wedge \delta \chi
	\, d\mu_{\spherem}
	\,.
\end{equation}
If $F$ and $G$ are associated with conserved functionals, the Poisson brackets thereof can be calculated using \eqref{5III23.1}. 
	
We wish to calculate the Noether charge associated with translations of the light-cones in $u$  for a conformally-covariant scalar field, thus with Lagrangian
\begin{equation}\label{30VII22.8}
		\mcL =- \frac 12  \sqrt{|\det g|}
		\big(
		g^{\mu\nu}\partial_\mu \SField \, \partial_\nu \SField
		+
		\underbrace{m^2}_{2 \alpha^2} \SField^2
		\big)
		\,.
	\end{equation}
	One expects a formula of the kind
	\begin{eqnarray}
		\nonumber 
		\newH &=& \int_{\mcC_u\cup \scrip_u} \mcH^\mu{[\partial_u]} dS_\mu \\
		&=&
		\int_{\mcC_u}
		\mcH^u{[\partial_u]}
		dS_u
		-
		\int_{\scrip_u} \mcH^x{[\partial_u]}
		\, dS_x
		\,,
		\label{20II23.21}
	\end{eqnarray}
	where the minus sign in front of the second integral arises again from the fact that $\partial_x$ is past-directed.
	However, the individual integrals diverge, so some care must be taken.
	For instance, assuming $\alpha \ne 0$, on the level sets of $x$ the integrand $\mcH^x {[\partial_u]} \equiv\mcH^x {[\mathcal{T}]}  $ is,
	\ptcheck{8VIII22,  with Maple }
	\begin{eqnarray}
		\mcH^x {[\partial_u]}
		& = &
		\sqrt{|\det g|}
		\Big(- \nabla^x \SField \,\Lie_\mathcal{T} \SField
		+
		\frac 12
		\big(\nabla^\alpha \SField \nabla_\alpha \SField \pmass \SField^2
		\big)
		\mathcal{T}^x
		\Big)
		\nonumber
		\\
		& = &
		(\alpha\xdS)^{-4}
		\sqrt{\frac { \det \spherem } {1-\xdS^2} }
		\Big( -\alpha^3 x^2  (1-\xdS^2)^{3/2} \partial_x \SField
		\big[ \xdS
		\cos \psi \partial_{\xdS} \SField + \sin \psi \partial_{\psi} \SField\big]
		\nonumber
		\\
		&&
		-\frac    { \alpha^3 \xdS ^2}  2 \sqrt{ {1-\xdS ^2}}
		\xdS
		\cos \psi
		\big( - (1-\xdS^2) (\partial_\xdS \SField)^2  + |\sD   \SField|^2_\spherem  + m^2(\alpha\xdS)^{-2} \SField^2
		\big)
		\Big) \nonumber
		\\
		&= &
		-
		\alpha^{-1}\xdS ^{-2}
		\sqrt{  \det \spherem }
		\Big(  (1-\xdS^2)
		\sin \psi \partial_x \SField \partial_{\psi} \SField
		\nonumber
		\\
		&&
		+\frac    { 1}  2
		\xdS
		\cos \psi
		\big( (1-\xdS^2) (\partial_\xdS \SField)^2  + |\sD   \SField|^2_\spherem  + m^2(\alpha\xdS)^{-2} \SField^2
		\big)
		\Big)
		\,,
		\label{30VII22.10}
	\end{eqnarray}
	where $\sD$
	is the covariant derivative associated with $\spherem$.
	Inspection of \eqref{30VII22.10} reveals terms which diverge as $x\to 0$  with this asymptotics and which, using $m^2 = 2\alpha^2$, can be collected into a divergence as follows:

	\begin{eqnarray}
		\mcH^x {[\partial_u]}
		& = &
		-
		\alpha^{-1}\xdS ^{-2}
		\sqrt{  \det \spherem }
		\Big\{  (1-\xdS^2)
		\sin \psi \partial_\xdS (\xdS\cphi) \partial_{\psi} (\xdS\cphi)
		\nonumber
		\\
		&&
		+\frac    { 1}  2
		\xdS
		\cos \psi
		\big(  (1-\xdS^2) (\partial_\xdS (\xdS\cphi))^2  + |\sD   (\xdS\cphi)|^2_\spherem  + 2    \cphi ^2
		\big)
		\Big\}
		\nonumber
		\\
		& = &
		-
		\alpha^{-1}\xdS ^{-2}
		\sin^2\psi \sin\theta
		\Big\{  \xdS (1-\xdS^2)
		\sin \psi (\cphi + \xdS\partial_\xdS \cphi) \partial_{\psi} \cphi
		\nonumber
		\\
		&&
		+\frac    { 1}  2
		\xdS
		\cos \psi
		\big(  (1-\xdS^2) (\cphi+  \xdS \partial_\xdS \cphi )^2  +  \xdS^2 |\sD   \cphi |^2_\spherem  + 2    \cphi ^2
		\big)
		\Big\}
		\nonumber
		\\
		& = &
		-
		\frac 12 \partial_\psi  \Big\{   \alpha^{-1}\xdS ^{-1}
		\sin^3\psi \sin\theta
		(1-\xdS^2)   \cphi^2\Big\}
		\nonumber
		\\
		&   &
		+ \frac 32
		\alpha^{-1}\xdS ^{-1}
		\sin^2\psi \cos \psi  \sin\theta
		(1-\xdS^2)   \cphi^2
		\nonumber
		\\
		&   &
		-
		\alpha^{-1}\xdS ^{-2}
		\sin^2\psi \sin\theta
		\Big\{ \xdS^2  (1-\xdS^2)
		\sin \psi \partial_\xdS  \cphi  \partial_{\psi}  \cphi
		\nonumber
		\\
		&&
		+\frac    { 1}  2
		\xdS
		\cos \psi
		\big(  (1-\xdS^2) (3 \cphi^2 + 2 \xdS\cphi  \partial_\xdS \cphi
		+   \xdS^2  (\partial_\xdS  \cphi)^2)
		+ { 2 x^2 \cphi^2 }
		+ \xdS^2 |\sD    \cphi |^2_\spherem
		\big)
		\Big\}
		\nonumber
		\\
		& = &
		-
		\partial_\psi
		\underbrace{
			\Big\{  \frac 12 \alpha^{-1}\xdS ^{-1}
			\sin^3\psi \sin\theta
			(1-\xdS^2)   \cphi^2\Big\}
		} _ {=:-\Hbd^x}
		\nonumber
		\\
		&   &
		-
		\alpha^{-1}
		\sqrt{  \det \spherem }
		\Big\{    (1-\xdS^2)
		\sin \psi \partial_\xdS  \cphi  \partial_{\psi}  \cphi
		\nonumber
		\\
		&&
		+\frac    { 1}  2
		\cos \psi
		\big(  (1-\xdS^2) (  2  \cphi  \partial_\xdS \cphi +   \xdS
		(\partial_\xdS \cphi)^2)
		{ + 2 \xdS \cphi^2 }
		+ \xdS  |\sD    \cphi |^2_\spherem
		\big)
		\Big\}
		\nonumber
		\\
		& =: &   \partial_\psi \Hbd^x + {\Hvol}^x
		\,.
		\phantom{xxxxxx}
		\label{1VIII22.2}
	\end{eqnarray}
	%
	It should be admitted that this way of handling the divergence is ambiguous, and could lead to a different finite part of the resulting boundary integral when, e.g., other coordinates are used.

	\subsubsection{Spacelike Cauchy surfaces}
	
	Consider, first, the Hamiltonian charge obtained by integrating \eqref{1VIII22.2} over a three-dimensional sphere of constant $\xdS$. Then there is no boundary term, and choosing the exterior orientation of the slices appropriately one is led to a finite charge equal to
	%
	\begin{eqnarray}
		\lefteqn{
			\alpha^{-1}  \int_{S^3}
			\Big \{
			(1-\xdS^2)
			\sin \psi \partial_\xdS  \cphi  \partial_{\psi}  \cphi
		}
		& &
		\nonumber
		\\
		& &
		+\frac    { 1}  2
		\cos \psi
		\big(  (1-\xdS^2) (  2  \cphi  \partial_\xdS \cphi +   \xdS
		(\partial_\xdS \cphi)^2 ) { + 2 \xdS \cphi^2 }  + \xdS  |\sD    \cphi |^2_\spherem
		\big)
		\Big\}
		d\mu_{\spherem}
		\,,
		\phantom{xxx}
		\label{1VIII22.3}
	\end{eqnarray}
	independently of $x$.
	In particular we can pass to the limit $x\to 0$ to define
	\begin{eqnarray}
		{\check E _\mcH [}\scrip]
		& := &
		\alpha^{-1}  \int_{S^3}
		\big(
		\sin \psi \partial_\xdS  \cphi  \partial_{\psi}  \cphi
		+
		\cos \psi
		\cphi  \partial_\xdS \cphi
		\big)|_{\xdS=0}
		\,
		d\mu_{\spherem}
		\nonumber
		\\
		& = &
		\alpha^{-1}  \int_{S^3}
		\big(
		\sin \psi \omone \cphi  \partial_{\psi} \ozero \cphi
		+
		\cos \psi
		\ozero \cphi \omone \cphi
		\big)
		d\mu_{\spherem}
		\nonumber
		\\
		& = &
		\alpha^{-1}  \int_{S^3}
		\omone \cphi  \partial_{\psi}
		(\sin \psi  \ozero \cphi)
		d\mu_{\spherem}
		\,.
		\label{1VIII22.4}
	\end{eqnarray}
	\ptcheck{ 23VIII22, up to here}
	Note that $\partial_u$ is tangent to $\{x=0\}$, and equals there
	$$
	\partial_u|_{x=0} = - \alpha \sin \psi \partial_\psi\,.
	$$
	
	Writing $\pi^x = - p^x \sqrt{\det \spherem}$,
	the symplectic form $\Omega$ is independent of $x$ on solutions of the field equations and reads 
	\begin{equation}\label{1IV23.2}
		\Omega = \int_{S^3} \delta p^x \wedge \delta \chi
		\, d\mu_\spherem
	\end{equation}
	The Legendre transformation leads to $p^x = -\alpha^{-2} \partial_x \chi$ and a Hamiltonian on $\scri$ equal to
	\begin{eqnarray}
		H
		&:= &
		\check E _\mcH [\scrip]
		\nonumber
		\\
		&
		=
		&
		\alpha \int_{S^3}
		p^x \partial_{\psi}
		(\sin \psi  \ozero \cphi)
		d\mu_{\spherem}
		\nonumber
		\\
		&
		\equiv
		&
		\alpha \int_{S^3}
		p^x \partial_{\psi}
		(\sin \psi  \ozero \cphi)
		\sin^2(\psi) \sin(\theta) d\psi d\theta d\varphi
		\,.
		\label{4VIII22.4}
	\end{eqnarray}
	The resulting
	Hamilton equations take the  simple form:
	\begin{equation}\label{1IV23.3}
		\frac{dp^x}{du} =  - \frac{\delta  H }{\delta \ozero\chi} =
		\frac{1}{\sin  \psi}\alpha \partial_\psi( \sin^2 \psi p^x)
		\,,
		\quad
		\frac{d\ozero\chi}{du} =  \frac{\delta  H }{\delta p^x } =  \alpha \partial_\psi (\sin \psi \ozero\chi)
		\,,
	\end{equation}
	with $p^x$ and $\ozero\chi$ evolving independently of each other.
	
	\subsubsection{Corner terms}
	\label{sss7XI22.1}
	
	We pass now to a Cauchy surface which is the union of a light-cone and the ``complementary part of $\scrip$''.
	
	Recall that $\mcC_u$ is a light-cone in  $\mcM$ with vertex at $r=0$. For $\alpha R>1$ we set
	$$
	\hyp_{\xdS,u} :=
	\hyp_\xdS \setminus I^+(\mcC_u)
	\,,
	\qquad
	\scrip_u:= \scrip \setminus I^+(\mcC_u)
	\,,
	$$
	see Figure~\ref{F23IV22.1}.
	The hypersurface $\scrip_u$ can be viewed as the limit, as $\xdS$ tends to $0^{+}$, of the $\hyp_{\xdS,u}$'s.
	Let $S_{\xdS,u}$ be the intersection of $\mcC_u$ with the $\hyp_\xdS$.
	Let $S_{0,u}$ be the intersection of $\mcC_u$ with the conformal boundary at infinity $\scrip$; thus the surface $S_{0,u}$ is a limit of $S_{\xdS,u}$ in which $\xdS$ tends to $0^{+}$.

	Since the Noether current $\mcH^\mu =\mcH^\mu[\partial_u]$ has vanishing divergence, we have for $x> x'> 0$
	\begin{equation}\label{4XI22.3b}
		\int_{\mcC_{u,R(x)}{\cup\hyp_{\xdS,u}}}\mcH^\mu  dS_\mu
		=
		\int_{\mcC_{u,R(x')}{\cup\hyp_{\xdS',u}}}\mcH^\mu dS_\mu
		\,.
	\end{equation}
	We can thus pass to the limit $x'\to0$ to obtain
	\begin{equation}\label{4XI22.3}
		\int_{\mcC_{u,R(x)}{\cup\hyp_{\xdS,u}}}\mcH^\mu  dS_\mu
		=
		\lim_{x\to 0} \int_{\mcC_{u,R(x )}\cup \hyp_{\xdS,u}}\mcH^\mu dS_\mu
		\,,
	\end{equation}
	in particular the limit exists and is finite.
	
	Recall that (see \eqref{1VIII22.2})
	\begin{equation}\label{4XI22.3a}
		\mcH^x = {\Hvol}^x + \partial_\psi \Hbd^x
		\,.
	\end{equation}
	so that
	\begin{equation}\label{4XI22.4}
		\int_{\hyp_{\xdS,u}}\mcH^\mu  dS_\mu
		=
		\int_{\hyp_{\xdS,u}}\Hvol^x   dS_x
		+  \mathcal{B}_{1}
		\,,
	\end{equation}
	where
	\begin{eqnarray}
		\mathcal{B}_{1}&:=&-
		\frac 12 \int_{S^3} \partial_\psi  \Big\{   \alpha^{-1}\xdS ^{-1}
		\sin^3\psi
		(1-\xdS^2)   \cphi^2\Big\}
		\, d \psi \, d \theta \, d \varphi
		\nonumber
		\\
		&=&-
		\frac 12 \int_{S_{x,u }}
		\Big\{   \alpha^{-1}\xdS ^{-1}
		\sin^3\psi
		(1-\xdS^2)   \cphi^2\Big\}
		\, d \theta \, d \varphi
		\nonumber
		\\
		&=&-
		\frac 1 {2 \alpha {
				\cosh(\alpha u   )^{3} }}
		\int_{S_{x,u }}
		\Big[\frac{( \ozero \cphi)^2 }{x}
		+ 2 \ozero \cphi \omone \cphi - 3 \sinh(\alpha u )
		(\ozero \cphi)^2
		\nonumber
		\\
		&&
		+ O(\xdS)
		\Big]
		\sin\theta \, d \theta \, d \varphi
		\, .
		\label{4XI22.1}
	\end{eqnarray}
	Similarly, setting
	\begin{equation}\label{4XI22.5}
		\mcH^u = {\Hvol}^u + \partial_r \Hbd^u
		\,,
	\end{equation}
	we can rewrite \eqref{24V21.t1as} as
	\begin{equation}\label{4XI22.6}
		\int_{\mcC_{u,R(x)}}\mcH^\mu  dS_\mu
		=
		\int_{\mcC_{u,R(x)} }{\Hvol}^u   dS_u
		+  \mathcal{B}_{2}
		\,,
	\end{equation}
	where
	\begin{eqnarray}
		\mathcal{B}_{2}&:=&
		\frac12
		\int_{\mcC_{u,R(x)}}
		\partial_{r}
		\Big[
		\big(r^{2} - \alpha^{2} r^{4}\big) \SField \big(\partial_{r} \SField\big)
		\sin \theta
		\Big]
		\, d r \, d \theta \, d \varphi
		\nonumber
		\\
		&=&
		\frac12
		\int_{S_{x,u }}
		\big(r^{2} - \alpha^{2} r^{4}\big) \SField  \partial_{r} \SField
		\sin \theta
		\, d \theta \, d \varphi
		\, .
		\label{4XI22.2}
	\end{eqnarray}%
	Thus
	\begin{equation}\label{4XI22.7}
		\int_{\mcC_{u,R(x)}{\cup\hyp_{\xdS,u}}}\mcH^\mu  dS_\mu
		=
		\int_{\hyp_{\xdS,u}}\Hvol^x   dS_x
		+ \int_{\mcC_{u,R(x)} }{\Hvol}^u   dS_u
		+  \mathcal{B}_{1}
		+  \mathcal{B}_{2}
		\,.
	\end{equation}

	Both volume integrals have a finite limit as $x\to 0$.
	It remains to analyze the divergent boundary terms in the energy on the family $S_{\xdS,u}$'s.
	This requires some changes of coordinates.
	Equations \eqref{27VII22.t1b}-\eqref{19VIII22.t1} together with $x=1/\cosh(\alpha\tau)$ can be inverted as
	\ptcheck{13X22, mathematica file, but needs inverting}
	\begin{eqnarray}
		r (\xdS,u  )&=& \alpha^{-1} \frac{\sqrt{1-\xdS^{2}}-\xdS \sinh (\alpha u ) }{\xdS \cosh (\alpha u )}
		\nonumber
		\\
		&=&\frac{ 1 -\xdS \sinh (\alpha u ) + O(\xdS^2) }{ \alpha  \xdS \cosh (\alpha u )}
		\, ,
		\label{21II23.t21}
		\\
		\sin \psi(\xdS,u )&=&\frac{\sqrt{1-\xdS^{2}}-\xdS \sinh (\alpha u ) }{\cosh (\alpha u )}
		\nonumber
		\\
		& = &
		\frac{1 -\xdS \sinh (\alpha u ) }{\cosh (\alpha u )} + O(x^2)
		\, .
	\end{eqnarray}
	Since we need to calculate derivatives of the fields along $\mcC_u$ in the new variables, we need instead $(x,\psi)$ as a function of $(r,u)$. For   $x>0$ and $\alpha r>0$ one finds:
	\ptcheck{3XI22}
	\begin{eqnarray}
		x(r,u) &=&\sqrt{\frac{1-\tanh (\alpha u)^2}
			{1+\alpha^2 r^2+2 \alpha r \tanh (\alpha u)}}
		\,,
		\\
		\sin\big(\psi(r,u)\big)
		&=&
		\alpha r \sqrt{\frac{1-\tanh (\alpha u)^2}{1+\alpha^2 r^2+2 \alpha r \tanh (\alpha u)}}
		\,.
	\end{eqnarray}
	This leads to
	\ptcheck{4XI22; but watch out if psi's beyond half cone are considered}
	\begin{eqnarray}
		\frac{\partial \psi  }{\partial r}\bigg|_{u=\text{const}} &=& \alpha x^2 \cosh (\alpha u)
		\,,
		\\
		\frac{\partial x  }{\partial r}\bigg|_{u=\text{const}}&=&-\alpha  x^2 \sqrt{1-x^2} \cosh (\alpha u)
		\,.
	\end{eqnarray}
	One then finds
	\ptcheck{4XI22; needs to be finished, and added to the summary section}
	\begin{eqnarray}
		\mathcal{B}_{2}
		&=&
		\frac12
		\int_{S_{x,u }}
		\big(r^{2} - \alpha^{2} r^{4}\big) \SField
		\Big(
		\frac{\partial \psi}{\partial r}  \partial_{\psi}
		+ \frac{\partial x}{\partial r} \partial_{x}
		\Big)
		\SField
		\sin \theta
		\, d \theta \, d \varphi
		\nonumber
		\\
		&=&
		\frac12
		\int_{S_{x,u }}
		\big(\alpha^{-2} \xdS^{-2} \sin^{2} \psi  - \alpha^{2} (\alpha^{-1} \xdS^{-1} \sin \psi )^{4}\big)   \xdS\cphi  \times
		\nonumber
		\\
		& &
		\alpha \xdS^2 \cosh (\alpha u) \bigg(\partial_{\psi}-\sqrt{1-\xdS^2}\partial_{x}\bigg)
		\xdS\cphi
		\sin \theta
		\, d \theta \, d \varphi
		\nonumber
		\\
		&=&
		\frac{1}{2 \alpha \cosh^3 (\alpha u) }
		\int_{S_{x,u }}
		\ozero \cphi
		\Big[
		\frac{\ozero \cphi}{ \xdS }
		-
		\partial_{\psi}\ozero \cphi+3 \oone \cphi
		-4 \sinh(\alpha u) \ozero \cphi
		\nonumber
		\\
		& &
		+O(\xdS)
		\Big]
		\sin \theta
		\, d \theta \, d \varphi
		\, .
		\label{4XI22.2a}
	\end{eqnarray}%

	We are ready now to compare \eqref{4XI22.1} and \eqref{4XI22.2a}. Note that each diverges when $\xdS$ tends to $0^{+}$, but   their sum $\Delta \mathcal{B}_{x,u }:=\mathcal{B}_{1}+\mathcal{B}_{2} \, ,$,  relevant for the total energy, is finite. Indeed, passing to the limit $\xdS \to 0^{+}$  one finds
	\ptcheck{20II23}
	\begin{eqnarray}
		\lefteqn{
			\Delta \mathcal{B}_{0,u }
			:=
			\lim\limits_{x \to 0^{+}}\Delta \mathcal{B}_{x,u }
		}
		&&
		\nonumber
		\\
		&   =
		&
		\frac{1}{2 \alpha \cosh^{3} (\alpha u ) }
		\int_{S_{0,u }}
		\ozero \cphi
		\Big(
		\omone \cphi
		-
		\partial_{\psi}\ozero \cphi
		-
		\ozero \cphi
		\sinh(\alpha u )
		\Big)
		\sin \theta
		\, d \theta \, d \varphi
		\, .
		\phantom{xxx}
		\label{5III23.6}
	\end{eqnarray}

	Summarising: in the phase space described above the dynamical system induced by translating in $u$  the tip of the light-cone is Hamiltonian, with Hamiltonian  equal to (compare \eqref{20II23.21} and \eqref{4XI22.3a}):
	\begin{eqnarray}
		\nonumber 
		\newH &: =&  \lim_{x\to 0} \int_{\mcC_{u,R(x )}\cup \hyp_{\xdS,u}}\mcH^\mu dS_\mu
		\\
		&=&
		\underbrace{  \int_{\mcC_{u }} {\Hvol}^u  dS_u
			+
			\Delta \mathcal{B}_{0,u }
		}_{=:\check E_\mcH[\mcC_u]}
		-
		\int_{\scrip_u} {\Hvol}^x dS_x
		\label{20II23.11}
	\end{eqnarray}
	(with the minus sign in the last integral arising from the fact that $\partial_x$ is past-directed),
	where now all the terms are finite.
	In this picture the ``leaking terms'' correspond   to an exchange of energy between the subsystem consisting of the field on the light cone and the field  on $\scrip_u$.

\subsection{Maxwell fields}
\label{secMaxwell}

The analysis for Maxwell fields is quite simpler than that for the conformally-covariant scalar field.
The phase space of Cauchy data on three-dimensional spheres of constant $x$ consists of smooth fields $(A_{\mu},\partial_{\nu} A_{\mu})$ with, loosely speaking,  symplectic form
\begin{eqnarray}
	\label{21II23.t1}
	\Omega  &= &
	-  \int_{x=\const} \delta \pi^{\mu x} \wedge \delta A_{\mu}
	\,
	dS_x
	\nonumber
	\\
	&= & 
	-
	\int_{x=\const} \delta \pi^{k x} \wedge \delta A_{k}
	\,
	dS_x
	\,,
\end{eqnarray}
where we used $(x^\mu) = (x,x^k)$, where the $x^k$'s are local coordinates on $\scrip$, as well as the fact that $\pi^{xx}=0$.
There exists a gauge in which all fields extend smoothly through $x=0$, so that we can write
\begin{eqnarray}
	\label{21II23.t4}
	A_{k}
	& = & \ozero A_{k}
	+
	\xdS
	\omone A_{k}
	+
	\cdots
	\, ,
	\\
	\label{7IX22.t0}
	F_{x k} & = &  \ozero F_{x k}+ \xdS \omone F_{x k} +\xdS^2 \omtwo F_{x k}   + \ldots
	\, ,
\end{eqnarray}
where the expansion coefficients are functions of $x^k$. Since $\Omega$ is conserved for variations satisfying the field equations it holds that
\begin{equation}\label{21II23.t2}
	\Omega = \frac{1}{4 \pi} \int_{x=0} \spherem^{k l}
	\delta \ozero F_{x l} \wedge \delta \ozero A_{l}
	\, d\mu_{\spherem}
	\,.
\end{equation}

The dynamics generated by the flow of $\partial_u$  is Hamiltonian, with
\begin{eqnarray}
	{E _\mcH [}S^{3}]  &:=&
	-
	\int_{S^{3}}  \mcH^x {[\partial_u]} dS _x
	\,.
	\phantom{xxx}
	\label{7III23.1a}
\end{eqnarray}
\ptcheck{7XI, all the section up to here}
Assuming $\alpha \ne 0$,
using \eqref{30VII12.2}-\eqref{30VII22.5} we find
\ptcheck{ 7XI, maple file M1 MaxwellinxPsi.mv}
\begin{eqnarray}
	\mcH^\xdS {[\partial_u]}
	& := &   \frac{\partial \mcL}{\partial A_{\beta,\xdS} } \myLie_{\mathcal{T}} A_{\beta}
	-
	\mcL
	\mathcal{T}^\xdS
	\nonumber
	\\
	&=&
	-\frac{1}{4 \pi} \sqrt{|-\det g|}
	\Big(F^{\xdS \beta} \mathcal{T}^{\alpha} F_{\alpha \beta}
	-
	\frac {1}{4}
	\big(
	F^{\nu \beta} F_{\nu\beta}
	\big)
	\mathcal{T}^\xdS
	\Big)
	\nonumber
	\\
	&=&
	\frac{(\alpha\xdS)^{-4}}{4 \pi}
	\sqrt{\frac { \det \spherem } {1-\xdS^2} }
	\Big\{
	F^{\xdS \beta}
	\alpha \sqrt{ {1-\xdS ^2}}
	\big[ \xdS
	\cos \psi F_{\xdS \beta} + \sin \psi F_{\psi \beta}\big]
	\nonumber
	\\
	& & -\frac{1}{4} \alpha \sqrt{ {1-\xdS ^2}} \xdS
	\cos \psi \big(
	F^{\nu \beta} F_{\nu\beta}
	\big)
	\Big\}
	\nonumber%
	\\
	&=&
	-\frac{1}{4 \pi} \alpha \sqrt{\det \spherem}
	\Big\{
	\frac{1}{2} \xdS (1-\xdS^2) \cos \psi F_{x k} F_{x l} \spherem^{k l}
	\nonumber
	\\
	& &
	+ (1-\xdS^2) \sin \psi F_{x k} F_{\psi l} \spherem^{k l}
	+\frac{1}{4} x \cos \psi F_{m k} F_{n l} \spherem^{m n} \spherem^{k l}
	\Big\}
	\, .
	\phantom{xxxxxx}
	\label{8III23.2}
\end{eqnarray}
Hence
\begin{eqnarray}
	{E _\mcH [}S^{3}]  &=&
	-\frac{\alpha }{4 \pi}\int_{S^{3}}  \Big[
	\sin \psi \, \spherem^{k l}\ozero F_{x k} \ozero F_{\psi l}
	+O(x)
	\Big]
	\, d\mu_{\spherem}
	\nonumber
	\\
	&=&-\frac{\alpha }{4 \pi}\int_{S^{3}}
	\sin \psi  \, \spherem^{k l}  \ozero F_{x k} \ozero F_{\psi l}
	\, d\mu_{\spherem}
	\, ,
	\phantom{xxx}
	\label{7III23.1}
\end{eqnarray}
\ptcheck{7XI, all the section up to here}
where in the last equality we used the fact that $ {E _\mcH [}S^{3}] $ does not depend upon $x$.

To take care of the leakage, for each $u$ we can consider the phase space consisting of the fields $ A_{\mu} $ on $\mcC_u$, and $(A_k, \partial_x A_k)$ on the set
$
\scrip_u
$
of \eqref{23II23.1},
equipped with the symplectic form
\begin{eqnarray}
	\Omega &=&
	\frac{1}{4 \pi}
	\int_{\mcC_u}
	\Big(
	r^2 \delta F_{u r} \wedge \delta A_{r}
	+
	\zzhTBW^{A B} \delta F_{r B} \wedge \delta A_{A}
	\Big)
	\,dr \,  d\mu_{\zzhTBW}
	\nonumber
	\\
	& &
	-
	\frac{1}{4 \pi} \int_{\scrip_u} \spherem^{k l}
	\delta \ozero F_{x k} \wedge \delta \ozero A_{l}
	\, d\mu_{\spherem}
	\,.
	\label{21II23.t3}
\end{eqnarray}

The Hamiltonian charge associated with moving the light-cones along the flow of the Killing vector $\partial_u\equiv \mathcal{T}$ decomposes as  in \eqref{20II23.21},
\begin{eqnarray}
	\newH &=&
	\int_{\mcC_u}
	\mcH^u{[\partial_u]}
	dS_u
	-
	\int_{\scrip_u} \mcH^x{[\partial_u]}
	\, dS_x
	\,,
	\label{20II23.21cf}
\end{eqnarray}
(where the minus sign in the second integral is again motivated by orientation considerations)
but now each integrand is finite without further due; hence no corner contributions arise.

\appendix

\section{Killing fields in Minkowski, de Sitter and  anti-de Sitter  spacetimes}
\label{A30IX21.1}

In order to determine the Noether charges in our formalism we will need the explicit form of the Killing vector fields in Bondi coordinates on the de Sitter, and anti-de Sitter and Minkowski spacetimes.

\subsection{Killing fields in de Sitter spacetime}
\label{s30VII21.5}

We use the following basis of the space of Killing vectors  in de Sitter spacetime
\begin{eqnarray}
	\mathcal{T}&=&\partial_{u} \, ,
	\label{24VI21.t1}
	\\
	\mathcal{R} &=& \varepsilon^{B A} \zspaceD_{A}\big(R_{i} \wtx ^i \big) \partial_{B}\, ,
	\\
	\LKmom&=&e^{\alpha u}\Big[\LKmcon_{i} \wtx ^i \partial_{u}-\big(\alpha r +1\big)\LKmcon_{i} \wtx ^i \partial_{r}-\frac{\alpha r +1}{r} \zspaceD^{A}(\LKmcon_{i} \wtx ^i) \partial_{A} \Big]\, ,
	\label{24VI21.t2a}
	\\
	\LKbst&=&e^{-\alpha u}\Big[\LKbcon_{i} \wtx ^i \partial_{u}+\big(\alpha r -1\big)\LKbcon_{i} \wtx ^i \partial_{r}+\frac{\alpha r -1}{r} \zspaceD^{A}(\LKbcon_{i} \wtx ^i) \partial_{A} \Big]\, ,\phantom{xx}
	\label{24VI21.t2}
\end{eqnarray}
where
$R_{i},\LKmcon_{i}$ and $\LKbcon_{i}$ are constants.
\ptcheck{5VIII21}
Using the following coordinate change
\begin{eqnarray}
	x^{0}&=&\frac{\big(\sinh (\alpha u)-r^2 \alpha^2 \sinh (\alpha u)-\alpha r\big) \cosh (\alpha u)+\alpha r-\sinh (\alpha u)}{\alpha\big(\big(\alpha^2 r^2-1\big) \cosh (\alpha u)^2+2 \cosh (\alpha u)-\alpha^2 r^2-1\big)} \, , \phantom{xxx} \\
	x^{1}&=&\frac{\big(\cosh (\alpha u)-\sinh (u \alpha) \alpha r-1\big) r \sin \theta \cos \phi}{\big(\alpha^2 r^2-1\big) \cosh (\alpha u)^2+2 \cosh (\alpha u)-\alpha^2 r^2-1} \, , \\
	x^{2}&=&\frac{\big(\cosh (\alpha u)-\sinh (u \alpha) \alpha r-1\big) r \sin \theta \sin \phi}{\big(\alpha^2 r^2-1\big) \cosh (\alpha u)^2+2 \cosh (\alpha u)-\alpha^2 r^2-1} \, , \\
	x^{3}&=&\frac{\big(\cosh (\alpha u)-\sinh (u \alpha) \alpha r-1\big) r \cos \theta}{\big(\alpha^2 r^2-1\big) \cosh (\alpha u)^2+2 \cosh (\alpha u)-\alpha^2 r^2-1} \, ,
\end{eqnarray}
the de Sitter metric \eqref{23IV22.1} transforms into conformally Minkowskian form
\begin{equation}
	g=\frac{4}{\big(1+s^2 \alpha^2\big)^2} \eta_{\mu \nu} d x^{\mu} d x^{\nu} \, ,
\end{equation}
where $\eta_{\mu \nu}=\mathrm{diag} (-1,1,1,1)$ and $s^2=\eta_{\mu \nu} x^{\mu} x^{\nu}$. Defining
\ptcheck{25III23}
\begin{eqnarray}
	S_{\mu}&=&-\frac{1}{2} \partial_{\mu}-\frac{\alpha^2}{2} \big(2 \eta_{\mu \lambda} x^{\lambda} x^{\nu}-s^2 {\delta_\mu}^{\nu}\big) \partial_{\nu} \, ,\\
	L_{\mu \nu}&=&\eta_{\mu \lambda} x^\lambda \partial_\nu-\eta_{\nu \lambda} x^\lambda \partial_\mu \, ,
\end{eqnarray}
one finds
\begin{eqnarray}
	\mathcal{T}&=&S_{0} \, , \\
	\label{8II23.t1}
	\mathcal{R}&=&\widetilde{R}_{i} \epsilon^{i j k} L_{j k} \, ,\\
	\LKmom&=& \widetilde{\LKmcon}_{i} \eta^{i j} (\alpha L_{0 j}- S_{j})  \, , \\
	\LKbst&=& \widetilde{\LKbcon}_{i}  \eta^{i j} (\alpha L_{0 j}+ S_{j}) \, ,
\end{eqnarray}
where $\{i,j,k\} \in \{1,2,3\}\, ,\epsilon^{1 2 3}=1$. $\widetilde{R}_{i},\widetilde{\LKmcon}_{i},\widetilde{\LKbcon}_{i}$ are respectively linear combinations of $R_{i}, $. The following commutation relations hold
\begin{eqnarray}
	\big[L_{\alpha \beta}, L_{\rho \sigma} \big]&=&\eta_{\beta \rho} L_{\alpha \sigma}+\eta_{\alpha \sigma} L_{\beta \rho}-\eta_{\beta \sigma} L_{\alpha \rho}-\eta_{\alpha \rho} L_{\beta \sigma} \, , \\
	\big[L_{\mu \nu}, S_{\rho} \big]&=& \big(\eta_{\mu \lambda} \eta_{\rho \nu} - \eta_{\mu \rho} \eta_{\lambda \nu}\big) \eta^{\lambda \sigma} S_{\sigma} \, , \\
	\big[S_{\beta}, S_{\rho} \big]&=& \alpha^2 L_{\rho \beta} \, ,
\end{eqnarray}
which leads to
\begin{eqnarray}
	\big[\mathcal{T},\mathcal{R} \big]&=&0 \, , \\
	\big[\mathcal{T},\LKmom\big] &=&\alpha \LKmom \, , \\
	\big[\mathcal{T},\LKbst\big] &=&-\alpha \LKbst \, ,\\
	\big[\mathcal{R}_{I},\mathcal{R}_{II} \big]&=&\mathcal{R}_{III} \, ,
\end{eqnarray}
where $\mathcal{R}_{I},\mathcal{R}_{II},\mathcal{R}_{III}$are given by
\eqref{8II23.t1}.

\subsection{Killing fields in anti-de Sitter spacetime}
\label{ss17IX21.1}
Using for anti-de Sitter $\aldS:=-\imath \alpha$ , while simultaneously keeping $\LKmcon_{i}, \LKbcon_{i}$ real, the real and imaginary parts of $\LKmom$ and $\LKbst$ are
\begin{eqnarray}
	\LKmom(\LKmcon_{i})&=&\PadS(\LKmcon_{i})+ \imath \LadS(\LKmcon_{i})
	\label{28IV22.3}
	\\
	\LKbst(\LKbcon_{i})&=&\PadS(\LKbcon_{i})+ \imath \LadS(-\LKbcon_{i})
	\label{28IV22.4}
\end{eqnarray}
where
%
\begin{eqnarray}
	\PadS(\PadSc_{i})&=&\PadSc_{i}\Big[\wtx^{i}  {\cos (\aldS u)}  \partial_{u}+ \wtx^{i}(\aldS r  {\sin (\aldS u)}- {\cos (\aldS u)}) \partial_{r}
	\nonumber
	\\
	& &+\frac{(\aldS r  {\sin (\aldS u)}- {\cos (\aldS u)})}{r}\zspaceD^{A} \wtx^{i}\partial_{A}
	\Big]
	\, ,
	\label{28IV22.5}
	\\
	\LadS(\LadSc_{i})&=& \LadSc_{i}\Big[\wtx^{i}  {\sin (\aldS u)} \partial_{u}-\wtx^{i}( {\sin (\aldS u)}+\aldS r  {\cos (\aldS u)})\partial_{r}
	\nonumber
	\\
	& &
	-\frac{( {\sin (\aldS u)}+\aldS r  {\cos (\aldS u)})}{r} \zspaceD^{A} \wtx^{i}\partial_{A}
	\Big]
	\, .
	\label{28IV22.6}
\end{eqnarray}

\subsection{Killing fields in Minkowski spacetime}
The Killing fields in Minkowski spacetime will be labelled as
\begin{eqnarray}
	\mathcal{T}&=&\partial_{t} \, ,
	\label{10VII21.t3}
	\\
	\mathcal{R} &=&\epsilon^{ijk} R_i \delta_{j l}x^l \partial_k
	\, ,
	\\
	\mathcal{P}&=&P^k \partial_k \, , \\
	\mathcal{L} &=&L_i x^i  \partial_{t} +t L^i \partial_{i}
	\, ,
	\label{10VII21.t4}
\end{eqnarray}
where $P_i \equiv P^i$ , $L_i \equiv L^i$ and $R_i$  are all constants.

The coordinate transformation between Minkowskian and   Bondi coordinates
\begin{equation}
	\left( u=t-r  ,r,x^A \right)
	\label{10VII21.t5}
\end{equation}
gives
\begin{equation}
	\partial_t = \partial_u
	\,,
	\quad
	\partial_i = \wtx ^i \left(\partial_r-\partial_u \right)+\frac{1}{r}\zspaceD^{A} {\wtx^i}  \partial_{A}
	\, ,
\end{equation}
where  the fields
\begin{equation}
	\wtx ^{i}:=\frac{x^{i}}{r}
\end{equation}
form a basis of the space of $\ell=1$ spherical harmonics,
and thus $\wtx^i$ is viewed as a scalar on $S^2$ in formulae such as $\zspaceD^{A} {\wtx^i} $.
Under \eqref{10VII21.t5} the Killing vectors \eqref{10VII21.t3}-\eqref{10VII21.t4} become
\begin{eqnarray}
	\mathcal{T}&=&\partial_{u} \, ,
	\label{10VII21.t6}
	\\
	\mathcal{R} &=&\varepsilon^{A B} \zspaceD_{B}(R_i \wtx ^i)   \partial_{A}
	\, ,
	\\
	\mathcal{P}&=&P_i \Big(\wtx ^i \left(\partial_r-\partial_u \right)+\frac{1}{r}\zspaceD^{A} {\wtx^i}  \partial_{A}\Big) \, ,
	\label{10VII21.t7a}
	\\
	\mathcal{L} &=&
	L_i
	\Big(
	-u \wtx ^i  \partial_{u} +(u+r)  \wtx^ i \partial_r +\big(1+\frac{u}{r}\big)\zspaceD^{A} {\wtx^i}  \partial_{A}\big)
	\Big)
	\, .
	\label{10VII21.t7}
\end{eqnarray}
where $\varepsilon^{A B}$ is a two-dimensional Levi-Civita tensor;  in spherical coordinates $(\theta, \phi)$ we take the sign $\varepsilon^{\theta \phi}=\frac{1}{\sin \theta} \, .$

The Killing fields for Minkowski spacetime can be obtained as a limit of de Sitter spacetime.  Equations \eqref{24VI21.t2a}-\eqref{24VI21.t2} and  \eqref{10VII21.t7a} give
\begin{equation}
	\mathcal{P} = -
	\frac 12
	\lim\limits_{\alpha \to 0} \Big(\LKmom
	+\LKbst\Big) \, ;
	\label{28IV22.1}
\end{equation}
where one has to set  $l_i=p_i= P_i$.
Analogically, \eqref{24VI21.t2a}-\eqref{24VI21.t2} and  \eqref{10VII21.t7} lead  to
\begin{equation}
	\mathcal{L} =
	\frac 12
	\lim\limits_{\alpha \to 0} \Big(\frac{\LKbst-\LKmom}{\alpha}\Big)
	\,,
	\label{28IV22.2}
\end{equation}
where the parameters should be taken as  $\LKmcon_{i}=\LKbcon_{i}=L_i$.

\section{An example:  Blanchet-Damour-type solutions of the Maxwell equations}
\label{s17VI21}

An elegant class of linearised solutions of the Maxwell equations with $\Lambda =0$ can be constructed in analogy to the Blanchet--Damour solution for linearized gravity,  introduced in~\cite{BlanchetDamour}.
The electromagnetic potential $A_{\mu} d x^{\mu}$ in Lorenz gauge satisfies
\begin{equation}\label{17VI21.t1}
	\Box_\eta A_{\nu} = 0
	\,,
	\quad
	\partial_\mu A^{\mu} = 0
	\,.
\end{equation}
Here $\eta$ is the Minkowski metric, taken to be $-(dx^0)^2+(dx^1)^2+(dx^2)^2+(dx^3)^2$ in the coordinates of \eqref{17VI21.t1}, and $\Box_\eta$ the associated wave operator.
As in~\cite{BlanchetDamour} we start with an ansatz for the electromagnetic potential in Lorenz gauge: given a collection of smooth functions $I_{i}:\R\to \R$, the one-form
\begin{eqnarray}
	A_{t} &=&  \partial_j
	\big(
	\frac{\dot I_{j}(t-r) - \dot I_{j}(t+r)}{r}
	\big)
	\nonumber
	\\
	&=&
	-\big(
	\ddot I_{j}(t-r) +  \ddot I_{j}(t+r)
	\big)
	\frac{x^j }{r^2}
	+ O(r^{-2})
	\,,
	\\
	A_{j} &=&   \frac{\ddot I_{j}(t-r) - \ddot I_{j}(t+r)}{r}
	\,,
	\label{17VI21.t2}
\end{eqnarray}
where each dot represents a derivative with respect to the argument of $I_{i}$,
is a smooth tensor field on Minkowski spacetime solving \eqref{17VI21.t1}.

Since the operators appearing in \eqref{17VI21.t1} commute with partial differentiation, further solutions can be constructed by applying $\partial_{\mu_1}\cdots \partial_{\mu_\ell}$ to $A_{\mu}$, and by applying Poincar\'e transformations.

\bigskip

\noindent {\sc Acknowledgements:}
We are grateful to Jacek Jezierski and Jerzy Kijowski for useful discussions.

\bibliographystyle{amsplain}

\bibliography{ChruscielSmolka-minimal}

\providecommand{\bysame}{\leavevmode\hbox to3em{\hrulefill}\thinspace}
\providecommand{\MR}{\relax\ifhmode\unskip\space\fi MR }
\providecommand{\MRhref}[2]{%
  \href{http://www.ams.org/mathscinet-getitem?mr=#1}{#2}
}
\providecommand{\href}[2]{#2}
\begin{thebibliography}{10}

\bibitem{IBBBook}
I.~Bialynicki-Birula and Z.~Bialynicka-Birula, \emph{{Quantum
  Electrodynamics}}, Pergamon Press, 1975.

\bibitem{BlanchetDamour}
L.~Blanchet and T.~Damour, \emph{Radiative gravitational fields in general
  relativity. {I}. {G}eneral structure of the field outside the source},
  Philos.\ Trans.\ Roy.\ Soc.\ London Ser.\ A \textbf{320} (1986), 379--430.
  \MR{874095}

\bibitem{BrownHenneaux}
J.D. Brown and M.~Henneaux, \emph{{On the Poisson} brackets of differentiable
  generators in classical field theory}, Jour.\ Math.\ Phys. \textbf{27}
  (1986), 489--491.

\bibitem{ChHMS}
P.T. Chru\'{s}ciel, Sk~J. Hoque, M.~Maliborski, and T.~Smo{\l}ka, \emph{{On the
  canonical energy of weak gravitational fields with a cosmological constant
  $\Lambda\in \R$}}, Eur.\ Phys.\ Jour.\ C \textbf{81} (2021), 696 (48 pp.),
  arXiv:2103.05982v2 [gr-qc].

\bibitem{ChIfsits}
P.T. Chru\'{s}ciel and L.~Ifsits, \emph{{The cosmological constant and the
  energy of gravitational radiation}}, Phys.\ Rev.\ D \textbf{93} (2016),
  124075 (40 pp.), arXiv:1603.07018 [gr-qc].

\bibitem{CJK}
P.T. Chru\'{s}ciel, J.~Jezierski, and J.~Kijowski, \emph{{H}amiltonian field
  theory in the radiating regime}, Lect. Notes in Physics, vol. m70, Springer,
  Berlin, Heidelberg, New York, 2002. \MR{1903925}

\bibitem{Compere}
G.~Comp\`ere, A.~Fiorucci, and R.~Ruzziconi, \emph{{The $\Lambda$-BMS$_4$
  Charge Algebra}}, JHEP \textbf{10} (2020), 205, arXiv:2004.10769 [hep-th].

\bibitem{FischerDeSitter}
K.~Fischer, \emph{Interpretation of {E}instein's theory of gravitation
  including the cosmological term as a de {S}itter-invariant field theory on
  the de {S}itter space}, Z.\ Physik \textbf{229} (1969), 33--43. \MR{0255216}

\bibitem{Freidel:2021dxw}
L.~Freidel, \emph{{A canonical bracket for open gravitational system}},
  (2021), arXiv:2111.14747 [hep-th].

\bibitem{JKW}
J.~Jezierski, J.~Kijowski, and P.~Waluk, \emph{Gauge-invariant quadratic
  approximation of quasi-local mass and its relation with {Hamiltonian} for
  gravitational field}, Class.\ Quantum Grav. \textbf{38} (2021), 095006.

\bibitem{KijowskiGRG}
J.~Kijowski, \emph{A simple derivation of canonical structure and quasi-local
  {Hamiltonians} in general relativity}, Gen.\ Rel.\ Grav. \textbf{29} (1997),
  307--343. \MR{1439857 (97m:83029)}

\bibitem{KijowskiTulczyjew}
J.~Kijowski and W.M. Tulczyjew, \emph{A symplectic framework for field
  theories}, Lecture Notes in Physics, vol. 107, Springer, New York,
  Heidelberg, Berlin, 1979. \MR{549772 (81m:70001)}

\bibitem{PooleSkenderisTaylor}
A.~Poole, K.~Skenderis, and M.~Taylor, \emph{{Charges, conserved quantities,
  and fluxes in de {S}itter spacetime}}, Phys. Rev. D \textbf{106} (2022),
  no.~6, L061901, arXiv:2112.14210 [hep-th].

\bibitem{SolovyevI}
V.O. Solovyev, \emph{Boundary values as {H}amiltonian variables. {I}. {N}ew
  {P}oisson brackets}, Jour.\ Math.\ Phys. \textbf{34} (1993), 5747--5769.
  \MR{1246246}

\end{thebibliography}
\end{document}